\documentclass[aps,nofootinbib,notitlepage,twocolumn,superscriptaddress]{revtex4-1}
\usepackage{amsfonts,amsmath,amssymb,amsthm}
\usepackage{array,bm,color}
\usepackage{epsfig,graphicx,nomencl,revsymb4-1,upgreek,url}
\usepackage{hyperref}
\usepackage{algpseudocode}
\usepackage{xspace}
\usepackage{tabularx}
\usepackage[normalem]{ulem}
\usepackage{changes}
\usepackage{enumitem}
\usepackage{booktabs}
\usepackage{mathtools}

\DeclarePairedDelimiter{\floor}{\lfloor}{\rfloor}
\newcolumntype{Y}{>{\centering\arraybackslash}X}
\newcolumntype{C}{>{$}c<{$}}
\AtBeginDocument{
\heavyrulewidth=.08em
\lightrulewidth=.05em
\cmidrulewidth=.03em
\belowrulesep=.65ex
\belowbottomsep=0pt
\aboverulesep=.4ex
\abovetopsep=0pt
\cmidrulesep=\doublerulesep
\cmidrulekern=.5em
\defaultaddspace=.5em
}

\usepackage{pifont}

\graphicspath{{./figures/}}
\hypersetup{colorlinks=true, citecolor=blue, pdfauthor=Kevin C. Young, pdftitle=SimultaneousGST}

\newcommand*{\ket}[1]{\ensuremath{\left\vert{#1}\right\rangle}}

\newcommand*{\ketbra}[2]{\left| #1 \right\rangle\!\!\!\,\left\langle #2 \right|}

\newcommand*{\expect}[1]{\ensuremath{\left\langle{#1}\right\rangle}}

\newcommand{\rot}{\ensuremath{\theta_0}}
\newcommand{\qubit}[1]{Q{#1}}
\newcommand{\pygsti}{\texttt{pyGSTi}\xspace}
\newcommand{\gate}{\mathcal{G}}

\newcommand{\gateX}{\ensuremath{\mathsf{X}_{\pi/2}}\xspace}
\newcommand{\gateY}{\ensuremath{\mathsf{Y}_{\pi/2}}\xspace}
\newcommand{\gateI}{\ensuremath{\mathsf{I}}\xspace}

\renewcommand{\boxed}[1]{\fbox{\begin{minipage}[c][1.5em][c]{1.5em}\centering\ensuremath{#1}\end{minipage}}}

\newcommand{\cellcolor}[1]{} 

\newcommand{\note}[1]{{\color{red}[#1]}}
\newcommand{\kcy}[1]{{\color{red}[#1]\textsubscript{KCY}}}
\newcommand{\kmr}[1]{{\color{purple}[#1]\textsubscript{KMR}}}
\newcommand{\mdg}[1]{{\color{blue}[#1]\textsubscript{MDG}}}
\newcommand{\tjp}[1]{{\color{magenta}[#1]\textsubscript{TJP}}}
\renewcommand{\note}[1]{} 
\renewcommand{\kcy}[1]{}
\renewcommand{\kmr}[1]{}
\renewcommand{\mdg}[1]{}
\renewcommand{\tjp}[1]{}

\begin{document}
\title{Experimental Characterization of Crosstalk Errors with \texorpdfstring{\\}{} Simultaneous Gate Set Tomography}

\author{Kenneth Rudinger}
\affiliation{Quantum Performance Laboratory, Sandia National Laboratories, Albuquerque, NM 87185 and Livermore, CA 94550}
\author{Craig W. Hogle}
\affiliation{Sandia National Laboratories, Albuquerque, NM 87185}
\author{Ravi K. Naik}
\affiliation{Quantum Nanoelectronics Laboratory, Department of Physics, University of California at Berkeley, Berkeley, CA 94720}
\author{Akel Hashim}
\affiliation{Quantum Nanoelectronics Laboratory, Department of Physics, University of California at Berkeley, Berkeley, CA 94720}
\author{Daniel Lobser}
\affiliation{Sandia National Laboratories, Albuquerque, NM 87185}
\author{David I. Santiago}
\affiliation{Quantum Nanoelectronics Laboratory, Department of Physics, University of California at Berkeley, Berkeley, CA 94720}
\affiliation{Computational Research Division, Lawrence Berkeley National Lab, Berkeley, CA 94720}
\author{Matthew D. Grace}
\author{Erik Nielsen}
\author{Timothy Proctor}
\author{Stefan Seritan}
\affiliation{Quantum Performance Laboratory, Sandia National Laboratories, Albuquerque, NM 87185 and Livermore, CA 94550}
\author{Susan M. Clark}
\affiliation{Sandia National Laboratories, Albuquerque, NM 87185}
\author{Robin Blume-Kohout}
\affiliation{Quantum Performance Laboratory, Sandia National Laboratories, Albuquerque, NM 87185 and Livermore, CA 94550}
\author{Irfan Siddiqi}
\affiliation{Quantum Nanoelectronics Laboratory, Department of Physics, University of California at Berkeley, Berkeley, CA 94720}
\affiliation{Computational Research Division, Lawrence Berkeley National Lab, Berkeley, CA 94720}
\affiliation{Materials Sciences Division, Lawrence Berkeley National Lab, Berkeley, CA 94720}
\author{Kevin C. Young}
\email[Corresponding author: ]{kyoung@sandia.gov}
\affiliation{Quantum Performance Laboratory, Sandia National Laboratories, Albuquerque, NM 87185 and Livermore, CA 94550}

\date{\today}

\begin{abstract}
\noindent Crosstalk is a leading source of failure in multiqubit quantum information processors. It can arise from a wide range of disparate physical phenomena, and can introduce subtle correlations in the errors experienced by a device. Several hardware characterization protocols are able to detect the presence of crosstalk, but few provide sufficient information to distinguish various crosstalk errors from one another. In this article we describe how gate set tomography, a protocol for detailed characterization of quantum operations, can be used to identify and characterize crosstalk errors in quantum information processors. We demonstrate our methods on a two-qubit trapped-ion processor and a two-qubit subsystem of a superconducting transmon processor. 
\end{abstract}

\maketitle

\section{Introduction}
\label{sec:introduction}
\noindent
Quantum information processors have demonstrated one- and two-qubit quantum operations with error rates below the threshold required for fault-tolerant quantum computation \cite{Chow2012-ey, Barends2014-ap, Gaebler2016-rp, Ballance2016-ab, Blume-Kohout2017-ww, Barends2019-mt, Negirneac2020-eg, Srinivas2021-fm, Li2019-uf,Sung2020-jv}. One of the biggest obstacles to achieving similarly low error rates in large, integrated quantum processors is the appearance of a large class of errors known collectively as \emph{crosstalk} \cite{Brink2018-la, Debroy2019-nd, Gessner2020-vs, Huang2020-jo, Sarovar2019-gc}. Crosstalk can increase error rates on individual qubits, and can also cause errors on different qubits to become correlated with one another. These correlations are particularly damaging for error correction \cite{landahl2011fault, devitt2013quantum, fowler2012surface}, and optimizing the power of quantum error correction requires understanding and strictly controlling crosstalk errors.

The underlying physical causes of crosstalk errors in quantum information processors are diverse. Perhaps the most familiar source is pulse spillover, wherein a control pulse (i.e., laser, RF signal, etc.) on a target qubit unintentionally affects a neighboring qubit. But crosstalk errors can also occur due to, e.g., coherent coupling between qubits, shared quantum environments, or even shared classical environments that experience spatially correlated fluctuations. To reduce or mitigate crosstalk errors~\cite{Buterakos2018-sk,Carvalho2020-mi,Chen2019-hf,Crain2019-eq,Debroy2020-ui,Ding2020-bd,Hashim2020-ub,Ku2020-ix,Majumder2020-ut,Mundada2018-qm,Murali2020-ab,Seif2018-sj,Sheldon2016-js,Sundaresan2020-re,Watson2018-la,Winick2020-pm,Xu2020-rs} and enable fault-tolerant quantum computation, experimentalists need characterization methods that provide detailed information about the specific crosstalk errors that occur in their processors.

\begin{figure}[!ht]
  \includegraphics[keepaspectratio=true,width=0.79\columnwidth]{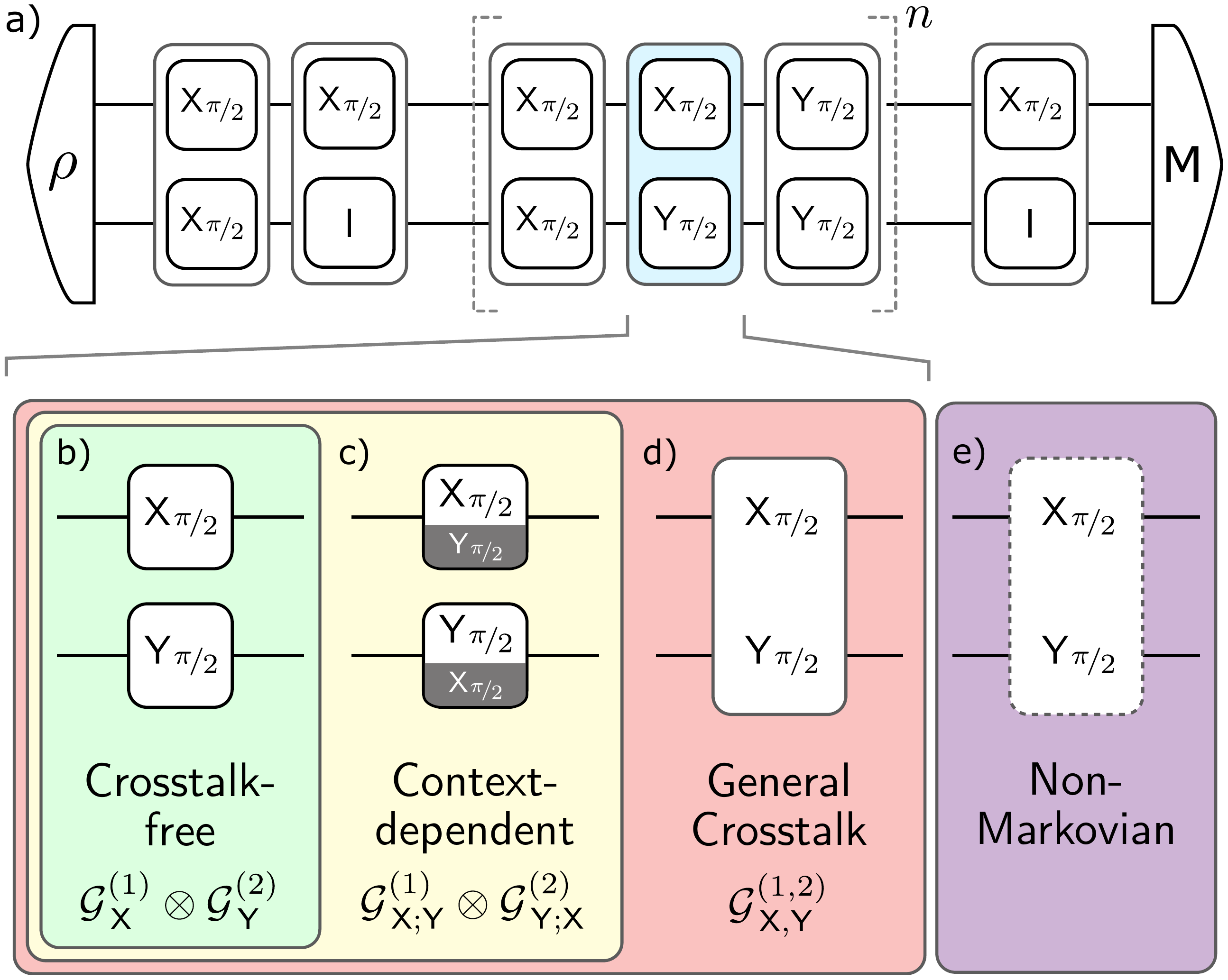}
  \caption{\textbf{Detecting and measuring crosstalk errors with gate set tomography.}  To probe crosstalk between two qubits, we execute a set of two-qubit quantum circuits \textbf{(a)} whose layers comprise parallel, single-qubit gates. These circuits consist of initialization into $\rho\approx\ketbra{00}{00}$, a short state-preparation operation, an $n$-fold repeated \emph{germ} operation, a short measurement-preparation operation, and finally measurement in the computational basis. The measured outcomes of these circuits are used to fit each of three models {\textbf{(b-d)}}: \textbf{(b)} Crosstalk-free models assume that each elementary gate can be described by a single-qubit process matrix. \textbf{(c)} Context-dependent models assume that each gate can be described by a single-qubit process matrix conditioned on the neighboring qubit operation. \textbf{(d)} General crosstalk models assume each two-qubit layer can be described by a full, two-qubit process matrix with no additional constraints. \textbf{(e)} Non-Markovian behavior cannot be described by two-qubit process matrices. We use tools from statistical model selection to decide which model offers the best balance between accuracy (describing the data) and simplicity (using the fewest parameters).}
  \label{fig:fig1}
\end{figure}

A number of techniques to characterize~\cite{Abrams2019-di,Balasiu2018-md,Kelly2014-qz,Krinner2020-ud,Piltz2014-bk,Gambetta2012-lt,McKay2020-jl,Erhard2019-wk,Harper2019-ab,Proctor2017-io,Kueng2016-kj,Huang2019-bh,Sarovar2019-gc,rudinger2019probing,Dumitrescu2016-so,mohseni2008quantum,merkel2013self,solgun2019simple} the impact of crosstalk have been developed and implemented. Randomized methods --- such as simultaneous randomized benchmarking (RB) \cite{Gambetta2012-lt}, correlated RB \cite{McKay2020-jl}, cycle benchmarking \cite{Erhard2019-wk}, and Pauli noise learning \cite{Harper2019-ab} --- are among the most popular, as they are generally simple to implement and analyze. However, these methods are typically sensitive to coherent errors at only second order \cite{Proctor2017-io,Kueng2016-kj,Huang2019-bh}, and rely on twirling techniques that obfuscate the underlying physical sources of observed errors. Model-free methods based on hypothesis testing of probability distributions \cite{Sarovar2019-gc, rudinger2019probing} can identify the presence of crosstalk errors, but cannot characterize those errors. Methods based on quantum process tomography, such as selective process tomography \cite{Dumitrescu2016-so} and direct characterization of quantum dynamics \cite{mohseni2008quantum}, can be adapted to provide insight into certain types of crosstalk (such as coherent two-qubit interactions) but are not designed to detect others (such as how gate errors depend on neighboring operations), and inherit many of the well-known problems of quantum process tomography \cite{merkel2013self}. Finally, specialized techniques, like cross-Rabi oscillations or direct capacitance measurements \cite{solgun2019simple} are useful for learning specific physical parameters, but they are generally unable to detect other crosstalk errors that may be present, or even dominant, in a system. 

In this article we demonstrate how to use gate set tomography (GST)
\cite{merkel2013self,Blume-Kohout2017-ww,Nielsen2020-th}
to perform a detailed investigation of crosstalk errors between two subsystems of a quantum information processor. GST is a protocol designed to provide detailed characterization of qubit dynamics by estimating a set of process matrices describing the various operations of a processor. So in principle, we could simply perform GST on a multi-qubit system, obtain multi-qubit process matrices describing the gates, and look for the presence of crosstalk errors in these process matrices. But such process matrices are large and unwieldy.  It is not clear what constitutes conclusive evidence for crosstalk errors, nor how to reliably distinguish real effects from statistical noise (finite-sample fluctuations).

Instead, we take a two-stage approach.  First, we construct three parameterized models that, by design, allow for different degrees and kinds of crosstalk errors (see Table~\ref{tab:models} and Fig.~\ref{fig:fig1} b-d).  We fit them to data, and we evaluate how well they explain the data \cite{Nielsen2021-Onion}.  We can infer a surprising amount of information just from this evaluation, because we know exactly what kinds of crosstalk each model can describe, and for each model we can quantify the amount of observed error that it failed to describe.  We also use this analysis to select the simplest model that fits the data well, estimate its parameters to obtain ``best fit'' process matrices describing the gates and their errors, and analyze those process matrices in detail to understand the nature of the crosstalk errors.  When the best-fit model only includes certain effects, we can apply customized analysis techniques that are tailored to those particular crosstalk errors.  Finally, by comparing the measured errors and their magnitudes against candidate physical device models, we draw conclusions about the underlying physical causes of the observed crosstalk errors --- and, potentially, how to mitigate them. 

In this work we focus on identifying and analyzing crosstalk induced by parallel single-qubit gates on two single-qubit subsystems. However, all of our methods are straightforwardly extensible to larger systems, e.g., to crosstalk between two-qubit subsystems, induced by entangling gates. We demonstrate our methods on two DOE-sponsored quantum computing testbed platforms --- the transmon-based Advanced Quantum Testbed (AQT) \cite{AQT-wp} and a prototype of the trapped-ion-based Quantum Scientific Computing Open User Testbed (QSCOUT)  \cite{Clark2020-jr,QSCOUT-wp} --- and we discuss and compare the errors observed on these two devices.

\section{Modeling crosstalk errors}
\label{sec:formal}
Crosstalk in quantum information processors can arise from a wide range of disparate physical mechanisms.  Our goal in this work is to characterize the crosstalk errors in quantum logical operations induced by those mechanisms. The standard model for any Markovian error, including crosstalk errors \cite{Sarovar2019-gc}, is a process matrix --- a completely positive, trace preserving map on density matrices.  Multi-qubit process matrices can describe a wide variety of crosstalk effects.  To disaggregate different categories of crosstalk errors, we construct a hierarchy of process matrix models, each of which can only represent certain kinds of crosstalk errors.  We will fit these models to data, compare their ability to describe the observations, select the best model, and use it to draw quantitative conclusions about the crosstalk errors present in the device.  In this section, we present (1) those models, (2) our methods for fitting them to data and evaluating fit quality, (3) how we analyze an estimated model, and (4) the data/model analysis pipeline used in our experiments.

\subsection{Families of crosstalk error models}

Reference \cite{Sarovar2019-gc} defines two conditions that must be satisfied for a quantum processor to be free of crosstalk:
\begin{enumerate}[label=(\roman*)]
    \item Locality of operations --- the process matrices decompose as tensor products;
    \item Independence of local operations --- each component in the tensor product depends only on what gate is acting on that subsystem. 
\end{enumerate}
Each condition defines a constraint on multi-qubit process matrices, which can otherwise describe many forms of crosstalk.  By systematically enforcing or relaxing constraints (i-ii), we can construct the three families of models shown in Table.~\ref{tab:models}. They describe increasingly complex crosstalk phenomena.

\setlength{\extrarowheight}{3pt}
\begin{table}[t!]
\begin{tabular*}{\columnwidth}{l@{\extracolsep{\fill}}cccc} 
 \toprule
   & & & \multicolumn{2}{c}{Constraints}    \\
     \cmidrule(r){4-5}
 Model  & Decomposition & $N_p$ & (i)  & (ii) \\
 \midrule
 Crosstalk-free         & $\gate_{\mathsf{A}}^{(1)}\otimes\gate_{\mathsf{B}}^{(2)}$                     & 86 & \checkmark & \checkmark \\
 Context-dependent      & $\gate_{{\mathsf{A}};{\mathsf{B}}}^{(1)}\otimes\gate_{\mathsf{B};{\mathsf{A}}}^{(2)}$  & 240   & \checkmark & \\
 General    & $\gate_{{\mathsf{A}},{\mathsf{B}} }^{(1,2)}$                                  & 1,683    & & \\
 \bottomrule
\end{tabular*}
\caption{\label{tab:models} Three nested families of process matrix models for quantum logic operations. Each successively larger model can capture richer forms of crosstalk between qubits than the previous ones. Also shown are the decompositions of the process matrices as tensor products of local operations when operations $\mathsf{A}$ and $\mathsf{B}$ are applied simultaneously to subsystems $1$ and $2$, respectively. $N_p$ is the number of free parameters required to describe the model.}
\end{table}

The crosstalk-free model cannot model any type of crosstalk in the system --- each gate is required to act locally and independently. The process matrix representing each layer of gates is required to have a tensor-product form, and the local process matrix on each subsystem is independent of its context (i.e., which operations are applied to other subsystems). 

The context-dependent model relaxes the independence constraint.  Each gate still acts locally --- a layer's process matrix must have tensor-product form --- but the local operations on a subsystem can vary from layer to layer (i.e., depend on context).  This model can capture pulse-spillover effects, where the operations on one subsystem are perturbed by operations on the other. But, like the crosstalk-free model, the context-dependent model cannot model entangling Hamiltonians or correlated errors that create correlations between the two subsystems. 

The general crosstalk model relaxes both constraints, and represents each parallel, single-qubit layer as a full two-qubit process matrix. This model can capture all Markovian crosstalk effects, including context dependence, entangling operations, and classical correlation.  

In principle, a complete description of the system's crosstalk errors could be extracted just by estimating the parameters of the general model.  In practice, analyzing those process matrices and deciding which effects are statistically significant is difficult and time-consuming.  We let the model-fitting process do that work for us.  If, e.g., the effects of crosstalk are not statistically significant, then model selection criteria (see below) will indicate that the data are consistent with a crosstalk-free model.  Conversely, if only the general model fits well, then this is unambiguous evidence of nonlocal interactions, which we can track down by detailed analysis of process matrices.

Because our models utilize process matrices, they are all explicitly Markovian. But real devices are often \emph{non-Markovian} --- experiments on them yield data inconsistent with the predictions of any process matrix model. This has implications for the use of process matrix models to diagnose crosstalk. As discussed in \cite{Sarovar2019-gc}, correlations induced by non-Markovianity can easily be mistaken for crosstalk errors, so finding some way to acknowledge and incorporate non-Markovian errors is extremely important! Later in this section we discuss techniques for quantifying non-Markovianity, and what to do when it (almost inevitably) appears. 

To predict data, our models also need to describe errors in state preparation and measurement (SPAM). We only consider systems that initialize and measure all qubits simultaneously.  In each of our models, the density matrices and POVM effects that represent SPAM operations are required to respect the same constraints (e.g. tensor product decomposition) as the gate operations.  Thus, both the crosstalk-free and context-dependent models incorporate initial states of the form $\rho=\rho^{(1)}\otimes\rho^{(2)}$ and POVM effects of the form $F_{i,j} = F^{(1)}_i\otimes F^{(2)}_j$. SPAM operations in the general crosstalk model are unconstrained, and can describe entangled or correlated initial states, and arbitrarily correlated measurements.

\subsection{Fitting models to data}

Fitting process matrix models to data is typically done using some variant of process tomography \cite{mohseni2008quantum}. While standard quantum process tomography can be used to characterize quantum gate operations, it is not \emph{self-consistent}, i.e., it assumes access to input states and measurements which are already highly accurately characterized \cite{merkel2013self}. We use GST, a protocol introduced to solve this self-consistency problem \cite{merkel2013self,Nielsen2020-th, Blume-Kohout2017-ww}. GST experiments include all the circuits required for state, measurement, and process tomography, plus a few additional circuits that establish a mutually consistent reference frame. This allows GST to fully characterize all state preparation, measurement, and gate operations on a processor, concurrently and self-consistently.

As illustrated in Fig.~\ref{fig:fig1}, GST relies on circuits of the form
\begin{equation}
\label{eq:1}
    \rho\;\rule[.2em]{10pt}{1pt}\boxed{\ensuremath{\mathsf{p}_i}}\rule[.2em]{10pt}{1pt}\boxed{\ensuremath{\mathsf{g}_j^n}}\rule[.2em]{10pt}{1pt}\boxed{\ensuremath{\mathsf{m}_k}}\rule[.2em]{10pt}{1pt} \; \ensuremath{\mathsf{M}}
\end{equation}
where $\mathsf{p}_i$ is a fiducial state preparation subcircuit, $\mathsf{m}_k$ is a fiducial measurement preparation subcircuit, and $\mathsf{g}_k$ is an $n$-fold repeated ``germ'' subcircuit. Data from running these circuits can be used to estimate a process matrix for each distinct circuit layer \cite{Nielsen2020-th}. In our experiments, we use and estimate the nine 2-qubit layers formed by all possible parallel combinations of a 3-element single-qubit gate set:
\begin{equation}
\mathbb{G} = \left\{a\otimes b : a, b \in \{\gateX, \gateY, \mathsf{I} \}\right\},
\end{equation}
where $\gateX$ and $\gateY$ are $\pi/2$ rotations around $X$ and $Y$, respectively, and $\mathsf{I}$ is an idle gate.
Our experiments use no entangling gates. As in standard GST \cite{Blume-Kohout2017-ww, Nielsen2020-th}, the fiducial operations are chosen to be informationally complete, and the germs are chosen to amplify all observable components of a full two-qubit process matrix model for each layer.  See Table~\ref{tab:sgst_circuits} for the specific subcircuits used in our experiments.

Once data has been obtained, we use maximum likelihood estimation to find a best-fit estimate for each of the three crosstalk models, by varying its parameters $\boldsymbol{\theta}$ to maximize the likelihood function $\mathcal{L}(\boldsymbol{\theta}) = \mathrm{Pr}(\mathrm{data}|\boldsymbol{\theta})$.  We denote the maximum likelihood estimate of the $i$th model by $\widehat{\boldsymbol{\theta}_i}$. We denote the maximum value of the likelihood function for the $i$th model by $\mathcal{L}^{(i)} = \mathcal{L}(\widehat{\boldsymbol{\theta}_i})$, and refer to it as the likelihood of model $(i)$.

To construct, fit, and analyze these models, we use \pygsti \cite{nielsen2019python,nielsen2020probing}, a Python implementation of GST that includes robust routines for fitting predictive models of quantum information processors to data, and analyzing the resulting estimates. 

\subsection{Comparing, selecting, and validating models}

How well a candidate model $(i)$ fits data is captured by its likelihood, $\mathcal{L}^{(i)}$.  Extracting useful information requires some simple manipulations.  We measure the quality of model $(i)$'s fit by the log-likelihood ratio between it and a ``maximal model'' that has no structure at all, and can assign an independent probability to each measurement outcome $a$ of each circuit $b$,
\begin{equation}
    \lambda^{(i)} = -2 \ln\left(\mathcal{L}^{(i)} / \mathcal{L}_{\rm max} \right), \label{eq:llr}
\end{equation}
where the maximum likelihood of the maximal model,
\begin{equation}
\mathcal{L}_{\rm{max}} = \prod_{a,b} f_{a,b}^{N_b f_{a,b}},
\end{equation}
is achieved by predicting the observed frequency $f_{a,b}$ after $N_b$ measurements of circuit $b$. The log-likelihood ratio is a standard hypothesis-testing statistic.  Wilks' theorem \cite{wilks1938large} states that when the data are actually generated by model $(i)$ with some parameters $\boldsymbol{\theta}$, $\lambda^{(i)}$ is a $\chi^2_k$ random variable, with $k$ equal to the difference between the number of free parameters ($N_p$) for model $(i)$ and the maximal model. Under this null hypothesis, $\left\langle\lambda^{(i)}\right\rangle=k$ and $\Delta \lambda^{(i)} = \sqrt{2k}$. 

The data are inconsistent with model $(i)$ if and only if $\lambda^{(i)}$ is inconsistent with a $\chi^2_k$ distribution.
In particular, when the data are not consistent with model $(i)$, $\lambda^{(i)}$ will be larger than its expected value under the null hypothesis, $k$. How much model $(i)$ is violated can be quantified by the number of standard deviations by which $\lambda^{(i)}$ exceeds its expected value under the null hypothesis, 
\begin{equation}
N_{\rm{\sigma}}^{(i)} = \frac{\lambda-k}{\sqrt{2k}}.
\end{equation}

Ideally, we would simply choose the smallest model that fits the data --- i.e., the smallest model for which $N_\sigma$ is negligible.  But in practice, most systems display enough non-Markovian behavior that no model --- not even the general (process matrix) model --- fits the data that well.  In this case, we need a criterion for identifying how much better (or worse) one model is than another.

To derive such a criterion \cite{Nielsen2021-Onion}, we observe that Wilks' theorem implies that removing exactly $n$ ``useless'' parameters from a model increases $\expect{\lambda}$ by exactly $n$.  So, if we consider two equally valid models $(i)$ and $(j)$, with $(j)$ nested within $(i)$ and having $n$ fewer parameters than it, then we expect $\lambda^{(j)} - \lambda^{(i)} \approx n$.  If we observe $\lambda^{(j)} - \lambda^{(i)} \gg n$, this suggests that the extra parameters in the larger model $(i)$ are not useless --- i.e., they describe real effects. But Akaike's derivation of his eponymous AIC \cite{Akaike1974-xt} demonstrates a scenario where using the larger model to fit those effects actually \emph{decreases} predictive accuracy, unless $\lambda^{(j)} - \lambda^{(i)} \geq 2n$.  We conclude that although there are multiple criteria for deciding which model is ``better'' in a given situation, they share a simple form:  Is $\lambda^{(j)} - \lambda^{(i)} \geq \alpha n$ for some $\alpha$?

To compare two nested models\footnote{Our crosstalk models form a hierarchy, so they are strictly nested --- for any pair of them, one contains the other as a subset.}, we use a quantity that we call the \emph{evidence ratio} of $(i)$ against the smaller model $(j)$ \cite{Nielsen2021-Onion}:
\begin{equation}
    \gamma^{(i,j)}  = \left(\frac{\lambda^{(j)} - \lambda^{(i)}}{N_p^{(i)} - N_p^{(j)}}\right),
\end{equation}
where $N_p^{(i)}$ and $N_{p^{(j}}$ indicate the number of parameters in models $(i)$ and $(j)$, respectively.  If $\gamma^{(i,j)} \leq 1$, then there is no evidence against the smaller model $(j)$ --- the larger model $(i)$'s extra parameters are functionally useless --- and so we always choose $(j)$.  If $1 < \gamma^{(i,j)} \leq 2$, then the data provides weak evidence against the smaller model, but the AIC suggests its predictions would still be more accurate.  Even when $\gamma^{(i,j)}>2$, we may still choose the smaller model if we prioritize simplicity, but for any use case there will be some threshold beyond which the smaller model must be rejected.  In general, $\gamma^{(i,j)}$ normalizes the weight of evidence against the smaller model, on a per-parameter basis, and provides a quantitative measure for comparing two models.

\subsection{Quantifying unmodeled error}
\label{sec:wildcard}

Statistical measures of model violation like $N_\sigma^{(i)}$ and $\gamma^{(i,j)}$ quantify the amount of evidence for errors outside a given model.  They do not quantify the magnitude of those errors.  For example, they depend strongly on the amount of data taken.  In many circumstances, we care more about the size of unmodeled errors than about the amount of evidence that they exist. In this work, we quantify the size of unmodeled errors using \emph{wildcard error} \cite{blume2020wildcard}.

Wildcard error can quantify the per-gate deviation between a model's predictions and observed data. To do this, we assign a \emph{minimal wildcard model} to an estimate.  A wildcard model assigns to each estimated gate $g$ a number $w_g\geq0$, and to each circuit $C$ the total $w$ for all the gates in it:  $w_C = \sum_{g\in C}{w_g}$.  Adding a wildcard model explicitly relaxes the estimate's prediction for each circuit $C$:  if the estimate originally predicted outcome distribution $\vec{p}_C$, then the wildcard-augmented estimate predicts only that $C$'s outcomes will be drawn from \emph{some} $\vec{p}'_C$ such that $\|\vec{p}_C - \vec{p}'_C\|\leq w_C$.  A minimal wildcard model is an assignment $\{w_g\}$ that just barely makes the estimate statistically consistent with the data.  We only use single-parameter wildcard models that assign a single wildcard error rate ($W_{(i)}$) to all gates in the estimate of a model $(i)$.

The minimal amount of $W$ required to reconcile an estimate with data tells us whether unmodeled errors are dominant or negligible. In this work we use very simple wildcard models that assign a single wildcard error rate ($W_{(i)}$) to all gates in a model $(i)$. If the $W$ assigned to the gates in a model's estimate is significantly less than their average diamond error ($\bar \epsilon_\diamond$) \cite{kitaev1997quantum,watrous2009semidefinite,sanders2015bounding}, that model explains most of the observed error.  But if $W \geq \bar\epsilon_\diamond$, unmodeled errors may be dominant, and the model should probably be discarded or not taken seriously.

Unmodeled errors --- heralded by significant $W$ --- can appear in our analysis from two distinct causes.  If the general model cannot fit the data, then its unmodeled errors constitute some sort of non-Markovian dynamics, since the general model (by construction) can model all Markovian errors on the gates.  Any non-Markovian effect will also go unmodeled by the smaller models (crosstalk-free and context-dependent).  But certain crosstalk errors are also excluded by those models (again, by construction).

When data show evidence of non-Markovianity (as is often the case), none of the three models will fit the data well.  But we can use wildcard error analysis to roughly estimate the magnitude of crosstalk errors even in the presence of non-Markovianity.  To do so, we assign wildcard error to each model.  The general model serves as a baseline; only non-Markovian errors contribute to its $W$.  A smaller model's $W$ accounts for both non-Markovian errors \emph{and} the crosstalk errors excluded by that model.  If the smaller model's $W$ is significantly higher, that indicates the presence of crosstalk errors that are not dominated by non-Markovianity.  We will see examples of this scenario in the experimental data.

\subsection{Metrics}
\label{sec:metrics}

The analysis in the preceding section can provide extensive high-level information about whether whole classes of crosstalk error are present or absent, and about their overall magnitude.  But it also lets us select one of the three models as the best fit to the data --- i.e., the one that best balances simplicity and explanatory power.  Once this model has been selected, we examine the process matrices that it assigns to each gate.  We can extract detailed performance metrics, identify dominant error channels, and/or use the process matrices to predict the processor's performance on specific tasks and benchmarks. 

Reductive gate error metrics like diamond distance or entanglement infidelity provide rough summaries of system performance, but to probe the details of estimated error models we transform process matrices to \emph{error generators} \cite{Blume-Kohout2021-Taxonomy}:
\begin{equation}\label{eq:generator}
    \mathcal{G} = e^{\mathcal{L}} \circ e^{\mathcal{H} + \Delta\mathcal{H}}
\end{equation}
where $\mathcal{G}$ is a gate's process matrix, $\mathcal{H}$ is a Hamiltonian superoperator that generates a perfect unitary implementation of the gate, $\Delta\mathcal{H}$ is a Hamiltonian error generator that generates the gate's unitary errors\footnote{This is a slightly different error generator representation than the one presented in Ref.~\cite{Blume-Kohout2021-Taxonomy}; we use \emph{during-gate} generators for (only) the Hamiltonian sector, because it is more convenient.}, and $\mathcal{L}$ is a non-unitary error generator that describes all non-unitary errors in the gate (see Ref.~\cite{Blume-Kohout2021-Taxonomy} for extensive discussion).

Gate sets like the ones we analyze here have a gauge freedom \cite{Rudnicki2018-bn, Lin2019-ck, Nielsen2020-th}; some of their parameters have no physical consequences and are unobservable.  Gauge degrees of freedom appear in error generator representations as unobservable linear combinations of error generator coefficients.  When we construct and examine estimates in this article, we manifest the gauge freedoms explicitly as unobservable constant offsets, and we measure crosstalk errors using strictly gauge-invariant properties constructed as differences between two coefficients with identical gauge freedoms.  Gauge-invariant parameters of the Hamiltonian error generator include:
\begin{enumerate}
    \item The coefficient (rate) of any error generator that commutes with the target gate, including:
    \begin{enumerate}
        \item The entire $\Delta\mathcal{H}$ for an idle operation,
        \item Over/under rotation angles of any active gate,
    \end{enumerate}
    \item The angle between the rotation axes of any two active gates,
    \item The change in $\Delta \mathcal{H}$ between the same gate acting in two different contexts.
\end{enumerate}
These will be sufficient for our analysis.

\subsection{Testing quantum information processors for crosstalk}

The previous sections each discussed an important element of a robust method for identifying and characterizing crosstalk errors in a quantum information processor. In this section we outline the steps we took for end-to-end characterization of crosstalk errors in the two experimental platforms discussed in the next section. For each processor, we collected all data in one contiguous experiment (details are given below), but here we present a step-by-step procedure for clarity.

First, to obtain a rough estimate of local and crosstalk error rates, we performed and analyzed a form of simultaneous RB \cite{Gambetta2012-lt}. Simultaneous RB involves three distinct RB experiments: running RB on subsystem (1) while idling subsystem (2); idling (1) while running RB on (2); and running RB on (1) and (2) simultaneously. This yields two error rates ($r_i$ and $r_s$, from the idle and driving contexts) for each subsystem. The change in each subsystem's RB error rate ($r_s-r_i$), when the other subsystem is driven instead of idled, provides an estimate of how much error the gates on one subsystem induce on its neighbor. Simultaneous RB can be implemented with any variant of RB; we used \emph{direct} RB (DRB), a variant of standard RB in which the Clifford RB circuits \cite{magesan2011scalable} are replaced by uncompiled circuits over a system's native gates \cite{proctor2019direct}. 

Next, we analyzed GST data. We fit our three models to this data (using \pygsti), and evaluated their fit quality using $N_{\sigma}$, evidence ratios, and wildcard error.  We used this information to deduce which forms of crosstalk were present, and to estimate their magnitude. In each use case, we then selected the model that best explained the observed data, for further analysis using the techniques of Section~\ref{sec:metrics}. The details of this analysis depends on the model:
\begin{itemize}
    \item Crosstalk-free model: If this model fits, there is no evidence for crosstalk.  Each single-system gate is represented by a local process matrix, independent of context.  It can be evaluated in the usual way.
    \item Context-dependent model:  If this model is selected, the gates still act locally, but their action is context-dependent.  The model specifies the action of each single-system gate in several contexts, indexed by which operation is performed on the neighboring subsystem.  The most relevant object of study is the \emph{variation} in a gate's action between contexts, which is gauge-invariant and easily extracted from the local process matrices that describe it in different layers.
    \item General model:  If this model is selected, at least one layer is inducing nonlocal (e.g. entangling or correlated) errors.  Each process matrix must be analyzed to see if it produces correlated errors, and if so, what their kind and magnitude are.
\end{itemize}

\section{Experimental Demonstration}
\label{sec:experimental}

We used our nested crosstalk models to investigate and characterize crosstalk errors on the two DOE Quantum Testbed platforms, AQT \cite{AQT-wp} and the QSCOUT prototype \cite{Clark2020-jr,QSCOUT-wp}. AQT is a transmon-based platform housed at Lawrence Berkeley National Laboratory and UC Berkeley. QSCOUT is a trapped-ion quantum computing platform housed at Sandia National Laboratories.  By characterizing both devices, we are able test the performance of our methods against vastly different physical error sources and experimental limitations. 

We report experimentally measured/estimated values throughout this section.  When possible, uncertainties are given, using concise notation, e.g., $1.234(5)$ indicates $1.234\pm 0.005$.  All uncertainty intervals are $95\%\approx 2\sigma$ confidence intervals, obtained using either bootstrapping or likelihood ratio confidence interval methods.

\subsection{The Advanced Quantum Testbed}
\label{sec:aqt}
Experiments on the AQT platform were performed using an eight-qubit superconducting transmon processor (\texttt{AQT@LBNL Trailblazer8-v5.c2}). The qubits are encoded as the \ket{0} and \ket{1} states of the transmons, and are coupled to their nearest neighbors in a ring geometry. The demonstrations here focus on two next-nearest-neighbor transmons, labeled \qubit{4} and \qubit{6}, whose fundamental transition frequencies range from 5.2 to 5.5 GHz, with anharmonicities around 270 MHz. Each qubit in the device has its own control line for applying $\gateX$ and $\gateY$ gates, while any necessary $\mathsf{Z}$ gates are applied virtually through discrete phase shifts of subsequent $\gateX$ and $\gateY$ gates \cite{McKay2017-fq}. 

\subsubsection{Potential (and actual) sources of crosstalk in AQT}
\label{sec:aqt_sources}
Crosstalk is a well known problem in many transmon-based quantum processors. Two important crosstalk effects are coherent ZZ interactions, induced by shared microwave resonator modes, and pulse spillover. Microwave drive signals are often poorly localized on superconducting chips, so they rely on resonance mismatch to mitigate the impact of control spillover between qubits. However, this does not work perfectly, and the lingering interactions can manifest locally on spectator qubits as unwanted Rabi oscillations (if the drive is close to resonant with the spectator) or dispersive AC Stark shifts (if the drive is off-resonant). For neighboring qubits, such spillover crosstalk might even cause unwanted cross-resonance entangling interactions \cite{patterson2019calibration}. 

Using spectroscopic analysis we found that the \ket{0}--\ket{1} transition frequency of \qubit{4} is nearly resonant with the \ket{1}--\ket{2} transition of \qubit{6}. When microwave drive tones are applied to \qubit{4}, some of this power impinges on \qubit{6} and this can therefore cause an AC Stark shift in the \ket{0}--\ket{1} transition frequency of \qubit{6}. The size of this shift can be measured by driving \qubit{4} on resonance while monitoring the frequency of \qubit{6} via Ramsey spectroscopy. 

We can correct for this drive-dependent Stark shift by adding an explicit \emph{crosstalk compensation} pulse on \qubit{6} that interferes destructively with the spillover pulse. This compensation pulse is optimized by first identifying the phase shift for which the Stark shift of \qubit{6} is maximal, and then finding the relative amplitude that minimizes the error on \qubit{6}. This compensation tone is then built into each active operation that is applied to \qubit{4}.

\subsubsection{Experiment design}

To characterize crosstalk in the AQT platform, we ran simultaneous DRB and GST experiments, both with and without crosstalk compensation. The GST circuit family we used is summarized in Table \ref{tab:sgst_circuits}. We used circuits up to depth $\simeq32$, resulting in 20,577 GST circuits in total. Our DRB circuits were constructed by sampling 30 two-qubit, simultaneous DRB circuits at each of 8 exponentially spaced depths up to a maximum depth of 256. From each of these DRB circuits, we created two additional independent DRB circuits for which all of the gate operations on one or the other qubit were replaced with idles, yielding 710 unique DRB circuits in total (if a circuit was duplicated we simply gathered twice as much data to avoid undersampling).

We combined the GST and DRB circuit sets into a single list and randomized its order (effectively interleaving the RB and GST circuits). Because neither GST nor RB are designed to be reliable when there are large drifts in a processor's behaviour over the duration of the experiment, we gathered data in four batches. Crosstalk compensation was employed in alternate batches, and each batch consisted of running every circuit in the combined circuit list 500 times. Using simple statistical tests described in  \cite{rudinger2019probing}, we confirmed that the data in similar batches were statistically consistent. We then aggregated data from similar batches (with and without crosstalk compensation) into a single dataset containing 1000 shots per circuit. All data was taken over a period of approximately 10 hours. 

Our measurement was able to distinguish between $\ket{0}$, $\ket{1}$ and the leaked state $\ket{2}$. However, the GST and RB techniques used in this work are not designed to characterize leakage, so we discarded any measurements for which either qubit was found in the \ket{2} state. This required discarding approximately $0.3\%$ of all measurements when crosstalk compensation was used, and approximately $0.5\%$ when it was not. 

\subsubsection{Randomized benchmarking}

\begin{figure}
\includegraphics[keepaspectratio=true,width=\columnwidth]{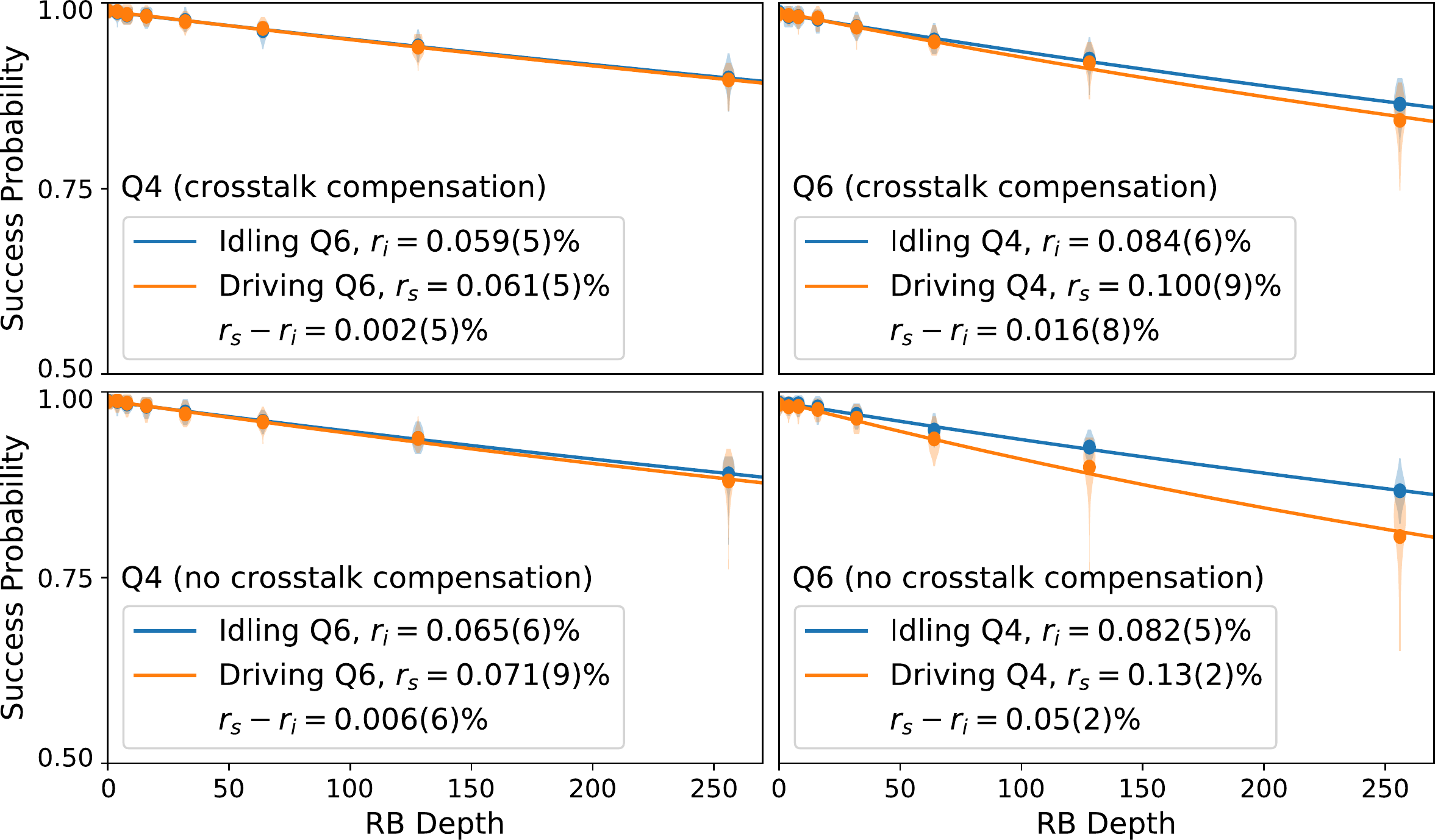}
\caption{\label{fig:aqt_sdrb} \textbf{Simultaneous direct randomized benchmarking (DRB) on the AQT platform.} We ran simultaneous DRB experiments (details in main text) on the AQT device's \qubit{4}-\qubit{6} subsystem both with (top) and without (bottom) crosstalk compensation enabled.  Estimates of each qubit's average error-per-gate were extracted in two contexts: ``idling spectator'' ($r_i$) and ``driving spectator'' ($r_s$).  The increase in each qubit's $r$ due to driving the spectator is a high-level measure of crosstalk-induced errors on that qubit.  Uncertainties are 95\% confidence intervals obtained from bootstrapping.  Violin plot regions around each point indicate the distribution of the 30 individual DRB circuits whose average success probability is represented by the point.  There is no statistically significant evidence of crosstalk-induced errors from \qubit{6}$\to$\qubit{4}.  However, there is significant evidence of crosstalk from \qubit{4}$\to$\qubit{6}, and it is reduced (though not eliminated) by crosstalk compensation.}
\end{figure}

We used simultaneous DRB on AQT's \qubit{4} and \qubit{6} to estimate (1) each qubit's typical error-per-gate, and (2) how much that error rate changed when the other qubit was left idle or driven with random gate sequences.  We call these ``contexts''.  Simultaneous DRB reveals each qubit's error-per-gate in two contexts, the ``spectator driven'' context ($r_s$) and the ``spectator idle'' context ($r_i$).  Its main result is the difference between these error rates ($r_s - r_i$), which we call the \emph{context-to-context variation} of a qubit's error-per-gate. Figure \ref{fig:aqt_sdrb} summarizes the results, showing success probability decay curves for each qubit, in both contexts, and the estimated values of each $r_s$ and $r_i$, with and without crosstalk compensation.

Without crosstalk compensation, we observe a striking asymmetry between the two qubits.  \qubit{4}'s error-per-gate is approximately $0.07\%$ regardless of context ($r_s\approx r_i$ and the estimated change of $0.006(6)\%$ is within error bars of zero). But \qubit{6}'s error-per-gate shows significant context-to-context variation --- it jumps from $0.082(5)\%$ when \qubit{4} is idled, to $0.13(2)\%$ when \qubit{4} is driven.  We conclude that there is significant crosstalk from \qubit{4} to \qubit{6}, but no evidence for crosstalk in the opposite direction.

Enabling crosstalk compensation reduces \qubit{6}'s ``spectator driven'' error-per-gate to $0.100(9)\%$, without significantly changing other decay rates.  So crosstalk compensation reduces the context-to-context variation of \qubit{6}'s error-per-gate from $0.05(2)\%$ to $0.016(8)\%$.

Simultaneous RB demonstrates that crosstalk is present, and that it is significant. It suggests that crosstalk effects are asymmetric (\qubit{4}$\to$\qubit{6}), and that crosstalk compensation reduces them.  But it is hard to draw further conclusions, because simultaneous RB does not reveal how each gate's action depends on context, nor does it distinguish context-dependent errors from entangling or correlated ones.  It is possible to extract some of this information from a more sophisticated analysis of RB data \cite{McKay2020-jl}, but we can answer these questions definitively with GST data and analysis.

\subsubsection{Comparing and selecting crosstalk models}
\label{sec:aqt_selecting}
We began our analysis of the GST data by fitting all three models --- crosstalk-free, context-dependent, and general --- to the two datasets (with and without crosstalk compensation), and evaluating their fit quality. Table~\ref{tab:aqt_results} displays the results. 

None of the three models fit either dataset perfectly.  We evaluated each model's fit quality by its loglikelihood ratio $\lambda$ with respect to the maximal model [see Eq.~\eqref{eq:llr}].  All six fits displayed $N_{\sigma} > 45$, i.e., at least $45\sigma$ of model violation.  This constitutes strong statistical evidence of non-Markovian behavior that cannot be modelled by two-qubit process matrices.  Under these circumstances, neither GST nor RB is guaranteed to be be reliable, and caution is required when interpreting results. 

However, large $N_{\sigma}$ does not imply that non-Markovian errors are dominant.  To compare the rates of modeled and unmodeled errors, we assigned a wildcard error (see Sec.~\ref{sec:wildcard}) to each model's best-fit estimate. The general model's wildcard error rate is less than $0.2\%$, both with and without crosstalk compensation. This is much smaller than the GST model's average diamond distance error rate ($\bar{\epsilon}_{\diamond} = 1.670(3)\%$ without crosstalk compensation, or $\bar{\epsilon}_{\diamond} = 1.097(3)\%$ with crosstalk compensation). We conclude that Markovian errors dominate, and are captured by the largest GST model.

Since the general GST model explains most of the errors observed in the GST data, we investigate whether smaller models --- context-dependent and crosstalk-free --- are equally consistent with the data. We can evaluate their fit relative to the general model using two criteria: evidence ratios, or the change in wildcard error.

The context-dependent model is conclusively accepted by both criteria --- it fits the data as well as the general model despite having more than 1400 fewer parameters.  The evidence ratio between them is $\gamma = 0.9$ ($\gamma = 0.58$) without (with) crosstalk compensation, and the wildcard error increases by only $0.014\%$.  We conclude there is \emph{no} evidence for entangling or correlated crosstalk errors; the context-dependent model's tensor product process matrices describe the observed data as well as possible by any (two-qubit) process matrices.

The crosstalk-free model, however, does not fit the data well.  In the absence of crosstalk compensation, it is overwhelmingly rejected by both criteria --- the evidence ratio between the context-dependent and crosstalk-free models is $\gamma = 495$, and the wildcard error required to reconcile it with the data is increased by more than $5\times$ ($W_{\mathrm{crosstalk-free}}=0.85\%\gg W_{\mathrm{context-dependent}}=0.15\%$).  This constitutes overwhelming evidence that crosstalk errors are present, and their magnitude is large (i.e., they make a substantial contribution to the total gate error rates).

When crosstalk compensation is enabled, the crosstalk-free model fits the data much better.  Its  wildcard error drops substantially ($W_{\mathrm{crosstalk-free}}=0.23\%\approx W_{\mathrm{context-dependent}}=0.20\%$), and the evidence ratio in favor of the context-dependent model drops to $\gamma = 15.6$.  This constitutes clear evidence that crosstalk is still present, but its magnitude and significance are greatly reduced by crosstalk compensation.

Our conclusions from this analysis are:
\begin{itemize}
    \item Crosstalk errors are clearly present, but only local context-dependent errors.
    \item Non-Markovian errors are present at the $0.2\%$/gate level, but are dominated by Markovian errors (and by Markovian crosstalk errors).
    \item Crosstalk compensation reduces crosstalk errors significantly, but it does not eliminate them.
    \item Both with and without crosstalk compensation, the best estimate (process matrices) to examine in detail is the context-dependent estimate.
\end{itemize}

\subsubsection{Extracting detailed crosstalk error rates}

We now examine the best GST estimates (with and without crosstalk compensation) in detail. A GST estimate of the context-dependent model specifies a 2-qubit process matrix for each of the 9 parallel-gate layers (e.g. $\gateX\otimes\mathsf{I}$, $\gateY\otimes\gateX$, etc).  Each 2-qubit process matrix is the tensor product of two 1-qubit process matrices.  Each 1-qubit process matrix describes one of 3 gates ($\gateX$, $\gateY$, or $\mathsf{I}$) acting on one of 2 qubits (\qubit{4} or \qubit{6}), in one of 3 contexts ($\gateX$, $\gateY$, or $\mathsf{I}$ applied to the other qubit).  We represent each 1-qubit process matrix using error generators (Sec.~\ref{sec:metrics}), and examine the context-to-context variation of the error generators for each of the 6 single-qubit gates ($\gateX$, $\gateY$, and $\mathsf{I}$ gates on \qubit{4} and \qubit{6}).

The non-unitary parts of each error generator --- e.g., rates of stochastic errors --- have almost no statistically significant context-to-context variation.  The very small number of variations that \emph{are} statistically significant are comparable in magnitude to the non-Markovian errors (measured by wildcard error at $W\approx0.2\%$).  In contrast, context-to-context variations in the unitary part of the gates' error generators ($\Delta \mathcal{H}$ in Eq.~\ref{eq:generator}) are both statistically significant and large.  

Table \ref{tab:aqt_results} presents the coefficients of coherent $X$, $Y$, and $Z$ Hamiltonians (in milliradians) for each gate in each context, both without (top) and with (bottom) crosstalk compensation enabled.  The target rotation angle $\theta_0 = \pi/2\times 10^3$ mrad is included where appropriate.  This gate set has 6 gauge freedoms (corresponding to 3-parameter unitary changes of basis on each qubit), which are reflected in this table by the unobservable constants $c_1\ldots c_6$.  The key to interpreting these error rates is that constants ($\theta_0$ and $c_i$) appear identically in each column.  So context-to-context variations in each gate --- i.e., differences between entries in the same row --- are gauge-invariant.

\begin{figure}[t!]
  \includegraphics[width=\columnwidth]{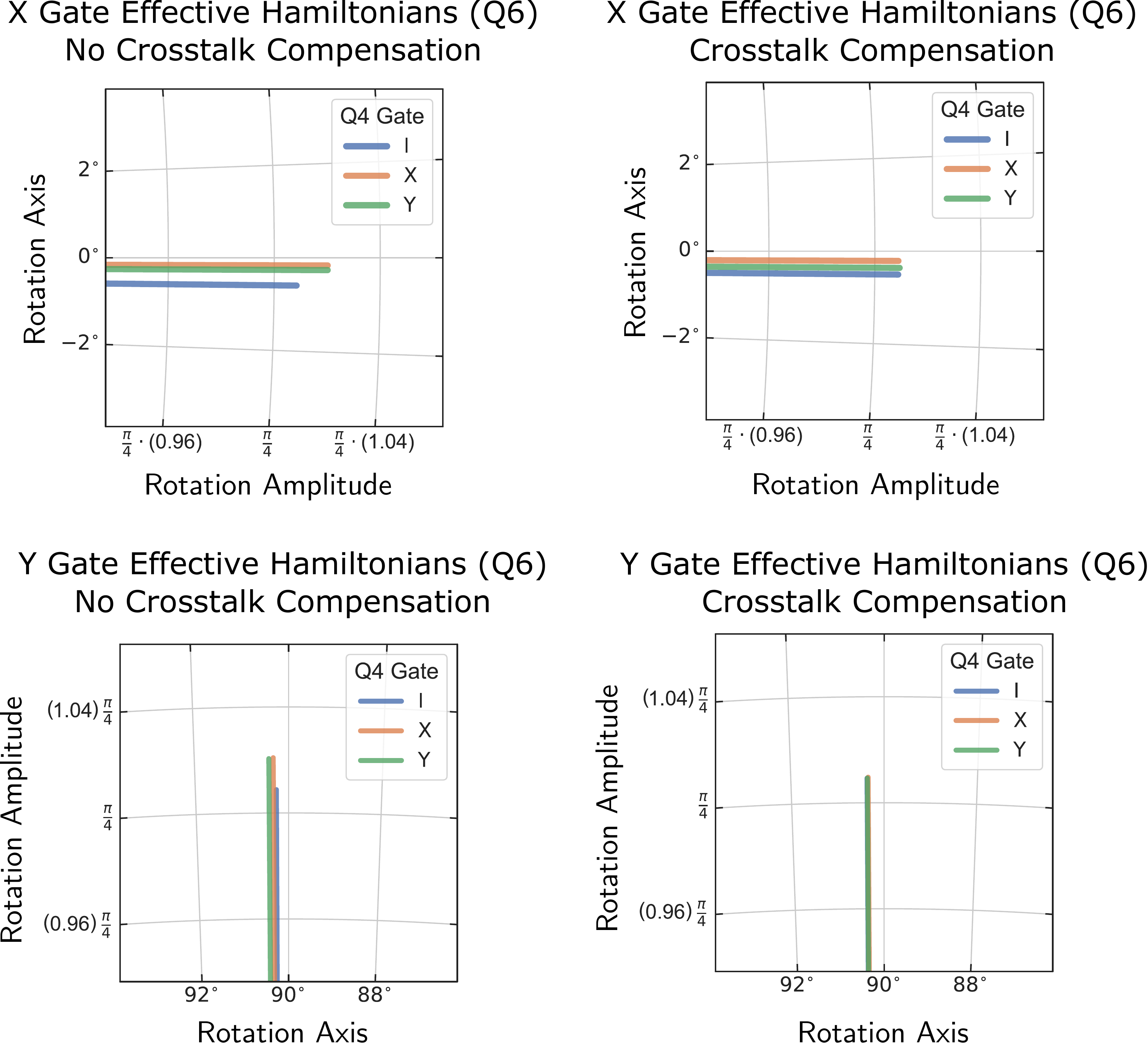}
  \caption{\textbf{Coherent errors in $\gateX$ and $\gateY$ gates on AQT \qubit{6} depend on context.}  We used GST to estimate 1-qubit process matrices describing the $\gateX$ \textbf{(top)} and $\gateY$ \textbf{(bottom)} gates on \qubit{6} conditional on $\{\mathsf{I},\gateX,\mathrm{\ and\ }\gateY\}$ gates performed at the same time on \qubit{4}, both without \textbf{(left)} and with \textbf{(right)} crosstalk compensation.  We extracted the effective Hamiltonian that generates the unitary part of each estimated process (see main text), represented it as $H_{\mathrm{eff}} = \vec{h}\cdot\vec{\sigma}$ [with $\vec{\sigma} = (X,Y,Z)$], and plotted the projection of each one into the $X$-$Y$ plane (all $h_Z$ components are negligible).  Each of the four panels shows a small region of the $X$-$Y$ plane; each effective Hamiltonian is represented by a line from the origin to $(h_X,h_Y)$.  Uncertainty regions are shown as ellipses, which are too small to be visible here.  Both rotation angles and axes vary significantly from context to context; crosstalk compensation reduces this effect, and essentially eliminates it for the $\gateY$ gate.}
  \label{fig:aqt}
\end{figure}

When no crosstalk compensation is applied, every gate displays \emph{some} statistically significant context-to-context variation.  However --- as suggested by simultaneous RB results --- the variations are significantly larger for \qubit{6}.  Its idle operation ($\mathsf{I}$) experiences phase ($Z$) errors that change by $13.1(2)$ mrad when the spectator qubit is driven.  Active gates ($\gateX$,$\gateY$) on \qubit{6} display similar variations in their rotation angles [e.g. $Y$ errors on the $\gateY$ gate, which vary by $9.5(1)$ mrad] or axes [e.g. $Y$ errors on the $\gateX$ gate, which vary by $14.4(1.0)$ mrad].  In contrast, the largest variation observed in a \qubit{4} gate is the rotation angle of the $\gateX$ gate, which varies by $4.6(5)$ mrad.  Figure \ref{fig:aqt} illustrates the variation in the effective Hamiltonians that generate $\gateX$ and $\gateY$ gates on \qubit{6}, in the $X$--$Y$ plane.

The context-to-context variations do not follow a simple pattern.  For the idle gates, only phase ($Z$) errors are observed, and they depend significantly only on whether the spectator qubit is driven or not (rather than on which gate is performed on the spectator).  For the active gates, however, both rotation angles and axes vary, and they depend not just on whether the spectator is driven, but on whether an $\gateX$ or $\gateY$ gate is performed on the spectator.  This detailed information about the nature of the crosstalk could in principle be compared to --- or used to inform --- physical models of gate context dependence, but we do not currently have such a model.

Enabling crosstalk compensation reduced the overall crosstalk significantly, but not uniformly.  It had little effect on \qubit{4}'s gates (which were already relatively good), but eliminated (a) essentially all context dependence for \qubit{6}'s $\gateY$ and $\mathsf{I}$ gates, and (b) essentially all variation in the rotation angle of \qubit{6}'s $\gateX$ gate.  Interestingly, variations in the $\gateX$ gate's rotation axis were not eliminated (see Fig.~\ref{fig:aqt}).

\subsubsection{Discussion of crosstalk in AQT}

The asymmetric crosstalk errors identified by our GST experiments are consistent with AC Stark shifts induced by the control fields. As discussed in Sec.~\ref{sec:aqt_sources}, spectroscopic data predicted that driving \qubit{6} should not influence \qubit{4} because of the large discrepancy in relevant transition frequencies. Conversely, that same analysis predicted \qubit{6} should experience phase shifts when \qubit{4} is driven, because the \qubit{4} drive tone is near-resonant with an excited state transition of \qubit{6}.  The crosstalk errors we observed are consistent with these predictions.

\setlength{\extrarowheight}{3pt}
\begin{table*}
\begin{tabular*}{\textwidth}{l@{\extracolsep{\fill}}ccccccccccc} 
 \toprule
    & & \multicolumn{5}{c}{Without crosstalk compensation}  & \multicolumn{5}{c}{With crosstalk compensation}  \\
    \cmidrule(r){3-7} \cmidrule(r){8-11}
 Model  & $N_p$ & $N_\sigma$ & $\gamma$ & $\lambda_{\rm LR}[10^3]$ & $W[10^{-3}]$ & $\bar\epsilon_\diamond [10^{-3}]$ & $N_\sigma$ & $\gamma$ & $\lambda_{\rm LR}[10^3]$ & $W[10^{-3}]$ & $\bar\epsilon_\diamond [10^{-3}]$ \\
 \midrule
 General     & 1,697 & \cellcolor{yellow!25}  46.28 & --- & 76.02 & $1.39$ & $16.70(3)$ & \cellcolor{yellow!25} 49.92 & --- & 77.21 & 1.89 & $10.97(3)$ \\
 Context-dependent      & 230   & \cellcolor{yellow!25}  45.90 & \cellcolor{green!25} 0.90 & 77.60 & $1.53$ & $16.54(2)$ & \cellcolor{yellow!25} 47.70 & \cellcolor{green!25} 0.58 & 78.23  & 2.03 & $10.62(3)$ \\
 Crosstalk-free         & 86    &  \cellcolor{orange!25} 248.27 & \cellcolor{red!25} 494.60 & 148.82 & $8.54$ & $15.94(1)$ & \cellcolor{yellow!25} 53.63 & \cellcolor{yellow!25} 15.58 & 80.48 & 2.29 & $10.21(2)$ \\
 \bottomrule
\end{tabular*}
\caption{\textbf{Fit quality metrics for three models on AQT device.} We tabulate metrics of fit quality and unmodeled error for three distinct models' when fit to GST data from the AQT platform (both with and without crosstalk compensation). The performance of these models can be compared in detail using the log-likelihood ratio score $\lambda_{\rm LR}$, the $N_\sigma$ of model violation, and the residual wildcard error $W$.  Each of these models further predicts an average diamond error $\bar\epsilon_\diamond$ for the gate set. Without crosstalk compensation, the context-dependent model is strongly preferred over the crosstalk-free model by the evidence ratio test $\gamma$, and the wildcard error is significantly reduced by moving to the larger model. The full Markovian model, however, requires nearly an order of magnitude more parameters, achieves only a small improvement in the likelihood ratio and the wildcard, and is rejected by the evidence ratio test. When crosstalk compensation in applied, the context-dependent model is again preferred by the evidence ratio test, but much more weakly, and the wildcard error nearly constant across models. Crosstalk compensation further results in a $\simeq35\%$ reduction in the average gate error. 
\label{tab:aqt_results}}
\end{table*}

\setlength{\extrarowheight}{3pt}
\setlength{\belowrulesep}{0mm}
\begin{table*}
\resizebox{.7\textwidth}{!}{
\begin{tabular*}{.8\textwidth}{c@{\extracolsep{\fill}}ccc} 
 \toprule
    & \multicolumn{3}{c}{Gate on spectator qubit (Context)} \\
    \cmidrule(r){2-4} 
 Gate on target qubit & $\mathsf{I}$  & $\gateX$ & $\gateY$ \\
 \midrule \midrule
    Without crosstalk compensation &&& \\
    \midrule  
  & $0.1(0.1)$ & $0.2(0.1)$ & $0.3(0.1)$ \\ 
 $\mathsf{I}^{(4)}$ & $0.1(0.1)$ & $0.1(0.1)$ & $0.1(0.1)$ \\ 
  & $0.9(0.2)$ & $1.2(0.2)$ & $1.7(0.2)$ \\ 
  \midrule
  & $\rot+3.6(0.2)$ & $\rot+3.7(0.1)$ & $\rot+3.7(0.1)$ \\ 
 $\gateX^{(4)}$ & $4.4(0.4)+c_1$ & $0.6(0.4)+c_1$ & $-0.2(0.4)+c_1$ \\ 
  & $1.0(4.1)+c_2$ & $-1.9(4.1)+c_2$ & $0.9(4.1)+c_2$ \\ 
  \midrule
  & $3.3(0.4)-c_1$ & $0.3(0.4)-c_1$ & $1.3(0.4)-c_1$ \\ 
 $\gateY^{(4)}$ & $\rot+3.6(0.2)$ & $\rot+3.5(0.1)$ & $\rot+3.2(0.1)$ \\ 
  & $-2.8(5.6)+c_3$ & $1.6(5.6)+c_3$ & $1.2(5.6)+c_3$ \\ 
  \midrule
  \midrule
  & $-0.1(0.1)$ & $0.2(0.1)$ & $0.2(0.1)$ \\ 
 $\mathsf{I}^{(6)}$ & $-0.1(0.1)$ & $-0.3(0.2)$ & $-0.2(0.2)$ \\ 
  & $3.0(0.2)$ & $-10.2(0.1)$ & $-10.1(0.1)$ \\ 
  \midrule
  & $\rot+6.6(0.1)$ & $\rot+15.8(0.1)$ & $\rot+15.8(0.1)$ \\ 
 $\gateX^{(6)}$ & $-8.2(0.3)+c_4$ & $-2.2(0.3)+c_4$ & $-3.6(0.3)+c_4$ \\ 
  & $8.1(0.7)+c_5$ & $-1.8(0.7)+c_5$ & $-6.3(0.8)+c_5$ \\ 
  \midrule
  & $-3.6(0.3)-c_4$ & $-4.6(0.3)-c_4$ & $-5.9(0.3)-c_4$ \\ 
 $\gateY^{(6)}$ & $\rot+6.9(0.1)$ & $\rot+16.4(0.1)$ & $\rot+16.1(0.1)$ \\ 
  & $7.3(12.1)+c_6$ & $-3.5(12.2)+c_6$ & $-3.8(12.2)+c_6$ \\ 
  \midrule
  \midrule
  With crosstalk compensation &&& \\
  \midrule
  & $0.0(0.1)$ & $0.1(0.1)$ & $0.3(0.1)$ \\ 
 $\mathsf{I}^{(4)}$ & $0.1(0.1)$ & $0.1(0.1)$ & $0.0(0.1)$ \\ 
  & $1.3(0.2)$ & $1.8(0.2)$ & $1.8(0.2)$ \\ 
  \midrule
  & $\rot-3.5(0.2)$ & $\rot-3.6(0.1)$ & $\rot-3.4(0.1)$ \\ 
 $\gateX^{(4)}$ & $4.2(0.3)+c_7$ & $0.1(0.3)+c_7$ & $-0.8(0.3)+c_7$ \\ 
  & $0.9(1.6)+c_8$ & $-1.5(1.6)+c_8$ & $0.6(1.6)+c_8$ \\ 
  \midrule
  & $2.9(0.3)-c_7$ & $0.1(0.3)-c_7$ & $0.5(0.3)-c_7$ \\ 
 $\gateY^{(4)}$ & $\rot-3.6(0.2)$ & $\rot-3.6(0.1)$ & $\rot-4.1(0.1)$ \\ 
  & $-2.6(3.3)+c_9$ & $1.4(3.3)+c_9$ & $1.3(3.3)+c_9$ \\ 
  \midrule
  \midrule
  & $0.1(0.1)$ & $0.1(0.1)$ & $-0.1(0.1)$ \\ 
 $\mathsf{I}^{(6)}$ & $-0.2(0.1)$ & $-0.2(0.1)$ & $-0.3(0.1)$ \\ 
  & $2.4(0.2)$ & $1.6(0.2)$ & $1.3(0.2)$ \\ 
  \midrule
  & $\rot+6.9(0.3)$ & $\rot+7.0(0.2)$ & $\rot+7.3(0.2)$ \\ 
 $\gateX^{(6)}$ & $-6.9(0.3)+c_{10}$ & $-2.9(0.3)+c_{10}$ & $-5.0(0.3)+c_{10}$ \\ 
  & $3.3(7.1)+c_{11}$ & $0.6(7.1)+c_{11}$ & $-3.9(7.1)+c_{11}$ \\ 
  \midrule
  & $-5.1(0.3)-c_{10}$ & $-4.7(0.3)-c_{10}$ & $-5.0(0.3)-c_{10}$ \\ 
 $\gateY^{(6)}$ & $\rot+7.3(0.2)$ & $\rot+7.6(0.2)$ & $\rot+7.3(0.2)$ \\ 
  & $1.5(6.7)+c_{12}$ & $-0.8(6.7)+c_{12}$ & $-0.7(6.7)+c_{12}$ \\
 \bottomrule
\end{tabular*}}
\caption{\textbf{Estimated rates of all Hamiltonian (unitary) errors, on all gates, in all contexts, for the context-dependent model of the AQT platform.} Each column of three values is the three components $\vec{h}$ of the effective Hamiltonian that generates the unitary part of the estimated process (see main text), represented as $H_{\mathrm{eff}} = \vec{h}\cdot\vec{\sigma}$ [with $\vec{\sigma} = (X,Y,Z)$]. This gate set has 6 gauge freedoms, which are reflected in this table by the unobservable constants $c_1\ldots c_6$. Any value, or linear combination of values, contain \emph{no} $c$'s is gauge invariant, and therefore physically meaningful. Units are mrad, and the target $\rot = \pi/4\,\mathrm{mrad} \simeq 785.4\,\mathrm{mrad}$.
and the target $\rot = \pi/4\,\mathrm{mrad} \simeq 785.4\,\mathrm{mrad}$.
\label{tab:aqt_hamiltonians}}
\end{table*}

\setlength{\belowrulesep}{0.65ex}

\subsection{The Quantum Scientific Computing Open User Testbed}
\label{sec:qscout}

Experiments on the QSCOUT platform were performed using a prototype system that is nearly identical to the deployed testbed. It was configured to use two qubits encoded as the hyperfine clock states of a pair of ${}^{171}{\mathrm{Yb}}^+$ ions~\cite{Olmschenk2007} that were held in a Sandia-fabricated HOA 2.0 surface ion trap~\cite{Maunz2016}. The ions were trapped together in a single pseudopotential and were spaced $4.5\;\rm{\mu m}$ apart. The trap frequencies were approximately 1 MHz axially and 2 MHz radially. The ions’ hyperfine states were manipulated using a two-photon Raman transition via a pair of phase-locked co-propagating frequency combs generated by a frequency-tripled Nd:YAG pulsed laser at {355\;\rm{nm}}~\cite{Islam2014}.  Each ion was individually addressed by first splitting a single laser beam into multiple beams, and sending each beam through a dedicated channel of a multichannel acousto-optic modulator (AOM) which allows for independent frequency, phase, and amplitude control.  Each beam was then tightly focused using custom optics to a $0.8\;\rm{\mu m}$ axial waist radius, ideally impinging on only a single target ion. During the detection cycle, light from each ion was focused to a different core of a multicore fiber. Each core was then sent to its own photomultiplier tube (PMT), allowing for distinguishable detection of the ions. 

\subsubsection{Potential (and actual) sources of crosstalk in QSCOUT}

Several physical phenomena are expected to manifest as crosstalk errors in the QSCOUT hardware. The first is straightforward: control lasers targeted at one ion have a non-zero beam waist, allowing some light to spill over onto neighboring ions. Because all ions in the system are at the same resonant frequency, this light can cause slow, coherent Rabi oscillations on the neighbor ions. The laser waist is relatively small relative to the ion spacing, so one might expect this effect to be relatively small. But its magnitude can be significant if the system is poorly aligned. A more subtle source of potential crosstalk is the internal dynamics of the multichannel AOM. The nonlinear optical crystals in the AOM are not perfectly isolated from one another, so acoustic drive tones applied to a target crystal can cause neighboring crystals to ring sympathetically. The neighboring channel will then be activated (or perturbed), and its target ion will experience an error. The signals that implement these drive tones can also electrically couple into neighboring channels. The wiring layout of the AOM predicts that next-nearest-neighbor correlations will be greatest. Correlated errors arising from common causes can also manifest as crosstalk. In the trapped-ion hardware we expect amplitude (or phase) fluctuations of the driving laser to result in correlated amplitude (or phase) errors at the qubits. Similarly, magnetic field fluctuations can result in correlated phase errors.

\subsubsection{Experiment design}
We used the same family of GST circuits for QSCOUT as for AQT (Table~\ref{tab:sgst_circuits}). Because trapped-ion operations are slower, we took less data on the QSCOUT device. The number of counts per circuit was reduced from 1000 to 80, and the GST circuits were limited to depth $\simeq8$ (rather than $\simeq32$). We ran simultaneous DRB circuits generated in exactly the same way as for AQT (maximum depth 256, 30  circuits at each of 8 logarithmically spaced depths), and interleaved them with the GST circuits in the same way. A total of 12,514 unique circuits were run in the QSCOUT experiment, including 11,813 GST circuits and 701 unique DRB circuits. Data was taken in 8 batches, each comprising 10 shots of every circuit. The device was recalibrated between each batch. Approximately one million individual circuit shots were performed over approximately four hours.

The circuits we ran on the QSCOUT platform were implemented using simple pulses, with no compensating pulse sequences of any kind. The QSCOUT hardware can implement gates using, e.g., BB1 composite pulse sequences \cite{Wimperis1994-fd} that can yield significantly reduced error rates.  However, they also require more time to implement, complicate the interpretation of unitary error generators, and can reduce the magnitude of certain crosstalk errors below detectable thresholds. So, for the purposes of this work, we restricted gate operations to simple, bare pulses.

\subsubsection{Randomized benchmarking}
\label{sec:qscout_drb}
\begin{figure}[t]
\includegraphics[keepaspectratio=true,width=\columnwidth]{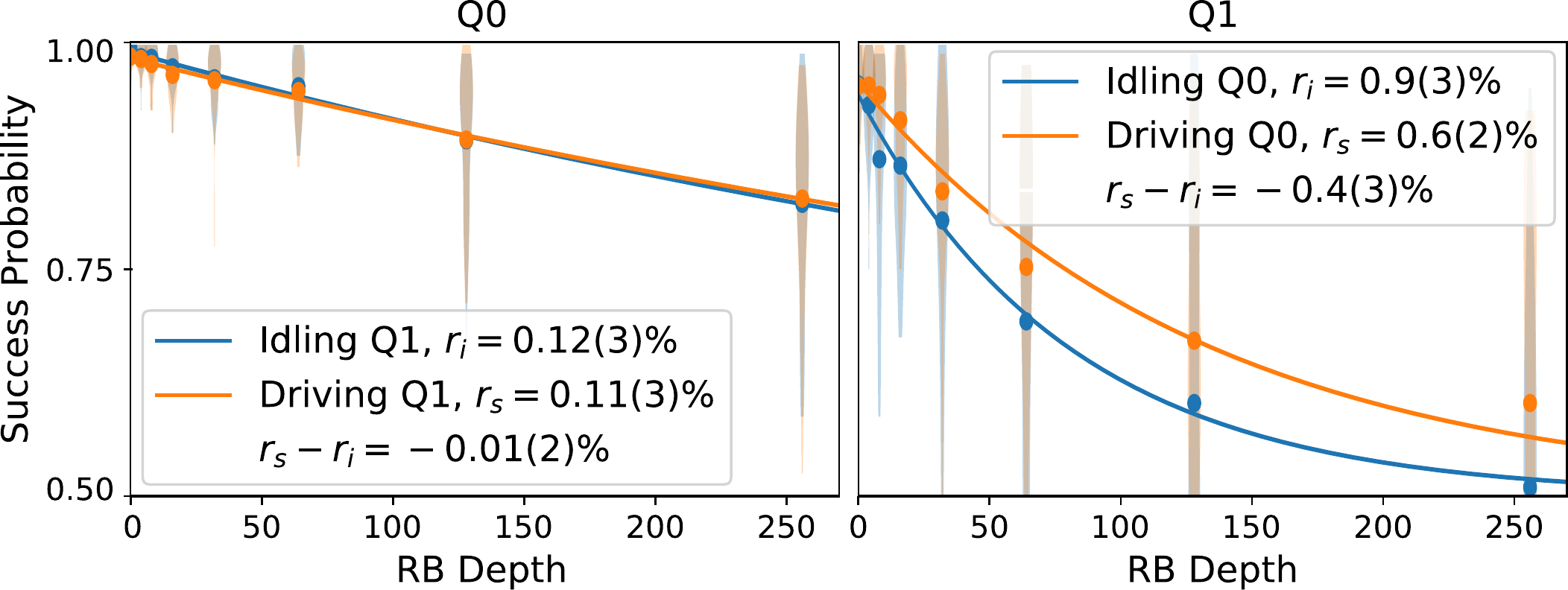} 
\caption{  \label{fig:qscout_sdrb} \textbf{Simultaneous direct randomized benchmarking (DRB) on the QSCOUT platform.} We ran simultaneous DRB experiments (see caption to Fig.~\ref{fig:aqt_sdrb}) on a 2-qubit QSCOUT processor.  No crosstalk compensation was used in this experiment. \qubit{0} performs well independent of context (i.e., whether \qubit{1} is driven).  In contrast, \qubit{1} performs at least 5$\times$ worse in both contexts, and its error-per-gate appears to depend on whether \qubit{0} is driven.  Oddly, however, driving \qubit{0} actually reduces \qubit{1}'s error rate, suggesting a negative rate of crosstalk-induced errors.  We identify the particular calibration protocol used as the probable cause (see main text).}
\end{figure}

We ran simultaneous DRB on the QSCOUT platform to estimate the context-to-context variation of each qubit's error-per-gate. Figure \ref{fig:qscout_sdrb} shows the results for both \qubit{0} and \qubit{1}. We observe an even more dramatic asymmetry than in the RB experiments on AQT; \qubit{0}'s error-per-gate is approximately $0.11(3)\%$ regardless of whether \qubit{1} is driven, but \qubit{1}'s error-per-gate is far higher [$0.6(2)\%-0.9(3)\%$], and varies dramatically depending on whether \qubit{0} is driven. Perhaps surprisingly, driving \qubit{0} actually \emph{improves} \qubit{1}'s performance.   We hypothesize that this is a consequence of the calibration procedure.  Only the $\gateX^{(0)}\gateX^{(1)}$ operation was calibrated. $\gateY$ gates use the same calibrated pulses, but phase-shifted by $\pi/2$, so effectively $\gateY^{(0)}\gateY^{(1)}$ was also calibrated.  This calibration procedure optimizes the $\gateX$ and $\gateY$ gates on each qubit for performance in the very specific context where the \emph{same} gate is performed on the other qubit. Other circuit layers --- e.g. $\gateX^{(0)}\gateY^{(1)}$ or $\gateX^{(0)}\gateI^{(1)}$ are not necessarily well-calibrated, and may experience different and/or larger crosstalk errors.  Independent DRB probes the performance of layers that are strictly uncalibrated (e.g. $\gateX^{(0)}\gateI^{(1)}$, $\gateY^{(0)}\gateI^{(1)}$, and $\gateI^{(0)}\gateI^{(1)}$ for \qubit{0}).  Approximately one third of the circuit layers used in simultaneous DRB have been calibrated, predicting that the simultaneous circuits should perform better --- which is what we observe. We shall see further consequences of this calibration protocol in the GST data.

\subsubsection{Comparing and selecting crosstalk models}

We fit all three gate set models to the data (just one dataset in this case, since no crosstalk compensation was performed) and evaluated their fit quality. Table~\ref{tab:qscout_results} displays the results.

No model fits the data perfectly.  The general model displays about $9\sigma$ of model violation, suggesting that the gates are somewhat non-Markovian.  However, the wildcard error assigned to the general model is almost zero, suggesting that this non-Markovianity is barely visible.  These results should not be compared directly to AQT --- far less data was taken on QSCOUT, so the experiment is less able to detect and quantify non-Markovianity.

The evidence ratio between the general model and the (smaller) context-dependent model is extraordinarily small ($\gamma = 0.29$), indicating no evidence whatsoever of entangling or correlating crosstalk errors between the qubits\footnote{Evidence ratios less than 1 are typically caused by extremely small sample counts, when many events appear 0 or 1 times, and stem from breakdown of the Gaussian ansatz used in deriving Wilks' theorem.  Since each circuit was repeated $N=80$ times, it is unsurprising to see this for the QSCOUT data.}.  The observed data can be described more or less perfectly by context-dependent local errors.

However, the evidence ratio between the context-dependent model and the crosstalk-free model clearly rejects the crosstalk-free model ($\gamma = 134$).  This constitutes clear and overwhelming evidence of crosstalk (as expected, given the simultaneous RB results).  We can also estimate the magnitude of the crosstalk errors from the wildcard error required to reconcile the best crosstalk-free estimate with the data ($W=2.4\%$).  Since the context-dependent model fits the data well, we can ascribe all crosstalk errors to context-to-context variation of local errors, which we proceed to examine in detail.

\subsubsection{Extracting detailed crosstalk error rates}

The GST estimate of the context-dependent model yields 1-qubit process matrices describing each of the 6 gates in 3 different contexts.  We analyze the error generators for these processes.  As in the AQT analysis, the non-unitary error generators reveal almost no significant context-to-context variation, so we focus on unitary (coherent) errors.  

Table \ref{tab:qscout_hamiltonians} presents the coefficients of coherent $X$, $Y$, and $Z$ Hamiltonians (in milliradians) for each gate in each context.  The target rotation angle $\theta_0 = \pi/2\cdot 10^3$ mrad is included where appropriate.  This gate set has 6 gauge freedoms (corresponding to 3-parameter unitary changes of basis on each qubit), which are reflected in this table by the constants $c_1\ldots c_6$, which are unobservable.  $\theta_0$ and $c_i$ appear identically in each column, so context-to-context variations in each gate --- i.e., differences between entries in the same row --- are automatically gauge-invariant.

\begin{figure}[t!]
  \includegraphics[keepaspectratio=true,width=\columnwidth]{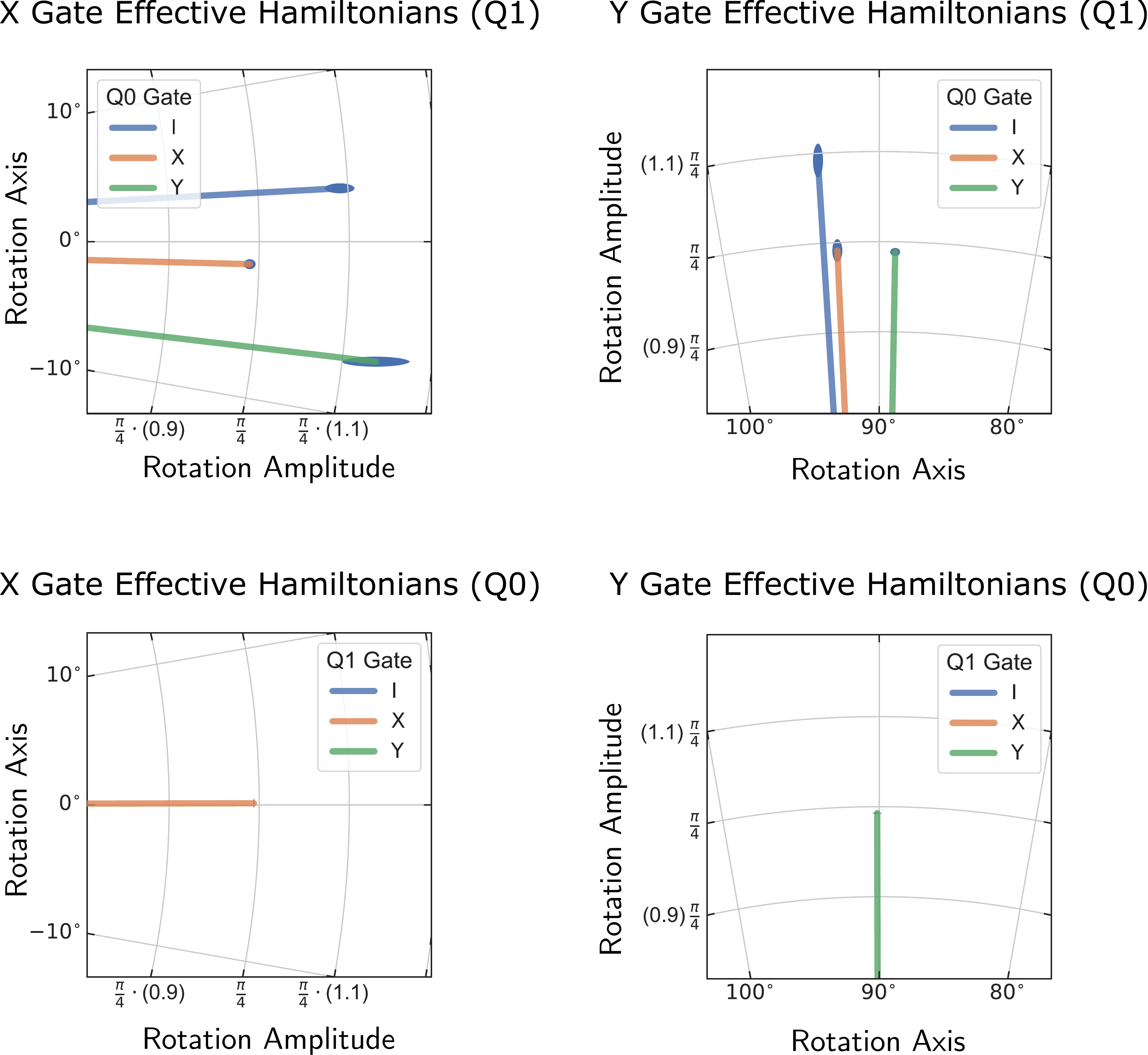}
  \caption{\textbf{Coherent errors in $\gateX$ and $\gateY$ gates on QSCOUT qubits depend on context.}  We used GST to estimate 1-qubit process matrices describing the $\gateX$ \textbf{(left)} and $\gateY$ \textbf{(right)} gates on \qubit{0} \textbf{(bottom)} and \qubit{1} \textbf{(top)}, conditional on $\{\mathsf{I},\gateX,\mathrm{\ and\ }\gateY\}$ gates performed at the same time on the other qubit, in the QSCOUT system.  Exactly as for Fig.~\ref{fig:aqt}, we extracted the effective Hamiltonian that generates the unitary part of each estimated process (see main text), represented it as $H_{\mathrm{eff}} = \vec{h}\cdot\vec{\sigma}$ [with $\vec{\sigma} = (X,Y,Z)$], and plotted the projection of each one into the $X$-$Y$ plane (all $h_Z$ components are negligible).  Each of the four panels shows a small region of the $X$-$Y$ plane; each effective Hamiltonian is represented by a line from the origin to $(h_X,h_Y)$.  Uncertainty regions are shown as ellipses.  \qubit{0}'s gates show essentially no context dependence.  In contrast, the rotation angles and axes of the $\gateX$ and $\gateY$ gates on \qubit{1} show extremely strong dependence on the gate performed on the spectator.  We also observed small context-dependent phase errors in the $\mathsf{I}$ gate (not shown here).}
\label{fig:qscout}
\end{figure}

We can immediately draw the following conclusions from Table \ref{tab:qscout_hamiltonians}:
\begin{itemize}
    \item Errors on both idle gates are small ($\leq3\cdot 10^{-3}$ radians) in all contexts, and almost exclusively indistinguishable from zero.  The $\mathsf{I}$ gate on \qubit{0} has no evident errors at all. The $\mathsf{I}$ gate on \qubit{1} shows barely significant context-dependent rotations by about 3 mrad.
    \item $Z$ Hamiltonian error rates on each qubit's $\gateX$ and $\gateY$ gates are also effectively negligible --- they are all $<4$ mrad in all contexts, and mostly indistinguishable from zero.
\end{itemize}
The remaining errors, with magnitudes of up to $0.1$ radians (100 mrad), fall into two categories:
\begin{enumerate}
    \item Over/under-rotation errors, i.e. $X$ Hamiltonian errors on $\gateX$ gates and $Y$ Hamiltonian errors on $\gateY$ gates.
    \item ``Tilt'' errors that change a gate's rotation axis, i.e. $Y$ Hamiltonian errors on $\gateX$ gates and $X$ Hamiltonian errors on $\gateY$ gates.
\end{enumerate}
All of these errors show significant context-to-context variation.  Figure \ref{fig:qscout} illustrates this variation by depicting each gate's angle and axis in the $X$-$Y$ plane, for each context.  Gates on \qubit{0} have rotation angles that vary by up to $33(7)$ mrad, and rotation axes that vary by up to $28(2)$ mrad.  For gates on \qubit{1}, rotation angles vary by up to $170(40)$ mrad, and rotation axes vary by up to $150(1)$ mrad.  The context dependence of errors on \qubit{1}'s gates are approximately 3-5$\times$ larger than \qubit{0}'s.  Because these errors are coherent, the error-per-gate observed in RB scales as $\theta^2$, and so in RB experiments we expect to see 10-25$\times$ more context-dependent error on \qubit{1} than \qubit{0}.  This is consistent with the observed results of simultaneous RB (Fig.~\ref{fig:qscout_sdrb}).

The GST estimate confirms the conjecture we stated in the discussion of simultaneous RB results:  errors are minimized when the same active gate ($\gateX$ or $\gateY$ is performed on both qubits.  These are the only layers that are explicitly calibrated.  Active gates performed in other contexts show clear and significant calibration errors in both their rotation angles and their rotation axes (relative to the calibrated operation).

\subsubsection{Discussion of QSCOUT results}

Simultaneous RB experiments clearly demonstrate the existence of \qubit{0}$\to$\qubit{1} crosstalk, but also demonstrate the counterintuitive result that gates on \qubit{1} perform better when \qubit{0} is driven.  GST experiments revealed what was actually happening: each active gate's behavior depends not just on the ``spectator driven'' and ``spectator idled'' contexts, but on exactly what gate is performed on the spectator.  Most of this effect is due to the calibration protocol, and the fact that only $\gateX\gateX$ and $\gateY\gateY$ gates were specifically calibrated.  However, Table \ref{tab:qscout_hamiltonians} shows additional variations between the other two contexts.

The crosstalk errors we observe in the QSCOUT system reflect the fact that the two qubits have identical energy splitting. Pulse spillover onto a spectator ion is therefore resonant with its qubit transition, leading to coherent Rabi oscillations around an axis on the equator of the Bloch sphere. In contrast, the AQT qubits have different frequencies, so spillover crosstalk acts very differently.  Instead of coherent Rabi oscillations, it induces an AC Stark shift on the spectator qubit, which manifests as coherent rotation about the Z axis. The asymmetry in the QSCOUT error rates (particularly the errors in the idle gates) support a hypothesis that the crosstalk arises from beam pointing errors, rather than internal AOM phenomena, which are more likely to result in crosstalk errors that are symmetric between qubits.

Restricting device calibration to parallel operations has serious, observable impacts on the performance on non-parallel layers. This suggests that the active gates' error rates could be reduced significantly \emph{and} made less context-dependent by explicitly calibrating all of the layers simultaneously. This would require a more complicated tune-up process.

\setlength{\extrarowheight}{3pt}
\begin{table*}
\begin{tabular*}{\textwidth}{l@{\extracolsep{\fill}}cccccc} 
 \toprule
 Model  & $N_p$ & $N_\sigma$ & $\gamma$ & $\lambda_{\rm LR}[10^3]$ & $W[10^{-3}]$ & $\bar\epsilon_\diamond [10^{-3}]$ \\
 \midrule
 General     & 1,697 & \cellcolor{green!25}  9.33 & --- & 36.17 & $0.02$ & $59.6(3)$ \\
 Context-dependent      & 230   & \cellcolor{green!25}  5.27 & \cellcolor{green!25} 0.29 & 36.61 & $0$ & $57.1(3)$ \\
 Crosstalk-free         & 86    &  \cellcolor{orange!25} 134.02 & \cellcolor{red!25} 139.38 & 70.99 & $24.1$ & $43.8(2)$ \\
 \bottomrule
\end{tabular*}
\caption{Details of the three best-fit gate models on the QSCOUT device. The performance of these models can be compared in detail using the log-likelihood ratio score $\lambda_{\rm LR}$, the $N_\sigma$ of model violation, and the residual wildcard error $W$.  Each of these models further predicts an average diamond error $\bar\epsilon_\diamond$ for the gate set. The context-dependent model is strongly preferred over the crosstalk-free model by the evidence ratio test $\gamma$, and the wildcard error is significantly reduced by moving to the larger model. The general crosstalk model, however, requires nearly an order of magnitude more parameters, achieves only a small improvement in the likelihood ratio and the wildcard, and is rejected by the evidence ratio test. Notably, $W$ is nonzero for the general model despite being zero for the smaller context-dependent model. This can occur because the wildcard error calculation requires models with more parameters to fit better that those with fewer. 
\label{tab:qscout_results}}
\end{table*}

\begin{table*}
\begin{tabular*}{0.8\textwidth}{c@{\extracolsep{\fill}}ccc} 
 \toprule
      & \multicolumn{3}{c}{Gate on spectator qubit (Context)}  \\
    \cmidrule(r){2-4} 
 Gate on target qubit & $\mathsf{I}$  & $\gateX$ & $\gateY$ \\
 \midrule
   & $0.8(1.8)$ & $-0.5(1.3)$ & $-0.6(1.5)$ \\ 
  $\mathsf{I}^{(0)}$               & $1.0(1.8)$ & $1.3(1.3)$ & $-0.4(1.5)$ \\ 
                & $1.5(2.5)$ & $0.4(2.0)$ & $1.9(2.2)$ \\ 
 \midrule
   & $\rot-12.0(1.4)$ & $\rot-4.6(1.2)$ & $\rot-37.4(6.4)$ \\
   $\gateX^{(0)}$              & $1.2(1.1)+c_1$ & $1.5(1.1)+c_1$ & $22.8(1.1)+c_1$ \\ 
                & $-0.3(1.1)+c_2$ & $-1.3(1.1)+c_2$ & $1.6(1.1)+c_2$ \\ 
 \midrule
    & $0.8(1.1)-c_1$ &  $26.6(1.1)-c_1$& $-1.8(1.1)-c_1$ \\
  $\gateY^{(0)}$              & $\rot-12.5(1.7)$ & $\rot-28.5(7.4)$ & $\rot-5.6(1.2)$ \\ 
                & $-2.8(1.1)+c_3$ & $1.2(1.1)+c_3$ & $1.6(1.1)+c_3$ \\ 
 \midrule
 \midrule
   & $0.7(1.7) $ & $-1.6(1.5)$& $-2.6(1.5)$\\ 
   $\mathsf{I}^{(1)}$  & $0.6(1.8) $ & $3.2(1.3) $& $-1.4(1.7)$ \\ 
                & $0.9(2.5) $ & $-0.5(2.1)$ &$ 0.3(2.4)$ \\ 
 \midrule
    & $\rot+70.2(13.5)$  &$\rot-14.3(5.8)$ & $\rot-101.5(30.3)$\\ 
 $\gateX^{(1)}$  & $ 46.5(1.0)+c_4 $   & $-19.7(1.0) +c_4 $&$-104.7(1.0) +c_4$ \\ 
                & $  1.1(1.0)+c_5$ & $  0.5(1.0)+c_5$&$ -1.7(1.0)+c_5$ \\ 
 \midrule
            & $-53.4(1.0)-c_4 $& $-36.5(1.0) -c_4$& $ 13.9(1.0) -c_4$\\ 
  $\gateY^{(1)}$ & $ \rot+70.7(15.4)$&$ \rot-7.9(10.7)$  & $ \rot-9.1(4.2)$\\ 
                & $ -2.5(1.0)+c_6$&$ -1.6(1.0)+c_6$   &$  3.9(1.0)+c_6$\\
 \bottomrule
\end{tabular*}
\caption{\textbf{Estimated rates of all Hamiltonian (unitary) errors, on all gates, in all contexts, for the context-dependent model of the QSCOUT platform.} Each column of three values is the three components $\vec{h}$ of the effective Hamiltonian that generates the unitary part of the estimated process (see main text), represented as $H_{\mathrm{eff}} = \vec{h}\cdot\vec{\sigma}$ [with $\vec{\sigma} = (X,Y,Z)$]. This gate set has 6 gauge freedoms, which are reflected in this table by the unobservable constants $c_1\ldots c_6$. Any value, or linear combination of values, that does \emph{not} contain any $c$'s is gauge invariant, and therefore physically meaningful. Units are mrad, and values have been shifted relative to the target value, $\rot = \pi/4\,\mathrm{mrad} \simeq 785.4\,\mathrm{mrad}$, where appropriate. 
\label{tab:qscout_hamiltonians}}
\end{table*}

\section{Discussion}
\label{sec:discussion}

This paper presents both a novel device characterization method, and the results of its deployment on two experimental platforms.  We separately discuss our conclusions about this method and our insights into the experimental results.

\subsection{Protocols and methods}

GST is best known as a replacement for process tomography, and for its use in constructing full-dimensional process matrices that represent gates' action on their target space.  But GST is inherently flexible \cite{Nielsen2021-Onion}, and we used that flexibility here to test \emph{multiple} error models --- ranging from full-dimensional process matrices down to a highly restricted crosstalk-free model --- to data generated using parallel gates on two subsystems.  This targeted adaptation of GST to probe crosstalk allowed us to extract a lot of information at many levels of detail, ranging from ``Yes, there's crosstalk, but it's not entangling'' down to the exact details of how much over-rotation each gate induced on the spectator.

Our results should be easy to reproduce, extend, and deploy in many experimental systems.  They are explicitly platform-agnostic, and they require only user-level access to a device and the ability to run simple circuits composed of device-native quantum operations. The data analysis routines are all built from free, widely available software tools, and can often be implemented with just a few lines of Python code (e.g., \pygsti). All of the experimental data and analysis code necessary to reproduce the results shown here are available upon request as supplementary material. 

The investigation and results presented here highlight the complexity of crosstalk errors.  If we had stopped after running simultaneous RB, we would have concluded simply that (1) the AQT qubits had a little bit of crosstalk and (2) the QSCOUT qubits had a small \emph{negative} amount of crosstalk.  The results of our detailed GST investigation do not contradict those findings --- but they illustrate that crosstalk is not described by a single number.  Crosstalk \emph{changes} errors --- e.g., depending on their context --- which can cause unexpected harm irrespective of which context (e.g. idle neighbor or driven neighbor) induces the worse error.  Eliminating crosstalk means removing all forms of context-dependence entirely, not just ensuring that ``idle'' and ``driven'' randomized benchmarking experiments yield the same error-per-gate.

The approach presented here complements high-level crosstalk benchmarks, such as simultaneous RB, by constructing detailed models that can identify specific errors. This low-level diagnostic information can elucidate the physics of quantum information processors, enable better calibration, and inform design of next-generation systems.  Detailed error models also enable more accurate estimates of device performance on real tasks, and noise-optimized decoders for quantum error correction. 

These advantages do have a cost.  Although GST experiments constituted the overwhelming majority of the circuits we ran, the simultaneous DRB experiments were actually more sensitive to certain errors.  GST and RB circuits are equally sensitive to the average stochastic error of a gate set, and they both amplify it proportional to circuit length $L$.  We ran GST circuits up to $L=8$ or $L=32$ (depending on platform), but RB circuits up to $L=256$.  As a result, the simultaneous RB analysis provides sharper information about the context-dependence of average stochastic error than GST estimates do; GST-based predictions of the simultaneous DRB results are within error bars, but those error bars are larger than the estimated effect. The GST estimates provide far more information about coherent errors (and individual stochastic error rates), but simultaneous RB extracts its single summary benchmark with unmatched efficiency.

Extensions of our methods to many-qubit systems and two-qubit gates are possible.  One obvious and immediately practical extension is to study the crosstalk induced by 2-qubit gates (on a 2-qubit subsystem) on neighboring 1- and 2-qubit subsystems.  This requires probing at most 4 qubits at once, which is feasible with existing analytic machinery.  Straightforwardly scaling our models to subsystems of $\gg2$ qubits will quickly become impractical.  Recent advances in many-qubit gate set tomography \cite{nielsen2020efficient} suggest a path to reducing this overhead and enabling GST for efficient crosstalk characterization of large-scale quantum systems. 

\subsection{Experimental results}

Our investigation revealed similar crosstalk errors in both platforms --- the gate errors were significantly context-dependent, but there was no evidence for entangling or correlated errors. And in both platforms, the dominant crosstalk errors are consistent with simple pulse spillover. 

Our experiments found no statistically significant evidence for entangling or correlated stochastic errors in either experiment. We confirmed using numerical simulations that our methodology is sensitive to entangling errors of this type (see Appendix), so our failure to detect them indicates their rate is below the detection threshold of this experiment.  GST circuits' sensitivity to coherent errors (of all types) increases proportional to their length, so in future work we intend to search for weak coherent ZZ errors using longer GST circuits.  More sensitive GST experiments could also reveal context dependence in the non-unitary errors, and/or correlated stochastic noise.

In contrast, the absence of entangling errors in the {QSCOUT} platform is unsurprising.  This platform's single-qubit gates are not expected to induce the ion/phonon couplings required to couple two ion qubits.  Correlated stochastic errors are plausible and could be produced by a wide array of environment effects, but we saw no evidence for such errors in the data. However, this experiment's detection threshold was relatively high --- the trapped-ion hardware has a relatively low data rate that limited both the circuit depth ($L=8$) and the number of counts ($N=80$).  As a result, our estimates of stochastic error rates have low precision, and correlated errors could be hiding in the noise.  More precise probing of these errors will require streamlined experiments.

One of the most impactful results of our analysis is that --- in both platforms --- the dominant crosstalk errors are restricted to (1) context-dependent local errors, and (2) coherent unitary rotations.  These restrictions single out a very small subset of all the possible crosstalk errors. It is always easier to characterize and track errors that lie in a constrained, low-dimensional set.  Perhaps more importantly, context-dependent unitary errors are among the easiest to eliminate. Dynamical decoupling, optimal control, or simple context-dependent calibration can all remove such errors.  As demonstrated by AQT's crosstalk compensation pulses, even simple techniques to cancel pulse spillover can improve device performance.  Similarly, the dominant errors observed in QSCOUT experiments stemmed directly from how the gates were calibrated.  Minor changes to that protocol --- e.g., independent calibrations for ``idle neighbor'' and ``driven neighbor'' contexts --- could reduce crosstalk errors below detectable thresholds. 

Our analysis demonstrated the utility and performance of crosstalk compensation in the AQT system.  Simultaneous GST could be used to enable continued, iterative reduction of crosstalk errors (via iterative calibration), but we don't think this is the right idea.  Full simultaneous two-qubit GST requires too much overhead to be used in an active optimization loop. However, we believe it is possible to construct simpler, more targeted characterization protocols that focus on a particular type of crosstalk.  These will run much faster, and be suitable for inclusion in active feedback loops.  We propose that the role of ``heavy'' protocols like GST is to identify \emph{which} crosstalk errors are dominant, so that specialized ``lightweight'' protocols can be deployed to tame them.


\section{Acknowledgements}
Sandia National Laboratories is a multimission laboratory managed and operated by National Technology and Engineering Solutions of Sandia, LLC, a wholly owned subsidiary of Honeywell International, Inc., for the U.S. Department of Energy's National Nuclear Security Administration under contract DE-NA0003525. This material was funded in part by the by the U.S. Department of Energy, Office of Science, Office of Advanced Scientific Computing Research's Quantum Testbeds for Science, Quantum Testbed Pathfinder, and Early Career Research Programs, by Sandia National Laboratories' Laboratory Directed Research and Development Program, and by the Office of the Director of National Intelligence (ODNI), Intelligence Advanced Research Projects Activity (IARPA). All statements of fact, opinion or conclusions contained herein are those of the authors and should not be construed as representing the official views or policies of IARPA, the ODNI, the U.S. Department of Energy, or the U.S. Government.

\appendix

\section{Gate set tomography circuits used in our experiments}
\label{sec:sgst_circuits}
Following \cite{Blume-Kohout2017-ww} and as shown in Fig.~\ref{fig:fig1} and Eq.~\ref{eq:1}, GST circuits consist of:
\begin{enumerate}
    \item preparing the system in the all-zeros state,
    \item applying a short preparation fiducial subcircuit, $\mathsf{p}_i$,
    \item applying a short germ subcircuit (repeated $n$ times), $\mathsf{g}_j^n$,
    \item applying a short measurement fiducial subcircuit, $\mathsf{m}_k$,
    \item measuring the qubits in the computational basis.
\end{enumerate}
In Table~\ref{tab:sgst_circuits} we list all $\{\mathsf{p}_i\}$, $\{\mathsf{g}_j\}$, and $\{\mathsf{m}_k\}$ subcircuits for the GST circuits used in our experiments. For each germ $\mathsf{g}_j$, $n$ takes on the values $n=\floor{L/\text{len}(\mathsf{g}_j)}$ for $L\in\{1,2,4,8\cdots,L_{\max}\}$. For AQT, $L_{\max}^{\rm{(AQT)}}=32$, while for QSCOUT, $L_{\max}^{\rm{(QSCOUT)}}=8$.  The GST circuit list includes all possible combinations of subcircuits of the form of Eq.~\ref{eq:1}. 

The fiducials and germs were chosen numerically via pyGSTi \cite{nielsen2019python} such that the fiducials generate an informationally complete set of states and measurements, and the germs are sensitive to all parameters in the general crosstalk model. Additionally, the germs and fiducials contain the circuits necessary to run isolated, single-qubit GST on each component qubit.  This last requirement did not increase the number of fiducials required, but did necessitate the addition of two germs (the last two in Table~\ref{tab:sgst_circuits}).

\setlength{\extrarowheight}{5pt}
\begin{table*}
\begin{tabular*}{\textwidth}{c@{\extracolsep{\fill}}cc} 
\midrule
Preparation Fiducials & Germs & Measurement Fiducials \\
\midrule
 $\left\{\right\}$ & $\left(\mathsf{I}^{(j)}\mathsf{I}^{(k)}\right)$ & $\left\{\right\}$ \\ 
 $\gateX^{(k)}$ & $\gateX^{(k)}$ & $\gateX^{(k)}$ \\ 
 $\gateY^{(k)}$ & $\gateY^{(k)}$ & $\gateY^{(k)}$ \\ 
 $\gateX^{(k)}\gateX^{(k)}$ & $\gateX^{(j)}$ & $\gateX^{(k)}\gateX^{(k)}$ \\ 
 $\gateX^{(j)}$ & $\gateY^{(j)}$ & $\gateX^{(j)}$ \\ 
 $\left(\gateX^{(j)}\gateX^{(k)}\right)\left(\gateX^{(j)}\gateX^{(k)}\right)\left(\gateX^{(j)}\gateX^{(k)}\right)$ & $\left(\gateX^{(j)}\gateX^{(k)}\right)$ & $\gateY^{(j)}$ \\ 
 $\left(\gateX^{(j)}\gateY^{(k)}\right)$ & $\left(\gateY^{(j)}\gateY^{(k)}\right)$ & $\gateX^{(j)}\gateX^{(j)}$ \\ 
 $\left(\gateX^{(j)}\gateX^{(k)}\right)\gateX^{(k)}$ & $\left(\gateX^{(j)}\gateY^{(k)}\right)$ & $\left(\gateX^{(j)}\gateX^{(k)}\right)\left(\gateX^{(j)}\gateX^{(k)}\right)\left(\gateX^{(j)}\gateX^{(k)}\right)$ \\ 
 $\gateY^{(j)}$ & $\left(\gateY^{(j)}\gateX^{(k)}\right)$ & $\left(\gateX^{(j)}\gateY^{(k)}\right)$ \\ 
 $\left(\gateY^{(j)}\gateX^{(k)}\right)$ & $\left(\gateX^{(j)}\gateX^{(k)}\right)\left(\gateY^{(j)}\gateX^{(k)}\right)\left(\gateY^{(j)}\gateY^{(k)}\right)$ & $\left(\gateY^{(j)}\gateX^{(k)}\right)$ \\ 
 $\left(\gateY^{(j)}\gateY^{(k)}\right)\left(\gateY^{(j)}\gateY^{(k)}\right)\left(\gateY^{(j)}\gateY^{(k)}\right)$ & $\left(\gateX^{(j)}\gateX^{(k)}\right)\left(\gateX^{(j)}\gateY^{(k)}\right)\left(\gateY^{(j)}\gateY^{(k)}\right)$ & $\left(\gateY^{(j)}\gateY^{(k)}\right)\left(\gateY^{(j)}\gateY^{(k)}\right)\left(\gateY^{(j)}\gateY^{(k)}\right)$ \\ 
 $\left(\gateY^{(j)}\gateX^{(k)}\right)\gateX^{(k)}$ & $\gateY^{(j)}\left(\gateY^{(j)}\gateX^{(k)}\right)\left(\gateX^{(j)}\gateX^{(k)}\right)$ &  \\ 
 $\gateX^{(j)}\gateX^{(j)}$ & $\gateY^{(k)}\left(\gateX^{(j)}\gateY^{(k)}\right)\left(\gateX^{(j)}\gateX^{(k)}\right)$ &  \\ 
 $\left(\gateX^{(j)}\gateX^{(k)}\right)\gateX^{(j)}$ & $\left(\gateY^{(j)}\gateX^{(k)}\right)\gateX^{(k)}\left(\gateX^{(j)}\gateY^{(k)}\right)\gateX^{(j)}$ &  \\ 
 $\left(\gateX^{(j)}\gateY^{(k)}\right)\gateX^{(j)}$ & $\gateX^{(j)}\left(\gateY^{(j)}\gateY^{(k)}\right)\left(\gateX^{(j)}\gateY^{(k)}\right)$ &  \\ 
 $\left(\gateX^{(j)}\gateX^{(k)}\right)\left(\gateX^{(j)}\gateX^{(k)}\right)$ & $\gateX^{(k)}\left(\gateX^{(j)}\gateX^{(k)}\right)\left(\gateX^{(j)}\gateY^{(k)}\right)$ &  \\ 
  & $\gateY^{(j)}\left(\gateY^{(j)}\gateY^{(k)}\right)\gateY^{(k)}\gateX^{(j)}$ &  \\ 
  & $\left(\gateY^{(j)}\gateY^{(k)}\right)\left(\gateX^{(j)}\gateY^{(k)}\right)\left(\gateY^{(j)}\gateX^{(k)}\right)$ &  \\ 
  & $\gateY^{(j)}\left(\gateX^{(j)}\gateY^{(k)}\right)\left(\gateY^{(j)}\gateY^{(k)}\right)$ &  \\ 
  & $\gateY^{(k)}\left(\gateY^{(j)}\gateX^{(k)}\right)\gateX^{(j)}$ &  \\ 
  & $\gateX^{(k)}\gateY^{(k)}$ &  \\ 
  & $\left(\gateY^{(j)}\gateY^{(k)}\right)\left(\gateY^{(j)}\gateX^{(k)}\right)$ &  \\ 
  & $\gateX^{(j)}\gateY^{(j)}$ &  \\ 
  & $\gateX^{(j)}\gateX^{(j)}\gateY^{(j)}$ &  \\ 
  & $\gateX^{(k)}\gateX^{(k)}\gateY^{(k)} $ & \\
  \midrule
\end{tabular*}
\caption{Building blocks of the GST circuits used in our experiments to investigate crosstalk between two qubits $j$ and $k$. Operations are applied sequentially from left to right, and parentheses indicate operations on separate qubits that are intended to be applied simultaneously. The full circuits of all possible choices of a preparation fiducial, an $n$-fold repeated germ operation, and a measurement fiducial (see main text). The particular set of germs and fiducials listed here was selected to enable high-precision estimation of all physical parameters of a general crosstalk model. \label{tab:sgst_circuits}}
\end{table*}

\section{Simulation}
\label{sec:simulation}

For all of the experiments presented in the main text, context-dependent models happened to provide the best balance of explanatory power and simplicity. This is evidence that correlated and entangling errors, which are specifically excluded from context-dependent models, are unnecessary to explain the observed data. Of course, not all experiments can be explained with these models. For instance, as a superconducting processor, the AQT platform might have been expected to experience weak $ZZ$-type coupling between the qubits~\cite{Mundada2018-qm}. In this section, we provide numerical evidence that our methods are capable of detecting a coherent entangling error and selecting the general crosstalk model, as long as the magnitude of the error is above a certain threshold, for a given amount of data gathered.

\begin{figure}[t]
  \includegraphics[keepaspectratio=true,width=\columnwidth]{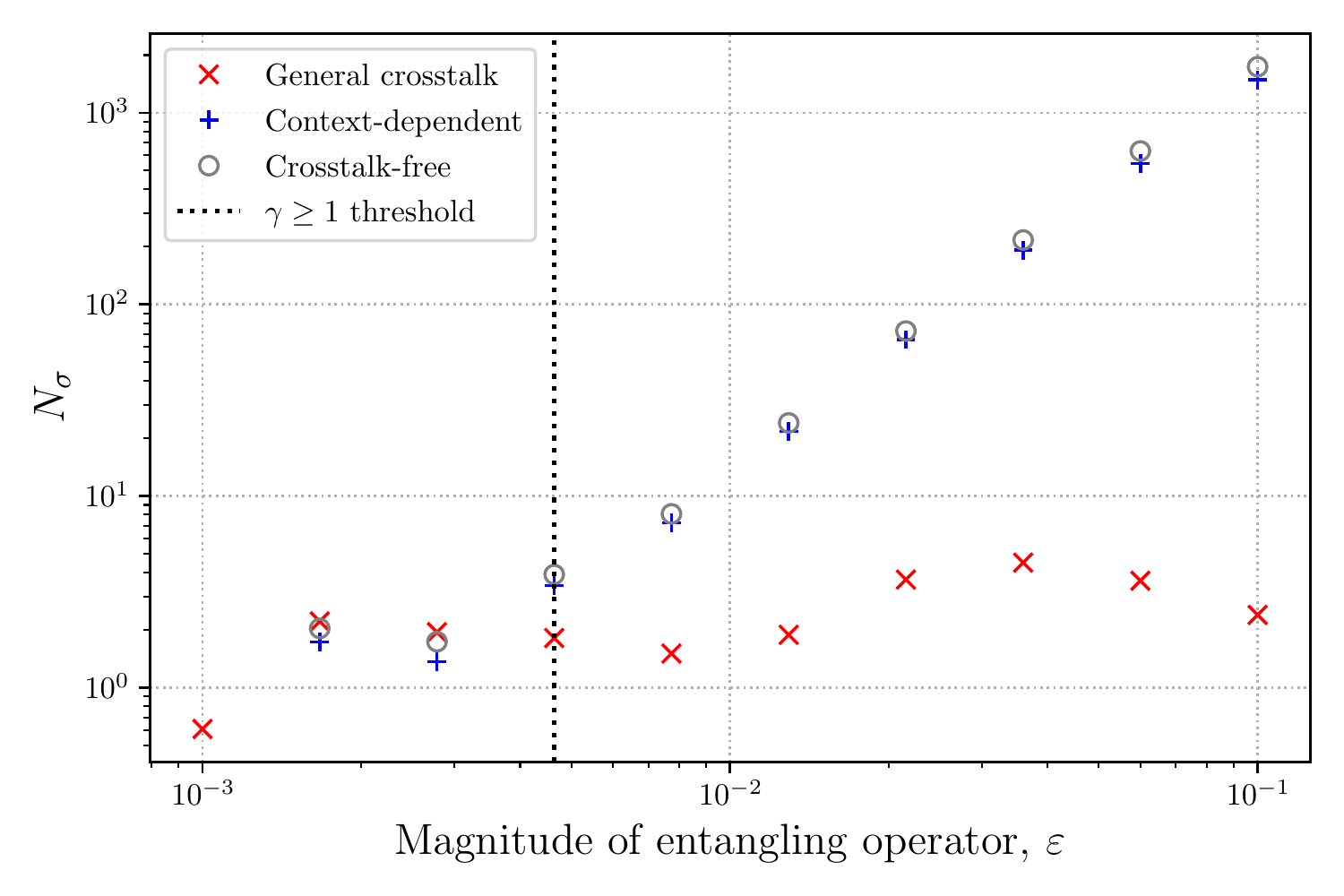}
  \caption{\textbf{Explanatory power of estimated crosstalk models vs. strength of entangling error (simulation).} Here, $N_{\sigma}$ quantifies fit violation for each model. When $\varepsilon < \varepsilon^* = 4.6 \times 10^{-3}$, the predictions of all three estimated models are statistically consistent with the simulated data. As $\varepsilon$ increases above this threshold, $N_\sigma$ for the context-dependent and crosstalk-free models increases rapidly, while for the general crosstalk model remains relatively constant. In this regime, the general crosstalk model is preferred by the evidence ratio test.}
  \label{fig:sim_log-likelihood}
\end{figure}

\begin{figure}[b!]
  \includegraphics[keepaspectratio=true,width=\columnwidth]{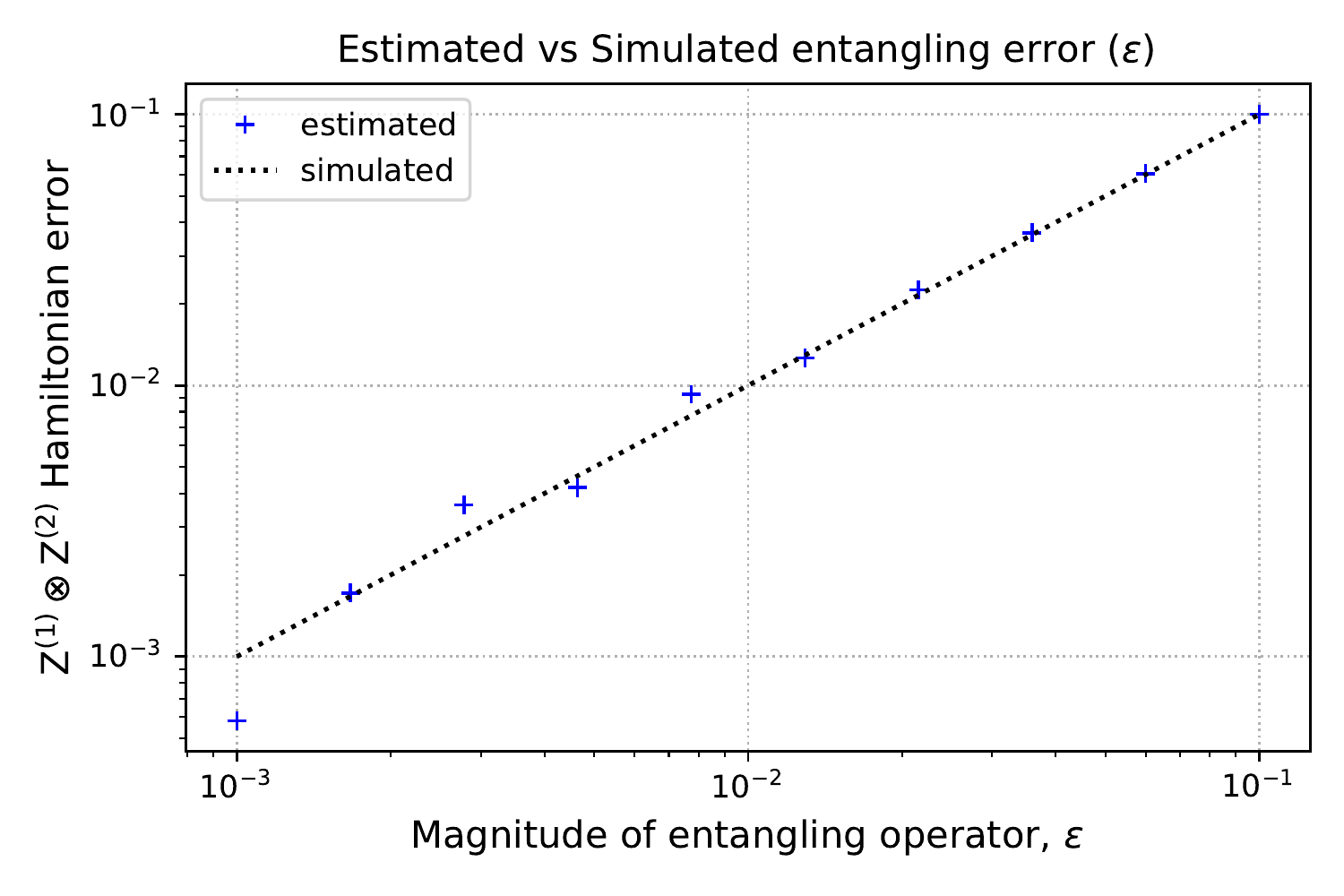}
  \caption{\textbf{Accuracy of the general crosstalk model's entangling entangling error estimate (simulation).} The general crosstalk model is able to accurately estimate the magnitude of a simulated entangling $\mathsf{Z}\otimes\mathsf{Z}$ crosstalk error.  Shown here is the magnitude of the estimate of this error term (extracted from the crosstalk-free model) for the idle gate versus the true magnitude of the entangling error.}
  \label{fig:sim_Hamiltonian_error}
\end{figure}

\begin{figure}[t]
  \includegraphics[keepaspectratio=true,width=\columnwidth]{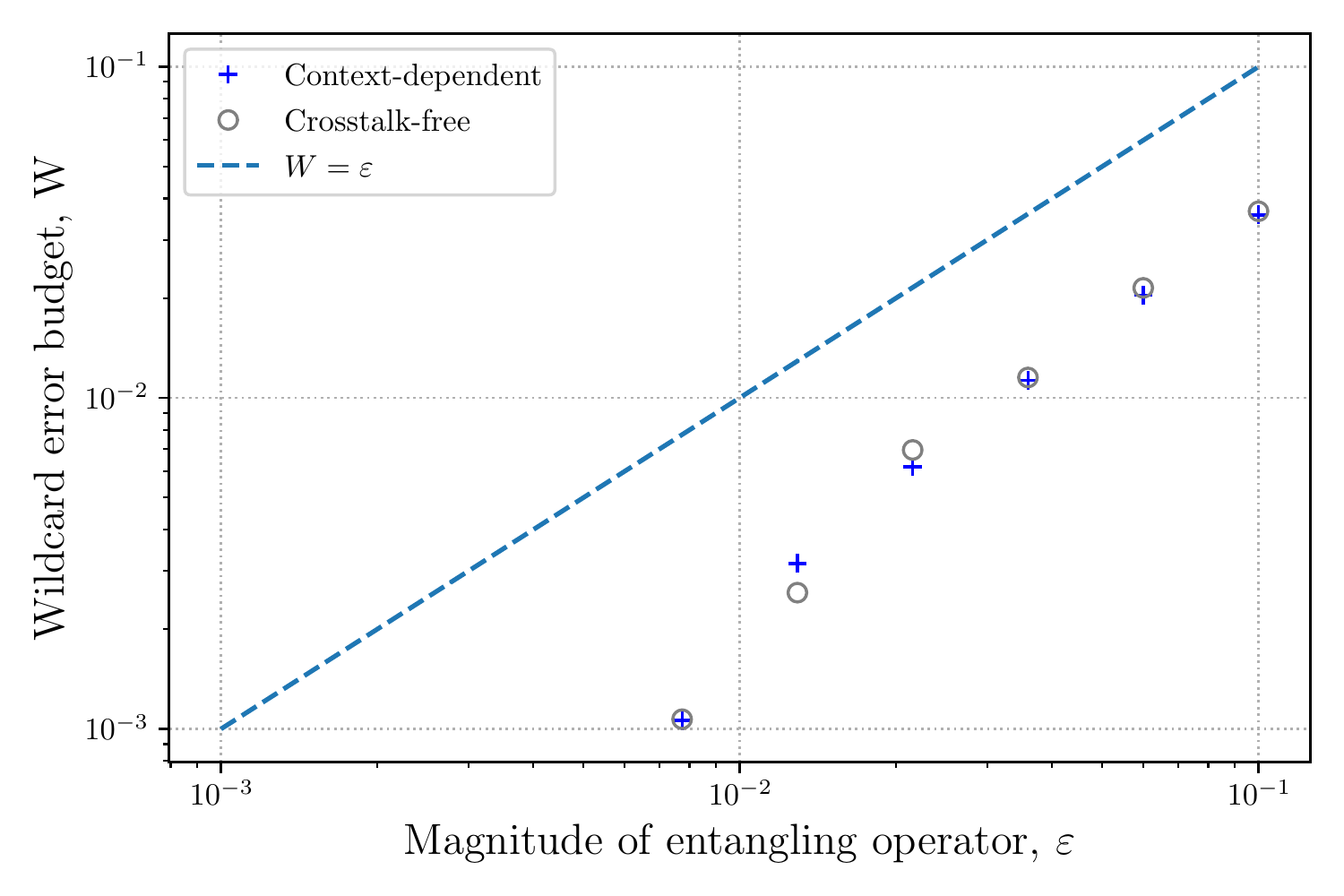}
  \caption{\textbf{Wildcard error for entangling-error-free models in the presence of entangling error (simulation).}  Here we show the wildcard error assigned to fits of the context-dependent and crosstalk-free models when an entangling error is present.  For $\varepsilon \leq \varepsilon^*=4.6 \times 10^{-3}$, the wildcard error budget for all three models is zero. However, for $\varepsilon > 4.6 \times 10^{-3}$, the wildcard error for the context-dependent and crosstalk-free models increase  approximately linearly with $\varepsilon$ in the regime shown. The wildcard error for the general crosstalk model is zero for all values of $\varepsilon$, as this model always properly fits the data.}
  \label{fig:sim_wildcard_error}
\end{figure}

For our simulations, we use the same target gate set and GST circuit family as in our experiments. We set $L_{\max}^{\rm{(sim)}} = 8$ and simulate 1000 measurements per circuit. Each circuit layer experiences an identical $Z^{(1)} \otimes Z^{(2)}$ entangling error of strength $\epsilon$, described by a Hamiltonian error generator (which is applied concurrent with each gate operation):
\begin{align}
    \Delta \mathcal{H}(\rho) &= -i \left[\frac{\varepsilon}{2}\,Z^{(1)}\otimes Z^{(2)}, \rho\right].
\end{align}
For each of 10 distinct, exponentially-spaced values of $\varepsilon$, we simulate data and fit our models. Results of these simulations are presented in Figs.~\ref{fig:sim_log-likelihood} -- \ref{fig:sim_wildcard_error}.  

\subsection{Analysis of simulations}
For the parameters used in our simulations, we identify a threshold value for the rate of the entangling error  $\varepsilon^* \simeq 4.6 \times 10^{-3}$, above which we successfully detect the injected crosstalk \emph{and} choose the crosstalk-containing model as the best fit. As shown in Fig.~\ref{fig:sim_log-likelihood}, for simulations with $\varepsilon \geq \varepsilon^*$,  $N_\sigma$ for the context-dependent and crosstalk-free models grows linearly with $\varepsilon$, but remains approximately constant for the general model. In these cases, the evidence ratio strongly (and correctly) selects the general model. Below the threshold, the context-dependent model is weakly preferred to the crosstalk-free model.  The precision of GST scales with the number of measurements $N$ and the maximum length $L_{\rm{max}}$ as $O(1/(L\sqrt{N}))$, and we expect similar scaling in the threshold $\varepsilon^*$.

From each fit to the general crosstalk model we can extract an estimate of the strength of the entangling $ZZ$ error present in any gate estimate.  We do this for the general crosstalk model's idle gate estimate, and plot it against $\epsilon$ in Fig.~\ref{fig:sim_Hamiltonian_error}.  We see in all cases considered that the general crosstalk model is able to accurately reconstruct this entangling error, whether or not the general crosstalk model's evidence ratio indicates its selection over the other models.

In Fig.~\ref{fig:sim_wildcard_error}, we show the wildcard error $W$ for the context-dependent and crosstalk-free models as a function of $\varepsilon$. For $\varepsilon>\varepsilon^*$, the wildcard error is proportional to, but non-trivially smaller than, the magnitude of the error.  This indicates that the wildcard is capturing some of the unmodeled error, but not all of it. In fact, this is entirely expected behavior. The wildcard error is intended to capture the extra error \emph{per gate} that is required for the model to be consistent with the data. For the wildcard to be approximately equal to the true value of the $ZZ$ coherent error rate, there would need to exist a circuit of depth $d$ for which the total variational distance between the observed and predicted probabilities was $\delta_{\rm TVD}(p_o, p_p) = d \epsilon$. No such circuit exists, though some get close.  For simulations with $\varepsilon<\varepsilon*$, all models are consistent with the data, so $W$ is zero. 

\clearpage
\bibliography{simultaneousGST.bib}

\begin{thebibliography}{77}%
\makeatletter
\providecommand \@ifxundefined [1]{%
 \@ifx{#1\undefined}
}%
\providecommand \@ifnum [1]{%
 \ifnum #1\expandafter \@firstoftwo
 \else \expandafter \@secondoftwo
 \fi
}%
\providecommand \@ifx [1]{%
 \ifx #1\expandafter \@firstoftwo
 \else \expandafter \@secondoftwo
 \fi
}%
\providecommand \natexlab [1]{#1}%
\providecommand \enquote  [1]{``#1''}%
\providecommand \bibnamefont  [1]{#1}%
\providecommand \bibfnamefont [1]{#1}%
\providecommand \citenamefont [1]{#1}%
\providecommand \href@noop [0]{\@secondoftwo}%
\providecommand \href [0]{\begingroup \@sanitize@url \@href}%
\providecommand \@href[1]{\@@startlink{#1}\@@href}%
\providecommand \@@href[1]{\endgroup#1\@@endlink}%
\providecommand \@sanitize@url [0]{\catcode `\\12\catcode `\$12\catcode
  `\&12\catcode `\#12\catcode `\^12\catcode `\_12\catcode `\%12\relax}%
\providecommand \@@startlink[1]{}%
\providecommand \@@endlink[0]{}%
\providecommand \url  [0]{\begingroup\@sanitize@url \@url }%
\providecommand \@url [1]{\endgroup\@href {#1}{\urlprefix }}%
\providecommand \urlprefix  [0]{URL }%
\providecommand \Eprint [0]{\href }%
\providecommand \doibase [0]{http://dx.doi.org/}%
\providecommand \selectlanguage [0]{\@gobble}%
\providecommand \bibinfo  [0]{\@secondoftwo}%
\providecommand \bibfield  [0]{\@secondoftwo}%
\providecommand \translation [1]{[#1]}%
\providecommand \BibitemOpen [0]{}%
\providecommand \bibitemStop [0]{}%
\providecommand \bibitemNoStop [0]{.\EOS\space}%
\providecommand \EOS [0]{\spacefactor3000\relax}%
\providecommand \BibitemShut  [1]{\csname bibitem#1\endcsname}%
\let\auto@bib@innerbib\@empty
\bibitem [{\citenamefont {Chow}\ \emph {et~al.}(2012)\citenamefont {Chow},
  \citenamefont {Gambetta}, \citenamefont {C{\'o}rcoles}, \citenamefont
  {Merkel}, \citenamefont {Smolin}, \citenamefont {Rigetti}, \citenamefont
  {Poletto}, \citenamefont {Keefe}, \citenamefont {Rothwell}, \citenamefont
  {Rozen}, \citenamefont {Ketchen},\ and\ \citenamefont
  {Steffen}}]{Chow2012-ey}%
  \BibitemOpen
  \bibfield  {author} {\bibinfo {author} {\bibfnamefont {J.~M.}\ \bibnamefont
  {Chow}}, \bibinfo {author} {\bibfnamefont {J.~M.}\ \bibnamefont {Gambetta}},
  \bibinfo {author} {\bibfnamefont {A.~D.}\ \bibnamefont {C{\'o}rcoles}},
  \bibinfo {author} {\bibfnamefont {S.~T.}\ \bibnamefont {Merkel}}, \bibinfo
  {author} {\bibfnamefont {J.~A.}\ \bibnamefont {Smolin}}, \bibinfo {author}
  {\bibfnamefont {C.}~\bibnamefont {Rigetti}}, \bibinfo {author} {\bibfnamefont
  {S.}~\bibnamefont {Poletto}}, \bibinfo {author} {\bibfnamefont {G.~A.}\
  \bibnamefont {Keefe}}, \bibinfo {author} {\bibfnamefont {M.~B.}\ \bibnamefont
  {Rothwell}}, \bibinfo {author} {\bibfnamefont {J.~R.}\ \bibnamefont {Rozen}},
  \bibinfo {author} {\bibfnamefont {M.~B.}\ \bibnamefont {Ketchen}}, \ and\
  \bibinfo {author} {\bibfnamefont {M.}~\bibnamefont {Steffen}},\ }\href
  {\doibase 10.1103/PhysRevLett.109.060501} {\bibfield  {journal} {\bibinfo
  {journal} {Phys. Rev. Lett.}\ }\textbf {\bibinfo {volume} {109}},\ \bibinfo
  {pages} {060501} (\bibinfo {year} {2012})}\BibitemShut {NoStop}%
\bibitem [{\citenamefont {Barends}\ \emph {et~al.}(2014)\citenamefont
  {Barends}, \citenamefont {Kelly}, \citenamefont {Megrant}, \citenamefont
  {Veitia}, \citenamefont {Sank}, \citenamefont {Jeffrey}, \citenamefont
  {White}, \citenamefont {Mutus}, \citenamefont {Fowler}, \citenamefont
  {Campbell}, \citenamefont {Chen}, \citenamefont {Chen}, \citenamefont
  {Chiaro}, \citenamefont {Dunsworth}, \citenamefont {Neill}, \citenamefont
  {O'Malley}, \citenamefont {Roushan}, \citenamefont {Vainsencher},
  \citenamefont {Wenner}, \citenamefont {Korotkov}, \citenamefont {Cleland},\
  and\ \citenamefont {Martinis}}]{Barends2014-ap}%
  \BibitemOpen
  \bibfield  {author} {\bibinfo {author} {\bibfnamefont {R.}~\bibnamefont
  {Barends}}, \bibinfo {author} {\bibfnamefont {J.}~\bibnamefont {Kelly}},
  \bibinfo {author} {\bibfnamefont {A.}~\bibnamefont {Megrant}}, \bibinfo
  {author} {\bibfnamefont {A.}~\bibnamefont {Veitia}}, \bibinfo {author}
  {\bibfnamefont {D.}~\bibnamefont {Sank}}, \bibinfo {author} {\bibfnamefont
  {E.}~\bibnamefont {Jeffrey}}, \bibinfo {author} {\bibfnamefont {T.~C.}\
  \bibnamefont {White}}, \bibinfo {author} {\bibfnamefont {J.}~\bibnamefont
  {Mutus}}, \bibinfo {author} {\bibfnamefont {A.~G.}\ \bibnamefont {Fowler}},
  \bibinfo {author} {\bibfnamefont {B.}~\bibnamefont {Campbell}}, \bibinfo
  {author} {\bibfnamefont {Y.}~\bibnamefont {Chen}}, \bibinfo {author}
  {\bibfnamefont {Z.}~\bibnamefont {Chen}}, \bibinfo {author} {\bibfnamefont
  {B.}~\bibnamefont {Chiaro}}, \bibinfo {author} {\bibfnamefont
  {A.}~\bibnamefont {Dunsworth}}, \bibinfo {author} {\bibfnamefont
  {C.}~\bibnamefont {Neill}}, \bibinfo {author} {\bibfnamefont
  {P.}~\bibnamefont {O'Malley}}, \bibinfo {author} {\bibfnamefont
  {P.}~\bibnamefont {Roushan}}, \bibinfo {author} {\bibfnamefont
  {A.}~\bibnamefont {Vainsencher}}, \bibinfo {author} {\bibfnamefont
  {J.}~\bibnamefont {Wenner}}, \bibinfo {author} {\bibfnamefont {A.~N.}\
  \bibnamefont {Korotkov}}, \bibinfo {author} {\bibfnamefont {A.~N.}\
  \bibnamefont {Cleland}}, \ and\ \bibinfo {author} {\bibfnamefont {J.~M.}\
  \bibnamefont {Martinis}},\ }\href {\doibase 10.1038/nature13171} {\bibfield
  {journal} {\bibinfo  {journal} {Nature}\ }\textbf {\bibinfo {volume} {508}},\
  \bibinfo {pages} {500} (\bibinfo {year} {2014})}\BibitemShut {NoStop}%
\bibitem [{\citenamefont {Gaebler}\ \emph {et~al.}(2016)\citenamefont
  {Gaebler}, \citenamefont {Tan}, \citenamefont {Lin}, \citenamefont {Wan},
  \citenamefont {Bowler}, \citenamefont {Keith}, \citenamefont {Glancy},
  \citenamefont {Coakley}, \citenamefont {Knill}, \citenamefont {Leibfried},\
  and\ \citenamefont {{Others}}}]{Gaebler2016-rp}%
  \BibitemOpen
  \bibfield  {author} {\bibinfo {author} {\bibfnamefont {J.~P.}\ \bibnamefont
  {Gaebler}}, \bibinfo {author} {\bibfnamefont {T.~R.}\ \bibnamefont {Tan}},
  \bibinfo {author} {\bibfnamefont {Y.}~\bibnamefont {Lin}}, \bibinfo {author}
  {\bibfnamefont {Y.}~\bibnamefont {Wan}}, \bibinfo {author} {\bibfnamefont
  {R.}~\bibnamefont {Bowler}}, \bibinfo {author} {\bibfnamefont {A.~C.}\
  \bibnamefont {Keith}}, \bibinfo {author} {\bibfnamefont {S.}~\bibnamefont
  {Glancy}}, \bibinfo {author} {\bibfnamefont {K.}~\bibnamefont {Coakley}},
  \bibinfo {author} {\bibfnamefont {E.}~\bibnamefont {Knill}}, \bibinfo
  {author} {\bibfnamefont {D.}~\bibnamefont {Leibfried}}, \ and\ \bibinfo
  {author} {\bibnamefont {{Others}}},\ }\href@noop {} {\bibfield  {journal}
  {\bibinfo  {journal} {Phys. Rev. Lett.}\ }\textbf {\bibinfo {volume} {117}},\
  \bibinfo {pages} {060505} (\bibinfo {year} {2016})}\BibitemShut {NoStop}%
\bibitem [{\citenamefont {Ballance}\ \emph {et~al.}(2016)\citenamefont
  {Ballance}, \citenamefont {Harty}, \citenamefont {Linke}, \citenamefont
  {Sepiol},\ and\ \citenamefont {Lucas}}]{Ballance2016-ab}%
  \BibitemOpen
  \bibfield  {author} {\bibinfo {author} {\bibfnamefont {C.~J.}\ \bibnamefont
  {Ballance}}, \bibinfo {author} {\bibfnamefont {T.~P.}\ \bibnamefont {Harty}},
  \bibinfo {author} {\bibfnamefont {N.~M.}\ \bibnamefont {Linke}}, \bibinfo
  {author} {\bibfnamefont {M.~A.}\ \bibnamefont {Sepiol}}, \ and\ \bibinfo
  {author} {\bibfnamefont {D.~M.}\ \bibnamefont {Lucas}},\ }\href {\doibase
  10.1103/PhysRevLett.117.060504} {\bibfield  {journal} {\bibinfo  {journal}
  {Phys. Rev. Lett.}\ }\textbf {\bibinfo {volume} {117}},\ \bibinfo {pages}
  {060504} (\bibinfo {year} {2016})}\BibitemShut {NoStop}%
\bibitem [{\citenamefont {Blume-Kohout}\ \emph {et~al.}(2017)\citenamefont
  {Blume-Kohout}, \citenamefont {Gamble}, \citenamefont {Nielsen},
  \citenamefont {Rudinger}, \citenamefont {Mizrahi}, \citenamefont {Fortier},\
  and\ \citenamefont {Maunz}}]{Blume-Kohout2017-ww}%
  \BibitemOpen
  \bibfield  {author} {\bibinfo {author} {\bibfnamefont {R.}~\bibnamefont
  {Blume-Kohout}}, \bibinfo {author} {\bibfnamefont {J.~K.}\ \bibnamefont
  {Gamble}}, \bibinfo {author} {\bibfnamefont {E.}~\bibnamefont {Nielsen}},
  \bibinfo {author} {\bibfnamefont {K.}~\bibnamefont {Rudinger}}, \bibinfo
  {author} {\bibfnamefont {J.}~\bibnamefont {Mizrahi}}, \bibinfo {author}
  {\bibfnamefont {K.}~\bibnamefont {Fortier}}, \ and\ \bibinfo {author}
  {\bibfnamefont {P.}~\bibnamefont {Maunz}},\ }\href {\doibase
  10.1038/ncomms14485} {\bibfield  {journal} {\bibinfo  {journal} {Nat.
  Commun.}\ }\textbf {\bibinfo {volume} {8}} (\bibinfo {year} {2017}),\
  10.1038/ncomms14485}\BibitemShut {NoStop}%
\bibitem [{\citenamefont {Barends}\ \emph {et~al.}(2019)\citenamefont
  {Barends}, \citenamefont {Quintana}, \citenamefont {Petukhov}, \citenamefont
  {Chen}, \citenamefont {Kafri}, \citenamefont {Kechedzhi}, \citenamefont
  {Collins}, \citenamefont {Naaman}, \citenamefont {Boixo}, \citenamefont
  {Arute}, \citenamefont {Arya}, \citenamefont {Buell}, \citenamefont
  {Burkett}, \citenamefont {Chen}, \citenamefont {Chiaro}, \citenamefont
  {Dunsworth}, \citenamefont {Foxen}, \citenamefont {Fowler}, \citenamefont
  {Gidney}, \citenamefont {Giustina}, \citenamefont {Graff}, \citenamefont
  {Huang}, \citenamefont {Jeffrey}, \citenamefont {Kelly}, \citenamefont
  {Klimov}, \citenamefont {Kostritsa}, \citenamefont {Landhuis}, \citenamefont
  {Lucero}, \citenamefont {McEwen}, \citenamefont {Megrant}, \citenamefont
  {Mi}, \citenamefont {Mutus}, \citenamefont {Neeley}, \citenamefont {Neill},
  \citenamefont {Ostby}, \citenamefont {Roushan}, \citenamefont {Sank},
  \citenamefont {Satzinger}, \citenamefont {Vainsencher}, \citenamefont
  {White}, \citenamefont {Yao}, \citenamefont {Yeh}, \citenamefont {Zalcman},
  \citenamefont {Neven}, \citenamefont {Smelyanskiy},\ and\ \citenamefont
  {Martinis}}]{Barends2019-mt}%
  \BibitemOpen
  \bibfield  {author} {\bibinfo {author} {\bibfnamefont {R.}~\bibnamefont
  {Barends}}, \bibinfo {author} {\bibfnamefont {C.~M.}\ \bibnamefont
  {Quintana}}, \bibinfo {author} {\bibfnamefont {A.~G.}\ \bibnamefont
  {Petukhov}}, \bibinfo {author} {\bibfnamefont {Y.}~\bibnamefont {Chen}},
  \bibinfo {author} {\bibfnamefont {D.}~\bibnamefont {Kafri}}, \bibinfo
  {author} {\bibfnamefont {K.}~\bibnamefont {Kechedzhi}}, \bibinfo {author}
  {\bibfnamefont {R.}~\bibnamefont {Collins}}, \bibinfo {author} {\bibfnamefont
  {O.}~\bibnamefont {Naaman}}, \bibinfo {author} {\bibfnamefont
  {S.}~\bibnamefont {Boixo}}, \bibinfo {author} {\bibfnamefont
  {F.}~\bibnamefont {Arute}}, \bibinfo {author} {\bibfnamefont
  {K.}~\bibnamefont {Arya}}, \bibinfo {author} {\bibfnamefont {D.}~\bibnamefont
  {Buell}}, \bibinfo {author} {\bibfnamefont {B.}~\bibnamefont {Burkett}},
  \bibinfo {author} {\bibfnamefont {Z.}~\bibnamefont {Chen}}, \bibinfo {author}
  {\bibfnamefont {B.}~\bibnamefont {Chiaro}}, \bibinfo {author} {\bibfnamefont
  {A.}~\bibnamefont {Dunsworth}}, \bibinfo {author} {\bibfnamefont
  {B.}~\bibnamefont {Foxen}}, \bibinfo {author} {\bibfnamefont
  {A.}~\bibnamefont {Fowler}}, \bibinfo {author} {\bibfnamefont
  {C.}~\bibnamefont {Gidney}}, \bibinfo {author} {\bibfnamefont
  {M.}~\bibnamefont {Giustina}}, \bibinfo {author} {\bibfnamefont
  {R.}~\bibnamefont {Graff}}, \bibinfo {author} {\bibfnamefont
  {T.}~\bibnamefont {Huang}}, \bibinfo {author} {\bibfnamefont
  {E.}~\bibnamefont {Jeffrey}}, \bibinfo {author} {\bibfnamefont
  {J.}~\bibnamefont {Kelly}}, \bibinfo {author} {\bibfnamefont {P.~V.}\
  \bibnamefont {Klimov}}, \bibinfo {author} {\bibfnamefont {F.}~\bibnamefont
  {Kostritsa}}, \bibinfo {author} {\bibfnamefont {D.}~\bibnamefont {Landhuis}},
  \bibinfo {author} {\bibfnamefont {E.}~\bibnamefont {Lucero}}, \bibinfo
  {author} {\bibfnamefont {M.}~\bibnamefont {McEwen}}, \bibinfo {author}
  {\bibfnamefont {A.}~\bibnamefont {Megrant}}, \bibinfo {author} {\bibfnamefont
  {X.}~\bibnamefont {Mi}}, \bibinfo {author} {\bibfnamefont {J.}~\bibnamefont
  {Mutus}}, \bibinfo {author} {\bibfnamefont {M.}~\bibnamefont {Neeley}},
  \bibinfo {author} {\bibfnamefont {C.}~\bibnamefont {Neill}}, \bibinfo
  {author} {\bibfnamefont {E.}~\bibnamefont {Ostby}}, \bibinfo {author}
  {\bibfnamefont {P.}~\bibnamefont {Roushan}}, \bibinfo {author} {\bibfnamefont
  {D.}~\bibnamefont {Sank}}, \bibinfo {author} {\bibfnamefont {K.~J.}\
  \bibnamefont {Satzinger}}, \bibinfo {author} {\bibfnamefont {A.}~\bibnamefont
  {Vainsencher}}, \bibinfo {author} {\bibfnamefont {T.}~\bibnamefont {White}},
  \bibinfo {author} {\bibfnamefont {J.}~\bibnamefont {Yao}}, \bibinfo {author}
  {\bibfnamefont {P.}~\bibnamefont {Yeh}}, \bibinfo {author} {\bibfnamefont
  {A.}~\bibnamefont {Zalcman}}, \bibinfo {author} {\bibfnamefont
  {H.}~\bibnamefont {Neven}}, \bibinfo {author} {\bibfnamefont {V.~N.}\
  \bibnamefont {Smelyanskiy}}, \ and\ \bibinfo {author} {\bibfnamefont {J.~M.}\
  \bibnamefont {Martinis}},\ }\href {\doibase 10.1103/PhysRevLett.123.210501}
  {\bibfield  {journal} {\bibinfo  {journal} {Phys. Rev. Lett.}\ }\textbf
  {\bibinfo {volume} {123}},\ \bibinfo {pages} {210501} (\bibinfo {year}
  {2019})}\BibitemShut {NoStop}%
\bibitem [{\citenamefont {Neg{\^\i}rneac}\ \emph {et~al.}(2020)\citenamefont
  {Neg{\^\i}rneac}, \citenamefont {Ali}, \citenamefont {Muthusubramanian},
  \citenamefont {Battistel}, \citenamefont {Sagastizabal}, \citenamefont
  {Moreira}, \citenamefont {Marques}, \citenamefont {Vlothuizen}, \citenamefont
  {Beekman}, \citenamefont {Haider}, \citenamefont {Bruno},\ and\ \citenamefont
  {DiCarlo}}]{Negirneac2020-eg}%
  \BibitemOpen
  \bibfield  {author} {\bibinfo {author} {\bibfnamefont {V.}~\bibnamefont
  {Neg{\^\i}rneac}}, \bibinfo {author} {\bibfnamefont {H.}~\bibnamefont {Ali}},
  \bibinfo {author} {\bibfnamefont {N.}~\bibnamefont {Muthusubramanian}},
  \bibinfo {author} {\bibfnamefont {F.}~\bibnamefont {Battistel}}, \bibinfo
  {author} {\bibfnamefont {R.}~\bibnamefont {Sagastizabal}}, \bibinfo {author}
  {\bibfnamefont {M.~S.}\ \bibnamefont {Moreira}}, \bibinfo {author}
  {\bibfnamefont {J.~F.}\ \bibnamefont {Marques}}, \bibinfo {author}
  {\bibfnamefont {W.}~\bibnamefont {Vlothuizen}}, \bibinfo {author}
  {\bibfnamefont {M.}~\bibnamefont {Beekman}}, \bibinfo {author} {\bibfnamefont
  {N.}~\bibnamefont {Haider}}, \bibinfo {author} {\bibfnamefont
  {A.}~\bibnamefont {Bruno}}, \ and\ \bibinfo {author} {\bibfnamefont
  {L.}~\bibnamefont {DiCarlo}},\ }\href@noop {} {\bibfield  {journal} {\bibinfo
   {journal} {arXiv}\ } (\bibinfo {year} {2020})},\ \Eprint
  {http://arxiv.org/abs/2008.07411} {arXiv:2008.07411 [quant-ph]} \BibitemShut
  {NoStop}%
\bibitem [{\citenamefont {Srinivas}\ \emph {et~al.}(2021)\citenamefont
  {Srinivas}, \citenamefont {Burd}, \citenamefont {Knaack}, \citenamefont
  {Sutherland}, \citenamefont {Kwiatkowski}, \citenamefont {Glancy},
  \citenamefont {Knill}, \citenamefont {Wineland}, \citenamefont {Leibfried},
  \citenamefont {Wilson}, \citenamefont {Allcock},\ and\ \citenamefont
  {Slichter}}]{Srinivas2021-fm}%
  \BibitemOpen
  \bibfield  {author} {\bibinfo {author} {\bibfnamefont {R.}~\bibnamefont
  {Srinivas}}, \bibinfo {author} {\bibfnamefont {S.~C.}\ \bibnamefont {Burd}},
  \bibinfo {author} {\bibfnamefont {H.~M.}\ \bibnamefont {Knaack}}, \bibinfo
  {author} {\bibfnamefont {R.~T.}\ \bibnamefont {Sutherland}}, \bibinfo
  {author} {\bibfnamefont {A.}~\bibnamefont {Kwiatkowski}}, \bibinfo {author}
  {\bibfnamefont {S.}~\bibnamefont {Glancy}}, \bibinfo {author} {\bibfnamefont
  {E.}~\bibnamefont {Knill}}, \bibinfo {author} {\bibfnamefont {D.~J.}\
  \bibnamefont {Wineland}}, \bibinfo {author} {\bibfnamefont {D.}~\bibnamefont
  {Leibfried}}, \bibinfo {author} {\bibfnamefont {A.~C.}\ \bibnamefont
  {Wilson}}, \bibinfo {author} {\bibfnamefont {D.~T.~C.}\ \bibnamefont
  {Allcock}}, \ and\ \bibinfo {author} {\bibfnamefont {D.~H.}\ \bibnamefont
  {Slichter}},\ }\href@noop {} {\bibfield  {journal} {\bibinfo  {journal}
  {arXiv}\ } (\bibinfo {year} {2021})},\ \Eprint
  {http://arxiv.org/abs/2102.12533} {arXiv:2102.12533 [quant-ph]} \BibitemShut
  {NoStop}%
\bibitem [{\citenamefont {Li}\ \emph {et~al.}(2019)\citenamefont {Li},
  \citenamefont {Castellano}, \citenamefont {Wang}, \citenamefont {Wu},
  \citenamefont {Gong}, \citenamefont {Yan}, \citenamefont {Rong},
  \citenamefont {Deng}, \citenamefont {Zha}, \citenamefont {Guo}, \citenamefont
  {Sun}, \citenamefont {Peng}, \citenamefont {Zhu},\ and\ \citenamefont
  {Pan}}]{Li2019-uf}%
  \BibitemOpen
  \bibfield  {author} {\bibinfo {author} {\bibfnamefont {S.}~\bibnamefont
  {Li}}, \bibinfo {author} {\bibfnamefont {A.~D.}\ \bibnamefont {Castellano}},
  \bibinfo {author} {\bibfnamefont {S.}~\bibnamefont {Wang}}, \bibinfo {author}
  {\bibfnamefont {Y.}~\bibnamefont {Wu}}, \bibinfo {author} {\bibfnamefont
  {M.}~\bibnamefont {Gong}}, \bibinfo {author} {\bibfnamefont {Z.}~\bibnamefont
  {Yan}}, \bibinfo {author} {\bibfnamefont {H.}~\bibnamefont {Rong}}, \bibinfo
  {author} {\bibfnamefont {H.}~\bibnamefont {Deng}}, \bibinfo {author}
  {\bibfnamefont {C.}~\bibnamefont {Zha}}, \bibinfo {author} {\bibfnamefont
  {C.}~\bibnamefont {Guo}}, \bibinfo {author} {\bibfnamefont {L.}~\bibnamefont
  {Sun}}, \bibinfo {author} {\bibfnamefont {C.}~\bibnamefont {Peng}}, \bibinfo
  {author} {\bibfnamefont {X.}~\bibnamefont {Zhu}}, \ and\ \bibinfo {author}
  {\bibfnamefont {J.-W.}\ \bibnamefont {Pan}},\ }\href {\doibase
  10.1038/s41534-019-0202-7} {\bibfield  {journal} {\bibinfo  {journal} {npj
  Quantum Information}\ }\textbf {\bibinfo {volume} {5}},\ \bibinfo {pages}
  {84} (\bibinfo {year} {2019})}\BibitemShut {NoStop}%
\bibitem [{\citenamefont {Sung}\ \emph {et~al.}(2020)\citenamefont {Sung},
  \citenamefont {Ding}, \citenamefont {Braum{\"u}ller}, \citenamefont
  {Veps{\"a}l{\"a}inen}, \citenamefont {Kannan}, \citenamefont {Kjaergaard},
  \citenamefont {Greene}, \citenamefont {Samach}, \citenamefont {McNally},
  \citenamefont {Kim}, \citenamefont {Melville}, \citenamefont {Niedzielski},
  \citenamefont {Schwartz}, \citenamefont {Yoder}, \citenamefont {Orlando},
  \citenamefont {Gustavsson},\ and\ \citenamefont {Oliver}}]{Sung2020-jv}%
  \BibitemOpen
  \bibfield  {author} {\bibinfo {author} {\bibfnamefont {Y.}~\bibnamefont
  {Sung}}, \bibinfo {author} {\bibfnamefont {L.}~\bibnamefont {Ding}}, \bibinfo
  {author} {\bibfnamefont {J.}~\bibnamefont {Braum{\"u}ller}}, \bibinfo
  {author} {\bibfnamefont {A.}~\bibnamefont {Veps{\"a}l{\"a}inen}}, \bibinfo
  {author} {\bibfnamefont {B.}~\bibnamefont {Kannan}}, \bibinfo {author}
  {\bibfnamefont {M.}~\bibnamefont {Kjaergaard}}, \bibinfo {author}
  {\bibfnamefont {A.}~\bibnamefont {Greene}}, \bibinfo {author} {\bibfnamefont
  {G.~O.}\ \bibnamefont {Samach}}, \bibinfo {author} {\bibfnamefont
  {C.}~\bibnamefont {McNally}}, \bibinfo {author} {\bibfnamefont
  {D.}~\bibnamefont {Kim}}, \bibinfo {author} {\bibfnamefont {A.}~\bibnamefont
  {Melville}}, \bibinfo {author} {\bibfnamefont {B.~M.}\ \bibnamefont
  {Niedzielski}}, \bibinfo {author} {\bibfnamefont {M.~E.}\ \bibnamefont
  {Schwartz}}, \bibinfo {author} {\bibfnamefont {J.~L.}\ \bibnamefont {Yoder}},
  \bibinfo {author} {\bibfnamefont {T.~P.}\ \bibnamefont {Orlando}}, \bibinfo
  {author} {\bibfnamefont {S.}~\bibnamefont {Gustavsson}}, \ and\ \bibinfo
  {author} {\bibfnamefont {W.~D.}\ \bibnamefont {Oliver}},\ }\href@noop {} {\
  (\bibinfo {year} {2020})},\ \Eprint {http://arxiv.org/abs/2011.01261}
  {arXiv:2011.01261 [quant-ph]} \BibitemShut {NoStop}%
\bibitem [{\citenamefont {Brink}\ \emph {et~al.}(2018)\citenamefont {Brink},
  \citenamefont {Chow}, \citenamefont {Hertzberg}, \citenamefont {Magesan},\
  and\ \citenamefont {Rosenblatt}}]{Brink2018-la}%
  \BibitemOpen
  \bibfield  {author} {\bibinfo {author} {\bibfnamefont {M.}~\bibnamefont
  {Brink}}, \bibinfo {author} {\bibfnamefont {J.~M.}\ \bibnamefont {Chow}},
  \bibinfo {author} {\bibfnamefont {J.}~\bibnamefont {Hertzberg}}, \bibinfo
  {author} {\bibfnamefont {E.}~\bibnamefont {Magesan}}, \ and\ \bibinfo
  {author} {\bibfnamefont {S.}~\bibnamefont {Rosenblatt}},\ }in\ \href
  {\doibase 10.1109/IEDM.2018.8614500} {\emph {\bibinfo {booktitle} {2018
  {IEEE} International Electron Devices Meeting ({{IEDM}})}}}\ (\bibinfo {year}
  {2018})\ pp.\ \bibinfo {pages} {6.1.1--6.1.3}\BibitemShut {NoStop}%
\bibitem [{\citenamefont {Debroy}\ \emph {et~al.}(2019)\citenamefont {Debroy},
  \citenamefont {Li}, \citenamefont {Huang},\ and\ \citenamefont
  {Brown}}]{Debroy2019-nd}%
  \BibitemOpen
  \bibfield  {author} {\bibinfo {author} {\bibfnamefont {D.~M.}\ \bibnamefont
  {Debroy}}, \bibinfo {author} {\bibfnamefont {M.}~\bibnamefont {Li}}, \bibinfo
  {author} {\bibfnamefont {S.}~\bibnamefont {Huang}}, \ and\ \bibinfo {author}
  {\bibfnamefont {K.~R.}\ \bibnamefont {Brown}},\ }\href@noop {} {\bibfield
  {journal} {\bibinfo  {journal} {arXiv}\ } (\bibinfo {year} {2019})},\ \Eprint
  {http://arxiv.org/abs/1910.08495} {arXiv:1910.08495 [quant-ph]} \BibitemShut
  {NoStop}%
\bibitem [{\citenamefont {Gessner}\ \emph {et~al.}(2020)\citenamefont
  {Gessner}, \citenamefont {Fabre},\ and\ \citenamefont
  {Treps}}]{Gessner2020-vs}%
  \BibitemOpen
  \bibfield  {author} {\bibinfo {author} {\bibfnamefont {M.}~\bibnamefont
  {Gessner}}, \bibinfo {author} {\bibfnamefont {C.}~\bibnamefont {Fabre}}, \
  and\ \bibinfo {author} {\bibfnamefont {N.}~\bibnamefont {Treps}},\
  }\href@noop {} {\bibfield  {journal} {\bibinfo  {journal} {arXiv}\ }
  (\bibinfo {year} {2020})},\ \Eprint {http://arxiv.org/abs/2004.07228}
  {arXiv:2004.07228 [quant-ph]} \BibitemShut {NoStop}%
\bibitem [{\citenamefont {Huang}\ \emph {et~al.}(2020)\citenamefont {Huang},
  \citenamefont {Ni}, \citenamefont {Zhang}, \citenamefont {Newman},
  \citenamefont {Ding}, \citenamefont {Gao}, \citenamefont {Wang},
  \citenamefont {Zhao}, \citenamefont {Wu}, \citenamefont {Zhang},
  \citenamefont {Deng}, \citenamefont {Ku}, \citenamefont {Chen},\ and\
  \citenamefont {Shi}}]{Huang2020-jo}%
  \BibitemOpen
  \bibfield  {author} {\bibinfo {author} {\bibfnamefont {C.}~\bibnamefont
  {Huang}}, \bibinfo {author} {\bibfnamefont {X.}~\bibnamefont {Ni}}, \bibinfo
  {author} {\bibfnamefont {F.}~\bibnamefont {Zhang}}, \bibinfo {author}
  {\bibfnamefont {M.}~\bibnamefont {Newman}}, \bibinfo {author} {\bibfnamefont
  {D.}~\bibnamefont {Ding}}, \bibinfo {author} {\bibfnamefont {X.}~\bibnamefont
  {Gao}}, \bibinfo {author} {\bibfnamefont {T.}~\bibnamefont {Wang}}, \bibinfo
  {author} {\bibfnamefont {H.-H.}\ \bibnamefont {Zhao}}, \bibinfo {author}
  {\bibfnamefont {F.}~\bibnamefont {Wu}}, \bibinfo {author} {\bibfnamefont
  {G.}~\bibnamefont {Zhang}}, \bibinfo {author} {\bibfnamefont
  {C.}~\bibnamefont {Deng}}, \bibinfo {author} {\bibfnamefont {H.-S.}\
  \bibnamefont {Ku}}, \bibinfo {author} {\bibfnamefont {J.}~\bibnamefont
  {Chen}}, \ and\ \bibinfo {author} {\bibfnamefont {Y.}~\bibnamefont {Shi}},\
  }\href@noop {} {\bibfield  {journal} {\bibinfo  {journal} {arXiv}\ }
  (\bibinfo {year} {2020})},\ \Eprint {http://arxiv.org/abs/2002.08918}
  {arXiv:2002.08918 [quant-ph]} \BibitemShut {NoStop}%
\bibitem [{\citenamefont {Sarovar}\ \emph {et~al.}(2020)\citenamefont
  {Sarovar}, \citenamefont {Proctor}, \citenamefont {Rudinger}, \citenamefont
  {Young}, \citenamefont {Nielsen},\ and\ \citenamefont
  {Blume-Kohout}}]{Sarovar2019-gc}%
  \BibitemOpen
  \bibfield  {author} {\bibinfo {author} {\bibfnamefont {M.}~\bibnamefont
  {Sarovar}}, \bibinfo {author} {\bibfnamefont {T.}~\bibnamefont {Proctor}},
  \bibinfo {author} {\bibfnamefont {K.}~\bibnamefont {Rudinger}}, \bibinfo
  {author} {\bibfnamefont {K.}~\bibnamefont {Young}}, \bibinfo {author}
  {\bibfnamefont {E.}~\bibnamefont {Nielsen}}, \ and\ \bibinfo {author}
  {\bibfnamefont {R.}~\bibnamefont {Blume-Kohout}},\ }\href@noop {} {\bibfield
  {journal} {\bibinfo  {journal} {Quantum}\ }\textbf {\bibinfo {volume} {4}},\
  \bibinfo {pages} {321} (\bibinfo {year} {2020})}\BibitemShut {NoStop}%
\bibitem [{\citenamefont {Landahl}\ \emph {et~al.}(2011)\citenamefont
  {Landahl}, \citenamefont {Anderson},\ and\ \citenamefont
  {Rice}}]{landahl2011fault}%
  \BibitemOpen
  \bibfield  {author} {\bibinfo {author} {\bibfnamefont {A.~J.}\ \bibnamefont
  {Landahl}}, \bibinfo {author} {\bibfnamefont {J.~T.}\ \bibnamefont
  {Anderson}}, \ and\ \bibinfo {author} {\bibfnamefont {P.~R.}\ \bibnamefont
  {Rice}},\ }\href@noop {} {\bibfield  {journal} {\bibinfo  {journal} {arXiv
  preprint arXiv:1108.5738}\ } (\bibinfo {year} {2011})}\BibitemShut {NoStop}%
\bibitem [{\citenamefont {Devitt}\ \emph {et~al.}(2013)\citenamefont {Devitt},
  \citenamefont {Munro},\ and\ \citenamefont {Nemoto}}]{devitt2013quantum}%
  \BibitemOpen
  \bibfield  {author} {\bibinfo {author} {\bibfnamefont {S.~J.}\ \bibnamefont
  {Devitt}}, \bibinfo {author} {\bibfnamefont {W.~J.}\ \bibnamefont {Munro}}, \
  and\ \bibinfo {author} {\bibfnamefont {K.}~\bibnamefont {Nemoto}},\
  }\href@noop {} {\bibfield  {journal} {\bibinfo  {journal} {Reports on
  Progress in Physics}\ }\textbf {\bibinfo {volume} {76}},\ \bibinfo {pages}
  {076001} (\bibinfo {year} {2013})}\BibitemShut {NoStop}%
\bibitem [{\citenamefont {Fowler}\ \emph {et~al.}(2012)\citenamefont {Fowler},
  \citenamefont {Mariantoni}, \citenamefont {Martinis},\ and\ \citenamefont
  {Cleland}}]{fowler2012surface}%
  \BibitemOpen
  \bibfield  {author} {\bibinfo {author} {\bibfnamefont {A.~G.}\ \bibnamefont
  {Fowler}}, \bibinfo {author} {\bibfnamefont {M.}~\bibnamefont {Mariantoni}},
  \bibinfo {author} {\bibfnamefont {J.~M.}\ \bibnamefont {Martinis}}, \ and\
  \bibinfo {author} {\bibfnamefont {A.~N.}\ \bibnamefont {Cleland}},\ }\href
  {https://journals.aps.org/pra/abstract/10.1103/PhysRevA.86.032324} {\bibfield
   {journal} {\bibinfo  {journal} {Physical Review A}\ }\textbf {\bibinfo
  {volume} {86}},\ \bibinfo {pages} {032324} (\bibinfo {year}
  {2012})}\BibitemShut {NoStop}%
\bibitem [{\citenamefont {Buterakos}\ \emph {et~al.}(2018)\citenamefont
  {Buterakos}, \citenamefont {Throckmorton},\ and\ \citenamefont
  {Das~Sarma}}]{Buterakos2018-sk}%
  \BibitemOpen
  \bibfield  {author} {\bibinfo {author} {\bibfnamefont {D.}~\bibnamefont
  {Buterakos}}, \bibinfo {author} {\bibfnamefont {R.~E.}\ \bibnamefont
  {Throckmorton}}, \ and\ \bibinfo {author} {\bibfnamefont {S.}~\bibnamefont
  {Das~Sarma}},\ }\href {\doibase 10.1103/PhysRevB.98.035406} {\bibfield
  {journal} {\bibinfo  {journal} {Phys. Rev. B Condens. Matter}\ }\textbf
  {\bibinfo {volume} {98}},\ \bibinfo {pages} {035406} (\bibinfo {year}
  {2018})}\BibitemShut {NoStop}%
\bibitem [{\citenamefont {Carvalho}\ \emph {et~al.}(2020)\citenamefont
  {Carvalho}, \citenamefont {Ball}, \citenamefont {Biercuk}, \citenamefont
  {Hush},\ and\ \citenamefont {Thomsen}}]{Carvalho2020-mi}%
  \BibitemOpen
  \bibfield  {author} {\bibinfo {author} {\bibfnamefont {A.~R.~R.}\
  \bibnamefont {Carvalho}}, \bibinfo {author} {\bibfnamefont {H.}~\bibnamefont
  {Ball}}, \bibinfo {author} {\bibfnamefont {M.~J.}\ \bibnamefont {Biercuk}},
  \bibinfo {author} {\bibfnamefont {M.~R.}\ \bibnamefont {Hush}}, \ and\
  \bibinfo {author} {\bibfnamefont {F.}~\bibnamefont {Thomsen}},\ }\href@noop
  {} {\bibfield  {journal} {\bibinfo  {journal} {arXiv}\ } (\bibinfo {year}
  {2020})},\ \Eprint {http://arxiv.org/abs/2010.08057} {arXiv:2010.08057
  [quant-ph]} \BibitemShut {NoStop}%
\bibitem [{\citenamefont {Chen}\ \emph {et~al.}(2019)\citenamefont {Chen},
  \citenamefont {Farahzad}, \citenamefont {Yoo},\ and\ \citenamefont
  {Wei}}]{Chen2019-hf}%
  \BibitemOpen
  \bibfield  {author} {\bibinfo {author} {\bibfnamefont {Y.}~\bibnamefont
  {Chen}}, \bibinfo {author} {\bibfnamefont {M.}~\bibnamefont {Farahzad}},
  \bibinfo {author} {\bibfnamefont {S.}~\bibnamefont {Yoo}}, \ and\ \bibinfo
  {author} {\bibfnamefont {T.-C.}\ \bibnamefont {Wei}},\ }\href {\doibase
  10.1103/PhysRevA.100.052315} {\bibfield  {journal} {\bibinfo  {journal}
  {Phys. Rev. A}\ }\textbf {\bibinfo {volume} {100}},\ \bibinfo {pages}
  {052315} (\bibinfo {year} {2019})}\BibitemShut {NoStop}%
\bibitem [{\citenamefont {Crain}\ \emph {et~al.}(2019)\citenamefont {Crain},
  \citenamefont {Cahall}, \citenamefont {Vrijsen}, \citenamefont {Wollman},
  \citenamefont {Shaw}, \citenamefont {Verma}, \citenamefont {Nam},\ and\
  \citenamefont {Kim}}]{Crain2019-eq}%
  \BibitemOpen
  \bibfield  {author} {\bibinfo {author} {\bibfnamefont {S.}~\bibnamefont
  {Crain}}, \bibinfo {author} {\bibfnamefont {C.}~\bibnamefont {Cahall}},
  \bibinfo {author} {\bibfnamefont {G.}~\bibnamefont {Vrijsen}}, \bibinfo
  {author} {\bibfnamefont {E.~E.}\ \bibnamefont {Wollman}}, \bibinfo {author}
  {\bibfnamefont {M.~D.}\ \bibnamefont {Shaw}}, \bibinfo {author}
  {\bibfnamefont {V.~B.}\ \bibnamefont {Verma}}, \bibinfo {author}
  {\bibfnamefont {S.~W.}\ \bibnamefont {Nam}}, \ and\ \bibinfo {author}
  {\bibfnamefont {J.}~\bibnamefont {Kim}},\ }\href {\doibase
  10.1038/s42005-019-0195-8} {\bibfield  {journal} {\bibinfo  {journal}
  {Communications Physics}\ }\textbf {\bibinfo {volume} {2}},\ \bibinfo {pages}
  {97} (\bibinfo {year} {2019})}\BibitemShut {NoStop}%
\bibitem [{\citenamefont {Debroy}\ and\ \citenamefont
  {Brown}(2020)}]{Debroy2020-ui}%
  \BibitemOpen
  \bibfield  {author} {\bibinfo {author} {\bibfnamefont {D.~M.}\ \bibnamefont
  {Debroy}}\ and\ \bibinfo {author} {\bibfnamefont {K.~R.}\ \bibnamefont
  {Brown}},\ }\href@noop {} {\bibfield  {journal} {\bibinfo  {journal} {arXiv}\
  } (\bibinfo {year} {2020})},\ \Eprint {http://arxiv.org/abs/2009.07752}
  {arXiv:2009.07752 [quant-ph]} \BibitemShut {NoStop}%
\bibitem [{\citenamefont {Ding}\ \emph {et~al.}(2020)\citenamefont {Ding},
  \citenamefont {Gokhale}, \citenamefont {Lin}, \citenamefont {Rines},
  \citenamefont {Propson},\ and\ \citenamefont {Chong}}]{Ding2020-bd}%
  \BibitemOpen
  \bibfield  {author} {\bibinfo {author} {\bibfnamefont {Y.}~\bibnamefont
  {Ding}}, \bibinfo {author} {\bibfnamefont {P.}~\bibnamefont {Gokhale}},
  \bibinfo {author} {\bibfnamefont {S.~F.}\ \bibnamefont {Lin}}, \bibinfo
  {author} {\bibfnamefont {R.}~\bibnamefont {Rines}}, \bibinfo {author}
  {\bibfnamefont {T.}~\bibnamefont {Propson}}, \ and\ \bibinfo {author}
  {\bibfnamefont {F.~T.}\ \bibnamefont {Chong}},\ }\href@noop {} {\bibfield
  {journal} {\bibinfo  {journal} {arXiv}\ } (\bibinfo {year} {2020})},\ \Eprint
  {http://arxiv.org/abs/2008.09503} {arXiv:2008.09503 [quant-ph]} \BibitemShut
  {NoStop}%
\bibitem [{\citenamefont {Hashim}\ \emph {et~al.}(2020)\citenamefont {Hashim},
  \citenamefont {Naik}, \citenamefont {Morvan}, \citenamefont {Ville},
  \citenamefont {Mitchell}, \citenamefont {Kreikebaum}, \citenamefont {Davis},
  \citenamefont {Smith}, \citenamefont {Iancu}, \citenamefont {O'Brien},
  \citenamefont {Hincks}, \citenamefont {Wallman}, \citenamefont {Emerson},\
  and\ \citenamefont {Siddiqi}}]{Hashim2020-ub}%
  \BibitemOpen
  \bibfield  {author} {\bibinfo {author} {\bibfnamefont {A.}~\bibnamefont
  {Hashim}}, \bibinfo {author} {\bibfnamefont {R.~K.}\ \bibnamefont {Naik}},
  \bibinfo {author} {\bibfnamefont {A.}~\bibnamefont {Morvan}}, \bibinfo
  {author} {\bibfnamefont {J.-L.}\ \bibnamefont {Ville}}, \bibinfo {author}
  {\bibfnamefont {B.}~\bibnamefont {Mitchell}}, \bibinfo {author}
  {\bibfnamefont {J.~M.}\ \bibnamefont {Kreikebaum}}, \bibinfo {author}
  {\bibfnamefont {M.}~\bibnamefont {Davis}}, \bibinfo {author} {\bibfnamefont
  {E.}~\bibnamefont {Smith}}, \bibinfo {author} {\bibfnamefont
  {C.}~\bibnamefont {Iancu}}, \bibinfo {author} {\bibfnamefont {K.~P.}\
  \bibnamefont {O'Brien}}, \bibinfo {author} {\bibfnamefont {I.}~\bibnamefont
  {Hincks}}, \bibinfo {author} {\bibfnamefont {J.~J.}\ \bibnamefont {Wallman}},
  \bibinfo {author} {\bibfnamefont {J.}~\bibnamefont {Emerson}}, \ and\
  \bibinfo {author} {\bibfnamefont {I.}~\bibnamefont {Siddiqi}},\ }\href@noop
  {} {\bibfield  {journal} {\bibinfo  {journal} {arXiv}\ } (\bibinfo {year}
  {2020})},\ \Eprint {http://arxiv.org/abs/2010.00215} {arXiv:2010.00215
  [quant-ph]} \BibitemShut {NoStop}%
\bibitem [{\citenamefont {Ku}\ \emph {et~al.}(2020)\citenamefont {Ku},
  \citenamefont {Xu}, \citenamefont {Brink}, \citenamefont {McKay},
  \citenamefont {Hertzberg}, \citenamefont {Ansari},\ and\ \citenamefont
  {Plourde}}]{Ku2020-ix}%
  \BibitemOpen
  \bibfield  {author} {\bibinfo {author} {\bibfnamefont {J.}~\bibnamefont
  {Ku}}, \bibinfo {author} {\bibfnamefont {X.}~\bibnamefont {Xu}}, \bibinfo
  {author} {\bibfnamefont {M.}~\bibnamefont {Brink}}, \bibinfo {author}
  {\bibfnamefont {D.~C.}\ \bibnamefont {McKay}}, \bibinfo {author}
  {\bibfnamefont {J.~B.}\ \bibnamefont {Hertzberg}}, \bibinfo {author}
  {\bibfnamefont {M.~H.}\ \bibnamefont {Ansari}}, \ and\ \bibinfo {author}
  {\bibfnamefont {B.~L.~T.}\ \bibnamefont {Plourde}},\ }\href@noop {}
  {\bibfield  {journal} {\bibinfo  {journal} {arXiv}\ } (\bibinfo {year}
  {2020})},\ \Eprint {http://arxiv.org/abs/2003.02775} {arXiv:2003.02775
  [quant-ph]} \BibitemShut {NoStop}%
\bibitem [{\citenamefont {Majumder}\ \emph {et~al.}(2020)\citenamefont
  {Majumder}, \citenamefont {de~Castro},\ and\ \citenamefont
  {Brown}}]{Majumder2020-ut}%
  \BibitemOpen
  \bibfield  {author} {\bibinfo {author} {\bibfnamefont {S.}~\bibnamefont
  {Majumder}}, \bibinfo {author} {\bibfnamefont {L.~A.}\ \bibnamefont
  {de~Castro}}, \ and\ \bibinfo {author} {\bibfnamefont {K.~R.}\ \bibnamefont
  {Brown}},\ }\href {\doibase 10.1038/s41534-020-0251-y} {\bibfield  {journal}
  {\bibinfo  {journal} {npj Quantum Information}\ }\textbf {\bibinfo {volume}
  {6}},\ \bibinfo {pages} {1} (\bibinfo {year} {2020})}\BibitemShut {NoStop}%
\bibitem [{\citenamefont {Mundada}\ \emph {et~al.}(2018)\citenamefont
  {Mundada}, \citenamefont {Zhang}, \citenamefont {Hazard},\ and\ \citenamefont
  {Houck}}]{Mundada2018-qm}%
  \BibitemOpen
  \bibfield  {author} {\bibinfo {author} {\bibfnamefont {P.~S.}\ \bibnamefont
  {Mundada}}, \bibinfo {author} {\bibfnamefont {G.}~\bibnamefont {Zhang}},
  \bibinfo {author} {\bibfnamefont {T.}~\bibnamefont {Hazard}}, \ and\ \bibinfo
  {author} {\bibfnamefont {A.~A.}\ \bibnamefont {Houck}},\ }\href {\doibase
  10.1103/PhysRevApplied.12.054023} {\bibfield  {journal} {\bibinfo  {journal}
  {Phys. Rev. Applied}\ }\textbf {\bibinfo {volume} {12}},\ \bibinfo {pages}
  {054023} (\bibinfo {year} {2018})},\ \Eprint
  {http://arxiv.org/abs/1810.04182} {arXiv:1810.04182 [quant-ph]} \BibitemShut
  {NoStop}%
\bibitem [{\citenamefont {Murali}\ \emph {et~al.}(2020)\citenamefont {Murali},
  \citenamefont {Mckay}, \citenamefont {Martonosi},\ and\ \citenamefont
  {Javadi-Abhari}}]{Murali2020-ab}%
  \BibitemOpen
  \bibfield  {author} {\bibinfo {author} {\bibfnamefont {P.}~\bibnamefont
  {Murali}}, \bibinfo {author} {\bibfnamefont {D.~C.}\ \bibnamefont {Mckay}},
  \bibinfo {author} {\bibfnamefont {M.}~\bibnamefont {Martonosi}}, \ and\
  \bibinfo {author} {\bibfnamefont {A.}~\bibnamefont {Javadi-Abhari}},\ }in\
  \href {\doibase 10.1145/3373376.3378477} {\emph {\bibinfo {booktitle}
  {Proceedings of the {Twenty-Fifth} International Conference on Architectural
  Support for Programming Languages and Operating Systems}}},\ \bibinfo {series
  and number} {ASPLOS '20}\ (\bibinfo  {publisher} {Association for Computing
  Machinery},\ \bibinfo {address} {New York, NY, USA},\ \bibinfo {year}
  {2020})\ pp.\ \bibinfo {pages} {1001--1016},\ \Eprint
  {http://arxiv.org/abs/2001.02826} {arXiv:2001.02826 [quant-ph]} \BibitemShut
  {NoStop}%
\bibitem [{\citenamefont {Seif}\ \emph {et~al.}(2018)\citenamefont {Seif},
  \citenamefont {Landsman}, \citenamefont {Linke}, \citenamefont {Figgatt},
  \citenamefont {Monroe},\ and\ \citenamefont {Hafezi}}]{Seif2018-sj}%
  \BibitemOpen
  \bibfield  {author} {\bibinfo {author} {\bibfnamefont {A.}~\bibnamefont
  {Seif}}, \bibinfo {author} {\bibfnamefont {K.~A.}\ \bibnamefont {Landsman}},
  \bibinfo {author} {\bibfnamefont {N.~M.}\ \bibnamefont {Linke}}, \bibinfo
  {author} {\bibfnamefont {C.}~\bibnamefont {Figgatt}}, \bibinfo {author}
  {\bibfnamefont {C.}~\bibnamefont {Monroe}}, \ and\ \bibinfo {author}
  {\bibfnamefont {M.}~\bibnamefont {Hafezi}},\ }\href {\doibase
  10.1088/1361-6455/aad62b} {\bibfield  {journal} {\bibinfo  {journal} {J.
  Phys. B At. Mol. Opt. Phys.}\ }\textbf {\bibinfo {volume} {51}},\ \bibinfo
  {pages} {174006} (\bibinfo {year} {2018})}\BibitemShut {NoStop}%
\bibitem [{\citenamefont {Sheldon}\ \emph {et~al.}(2016)\citenamefont
  {Sheldon}, \citenamefont {Magesan}, \citenamefont {Chow},\ and\ \citenamefont
  {Gambetta}}]{Sheldon2016-js}%
  \BibitemOpen
  \bibfield  {author} {\bibinfo {author} {\bibfnamefont {S.}~\bibnamefont
  {Sheldon}}, \bibinfo {author} {\bibfnamefont {E.}~\bibnamefont {Magesan}},
  \bibinfo {author} {\bibfnamefont {J.~M.}\ \bibnamefont {Chow}}, \ and\
  \bibinfo {author} {\bibfnamefont {J.~M.}\ \bibnamefont {Gambetta}},\ }\href
  {\doibase 10.1103/PhysRevA.93.060302} {\bibfield  {journal} {\bibinfo
  {journal} {Phys. Rev. A}\ }\textbf {\bibinfo {volume} {93}},\ \bibinfo
  {pages} {060302} (\bibinfo {year} {2016})}\BibitemShut {NoStop}%
\bibitem [{\citenamefont {Sundaresan}\ \emph {et~al.}(2020)\citenamefont
  {Sundaresan}, \citenamefont {Lauer}, \citenamefont {Pritchett}, \citenamefont
  {Magesan}, \citenamefont {Jurcevic},\ and\ \citenamefont
  {Gambetta}}]{Sundaresan2020-re}%
  \BibitemOpen
  \bibfield  {author} {\bibinfo {author} {\bibfnamefont {N.}~\bibnamefont
  {Sundaresan}}, \bibinfo {author} {\bibfnamefont {I.}~\bibnamefont {Lauer}},
  \bibinfo {author} {\bibfnamefont {E.}~\bibnamefont {Pritchett}}, \bibinfo
  {author} {\bibfnamefont {E.}~\bibnamefont {Magesan}}, \bibinfo {author}
  {\bibfnamefont {P.}~\bibnamefont {Jurcevic}}, \ and\ \bibinfo {author}
  {\bibfnamefont {J.~M.}\ \bibnamefont {Gambetta}},\ }\href@noop {} {\bibfield
  {journal} {\bibinfo  {journal} {arXiv}\ } (\bibinfo {year} {2020})},\ \Eprint
  {http://arxiv.org/abs/2007.02925} {arXiv:2007.02925 [quant-ph]} \BibitemShut
  {NoStop}%
\bibitem [{\citenamefont {Watson}\ \emph {et~al.}(2018)\citenamefont {Watson},
  \citenamefont {Philips}, \citenamefont {Kawakami}, \citenamefont {Ward},
  \citenamefont {Scarlino}, \citenamefont {Veldhorst}, \citenamefont {Savage},
  \citenamefont {Lagally}, \citenamefont {Friesen}, \citenamefont
  {Coppersmith}, \citenamefont {Eriksson},\ and\ \citenamefont
  {Vandersypen}}]{Watson2018-la}%
  \BibitemOpen
  \bibfield  {author} {\bibinfo {author} {\bibfnamefont {T.~F.}\ \bibnamefont
  {Watson}}, \bibinfo {author} {\bibfnamefont {S.~G.~J.}\ \bibnamefont
  {Philips}}, \bibinfo {author} {\bibfnamefont {E.}~\bibnamefont {Kawakami}},
  \bibinfo {author} {\bibfnamefont {D.~R.}\ \bibnamefont {Ward}}, \bibinfo
  {author} {\bibfnamefont {P.}~\bibnamefont {Scarlino}}, \bibinfo {author}
  {\bibfnamefont {M.}~\bibnamefont {Veldhorst}}, \bibinfo {author}
  {\bibfnamefont {D.~E.}\ \bibnamefont {Savage}}, \bibinfo {author}
  {\bibfnamefont {M.~G.}\ \bibnamefont {Lagally}}, \bibinfo {author}
  {\bibfnamefont {M.}~\bibnamefont {Friesen}}, \bibinfo {author} {\bibfnamefont
  {S.~N.}\ \bibnamefont {Coppersmith}}, \bibinfo {author} {\bibfnamefont
  {M.~A.}\ \bibnamefont {Eriksson}}, \ and\ \bibinfo {author} {\bibfnamefont
  {L.~M.~K.}\ \bibnamefont {Vandersypen}},\ }\href {\doibase
  10.1038/nature25766} {\bibfield  {journal} {\bibinfo  {journal} {Nature}\
  }\textbf {\bibinfo {volume} {555}},\ \bibinfo {pages} {633} (\bibinfo {year}
  {2018})}\BibitemShut {NoStop}%
\bibitem [{\citenamefont {Winick}\ \emph {et~al.}(2020)\citenamefont {Winick},
  \citenamefont {Wallman},\ and\ \citenamefont {Emerson}}]{Winick2020-pm}%
  \BibitemOpen
  \bibfield  {author} {\bibinfo {author} {\bibfnamefont {A.}~\bibnamefont
  {Winick}}, \bibinfo {author} {\bibfnamefont {J.~J.}\ \bibnamefont {Wallman}},
  \ and\ \bibinfo {author} {\bibfnamefont {J.}~\bibnamefont {Emerson}},\
  }\href@noop {} {\bibfield  {journal} {\bibinfo  {journal} {arXiv}\ }
  (\bibinfo {year} {2020})},\ \Eprint {http://arxiv.org/abs/2006.09596}
  {arXiv:2006.09596 [quant-ph]} \BibitemShut {NoStop}%
\bibitem [{\citenamefont {Xu}\ \emph {et~al.}(2020)\citenamefont {Xu},
  \citenamefont {Chu}, \citenamefont {Yuan}, \citenamefont {Qiu}, \citenamefont
  {Zhou}, \citenamefont {Zhang}, \citenamefont {Tan}, \citenamefont {Yu},
  \citenamefont {Liu}, \citenamefont {Li}, \citenamefont {Yan},\ and\
  \citenamefont {Yu}}]{Xu2020-rs}%
  \BibitemOpen
  \bibfield  {author} {\bibinfo {author} {\bibfnamefont {Y.}~\bibnamefont
  {Xu}}, \bibinfo {author} {\bibfnamefont {J.}~\bibnamefont {Chu}}, \bibinfo
  {author} {\bibfnamefont {J.}~\bibnamefont {Yuan}}, \bibinfo {author}
  {\bibfnamefont {J.}~\bibnamefont {Qiu}}, \bibinfo {author} {\bibfnamefont
  {Y.}~\bibnamefont {Zhou}}, \bibinfo {author} {\bibfnamefont {L.}~\bibnamefont
  {Zhang}}, \bibinfo {author} {\bibfnamefont {X.}~\bibnamefont {Tan}}, \bibinfo
  {author} {\bibfnamefont {Y.}~\bibnamefont {Yu}}, \bibinfo {author}
  {\bibfnamefont {S.}~\bibnamefont {Liu}}, \bibinfo {author} {\bibfnamefont
  {J.}~\bibnamefont {Li}}, \bibinfo {author} {\bibfnamefont {F.}~\bibnamefont
  {Yan}}, \ and\ \bibinfo {author} {\bibfnamefont {D.}~\bibnamefont {Yu}},\
  }\href@noop {} {\bibfield  {journal} {\bibinfo  {journal} {arXiv}\ }
  (\bibinfo {year} {2020})},\ \Eprint {http://arxiv.org/abs/2006.11860}
  {arXiv:2006.11860 [quant-ph]} \BibitemShut {NoStop}%
\bibitem [{\citenamefont {Abrams}\ \emph {et~al.}(2019)\citenamefont {Abrams},
  \citenamefont {Didier}, \citenamefont {Caldwell}, \citenamefont {Johnson},\
  and\ \citenamefont {Ryan}}]{Abrams2019-di}%
  \BibitemOpen
  \bibfield  {author} {\bibinfo {author} {\bibfnamefont {D.~M.}\ \bibnamefont
  {Abrams}}, \bibinfo {author} {\bibfnamefont {N.}~\bibnamefont {Didier}},
  \bibinfo {author} {\bibfnamefont {S.~A.}\ \bibnamefont {Caldwell}}, \bibinfo
  {author} {\bibfnamefont {B.~R.}\ \bibnamefont {Johnson}}, \ and\ \bibinfo
  {author} {\bibfnamefont {C.~A.}\ \bibnamefont {Ryan}},\ }\href {\doibase
  10.1103/PhysRevApplied.12.064022} {\bibfield  {journal} {\bibinfo  {journal}
  {Phys. Rev. Applied}\ }\textbf {\bibinfo {volume} {12}},\ \bibinfo {pages}
  {064022} (\bibinfo {year} {2019})},\ \Eprint
  {http://arxiv.org/abs/1908.11856} {arXiv:1908.11856 [quant-ph]} \BibitemShut
  {NoStop}%
\bibitem [{\citenamefont {Balasiu}(2018)}]{Balasiu2018-md}%
  \BibitemOpen
  \bibfield  {author} {\bibinfo {author} {\bibfnamefont {S.}~\bibnamefont
  {Balasiu}},\ }\emph {\bibinfo {title} {Characterization of {Multi-Qubit}
  Algorithms with Randomized Benchmarking}},\ \href@noop {} {Ph.D. thesis},\
  \bibinfo  {school} {ETH-Zurich} (\bibinfo {year} {2018})\BibitemShut
  {NoStop}%
\bibitem [{\citenamefont {Kelly}\ \emph {et~al.}(2014)\citenamefont {Kelly},
  \citenamefont {Barends}, \citenamefont {Campbell}, \citenamefont {Chen},
  \citenamefont {Chen}, \citenamefont {Chiaro}, \citenamefont {Dunsworth},
  \citenamefont {Fowler}, \citenamefont {Hoi}, \citenamefont {Jeffrey},
  \citenamefont {Megrant}, \citenamefont {Mutus}, \citenamefont {Neill},
  \citenamefont {O'Malley}, \citenamefont {Quintana}, \citenamefont {Roushan},
  \citenamefont {Sank}, \citenamefont {Vainsencher}, \citenamefont {Wenner},
  \citenamefont {White}, \citenamefont {Cleland},\ and\ \citenamefont
  {Martinis}}]{Kelly2014-qz}%
  \BibitemOpen
  \bibfield  {author} {\bibinfo {author} {\bibfnamefont {J.}~\bibnamefont
  {Kelly}}, \bibinfo {author} {\bibfnamefont {R.}~\bibnamefont {Barends}},
  \bibinfo {author} {\bibfnamefont {B.}~\bibnamefont {Campbell}}, \bibinfo
  {author} {\bibfnamefont {Y.}~\bibnamefont {Chen}}, \bibinfo {author}
  {\bibfnamefont {Z.}~\bibnamefont {Chen}}, \bibinfo {author} {\bibfnamefont
  {B.}~\bibnamefont {Chiaro}}, \bibinfo {author} {\bibfnamefont
  {A.}~\bibnamefont {Dunsworth}}, \bibinfo {author} {\bibfnamefont {A.~G.}\
  \bibnamefont {Fowler}}, \bibinfo {author} {\bibfnamefont {I.-C.}\
  \bibnamefont {Hoi}}, \bibinfo {author} {\bibfnamefont {E.}~\bibnamefont
  {Jeffrey}}, \bibinfo {author} {\bibfnamefont {A.}~\bibnamefont {Megrant}},
  \bibinfo {author} {\bibfnamefont {J.}~\bibnamefont {Mutus}}, \bibinfo
  {author} {\bibfnamefont {C.}~\bibnamefont {Neill}}, \bibinfo {author}
  {\bibfnamefont {P.~J.~J.}\ \bibnamefont {O'Malley}}, \bibinfo {author}
  {\bibfnamefont {C.}~\bibnamefont {Quintana}}, \bibinfo {author}
  {\bibfnamefont {P.}~\bibnamefont {Roushan}}, \bibinfo {author} {\bibfnamefont
  {D.}~\bibnamefont {Sank}}, \bibinfo {author} {\bibfnamefont {A.}~\bibnamefont
  {Vainsencher}}, \bibinfo {author} {\bibfnamefont {J.}~\bibnamefont {Wenner}},
  \bibinfo {author} {\bibfnamefont {T.~C.}\ \bibnamefont {White}}, \bibinfo
  {author} {\bibfnamefont {A.~N.}\ \bibnamefont {Cleland}}, \ and\ \bibinfo
  {author} {\bibfnamefont {J.~M.}\ \bibnamefont {Martinis}},\ }\href {\doibase
  10.1103/PhysRevLett.112.240504} {\bibfield  {journal} {\bibinfo  {journal}
  {Phys. Rev. Lett.}\ }\textbf {\bibinfo {volume} {112}},\ \bibinfo {pages}
  {240504} (\bibinfo {year} {2014})}\BibitemShut {NoStop}%
\bibitem [{\citenamefont {Krinner}\ \emph {et~al.}(2020)\citenamefont
  {Krinner}, \citenamefont {Lazar}, \citenamefont {Remm}, \citenamefont
  {Andersen}, \citenamefont {Lacroix}, \citenamefont {Norris}, \citenamefont
  {Hellings}, \citenamefont {Gabureac}, \citenamefont {Eichler},\ and\
  \citenamefont {Wallraff}}]{Krinner2020-ud}%
  \BibitemOpen
  \bibfield  {author} {\bibinfo {author} {\bibfnamefont {S.}~\bibnamefont
  {Krinner}}, \bibinfo {author} {\bibfnamefont {S.}~\bibnamefont {Lazar}},
  \bibinfo {author} {\bibfnamefont {A.}~\bibnamefont {Remm}}, \bibinfo {author}
  {\bibfnamefont {C.~K.}\ \bibnamefont {Andersen}}, \bibinfo {author}
  {\bibfnamefont {N.}~\bibnamefont {Lacroix}}, \bibinfo {author} {\bibfnamefont
  {G.~J.}\ \bibnamefont {Norris}}, \bibinfo {author} {\bibfnamefont
  {C.}~\bibnamefont {Hellings}}, \bibinfo {author} {\bibfnamefont
  {M.}~\bibnamefont {Gabureac}}, \bibinfo {author} {\bibfnamefont
  {C.}~\bibnamefont {Eichler}}, \ and\ \bibinfo {author} {\bibfnamefont
  {A.}~\bibnamefont {Wallraff}},\ }\href {\doibase
  10.1103/PhysRevApplied.14.024042} {\bibfield  {journal} {\bibinfo  {journal}
  {Phys. Rev. Applied}\ }\textbf {\bibinfo {volume} {14}},\ \bibinfo {pages}
  {024042} (\bibinfo {year} {2020})}\BibitemShut {NoStop}%
\bibitem [{\citenamefont {Piltz}\ \emph {et~al.}(2014)\citenamefont {Piltz},
  \citenamefont {Sriarunothai}, \citenamefont {Var{\'o}n},\ and\ \citenamefont
  {Wunderlich}}]{Piltz2014-bk}%
  \BibitemOpen
  \bibfield  {author} {\bibinfo {author} {\bibfnamefont {C.}~\bibnamefont
  {Piltz}}, \bibinfo {author} {\bibfnamefont {T.}~\bibnamefont {Sriarunothai}},
  \bibinfo {author} {\bibfnamefont {A.~F.}\ \bibnamefont {Var{\'o}n}}, \ and\
  \bibinfo {author} {\bibfnamefont {C.}~\bibnamefont {Wunderlich}},\ }\href
  {\doibase 10.1038/ncomms5679} {\bibfield  {journal} {\bibinfo  {journal}
  {Nat. Commun.}\ }\textbf {\bibinfo {volume} {5}},\ \bibinfo {pages} {4679}
  (\bibinfo {year} {2014})}\BibitemShut {NoStop}%
\bibitem [{\citenamefont {Gambetta}\ \emph {et~al.}(2012)\citenamefont
  {Gambetta}, \citenamefont {C{\'o}rcoles}, \citenamefont {Merkel},
  \citenamefont {Johnson}, \citenamefont {Smolin}, \citenamefont {Chow},
  \citenamefont {Ryan}, \citenamefont {Rigetti}, \citenamefont {Poletto},
  \citenamefont {Ohki}, \citenamefont {Ketchen},\ and\ \citenamefont
  {Steffen}}]{Gambetta2012-lt}%
  \BibitemOpen
  \bibfield  {author} {\bibinfo {author} {\bibfnamefont {J.~M.}\ \bibnamefont
  {Gambetta}}, \bibinfo {author} {\bibfnamefont {A.~D.}\ \bibnamefont
  {C{\'o}rcoles}}, \bibinfo {author} {\bibfnamefont {S.~T.}\ \bibnamefont
  {Merkel}}, \bibinfo {author} {\bibfnamefont {B.~R.}\ \bibnamefont {Johnson}},
  \bibinfo {author} {\bibfnamefont {J.~A.}\ \bibnamefont {Smolin}}, \bibinfo
  {author} {\bibfnamefont {J.~M.}\ \bibnamefont {Chow}}, \bibinfo {author}
  {\bibfnamefont {C.~A.}\ \bibnamefont {Ryan}}, \bibinfo {author}
  {\bibfnamefont {C.}~\bibnamefont {Rigetti}}, \bibinfo {author} {\bibfnamefont
  {S.}~\bibnamefont {Poletto}}, \bibinfo {author} {\bibfnamefont {T.~A.}\
  \bibnamefont {Ohki}}, \bibinfo {author} {\bibfnamefont {M.~B.}\ \bibnamefont
  {Ketchen}}, \ and\ \bibinfo {author} {\bibfnamefont {M.}~\bibnamefont
  {Steffen}},\ }\href {\doibase 10.1103/PhysRevLett.109.240504} {\bibfield
  {journal} {\bibinfo  {journal} {Phys. Rev. Lett.}\ }\textbf {\bibinfo
  {volume} {109}},\ \bibinfo {pages} {240504} (\bibinfo {year}
  {2012})}\BibitemShut {NoStop}%
\bibitem [{\citenamefont {McKay}\ \emph {et~al.}(2020)\citenamefont {McKay},
  \citenamefont {Cross}, \citenamefont {Wood},\ and\ \citenamefont
  {Gambetta}}]{McKay2020-jl}%
  \BibitemOpen
  \bibfield  {author} {\bibinfo {author} {\bibfnamefont {D.~C.}\ \bibnamefont
  {McKay}}, \bibinfo {author} {\bibfnamefont {A.~W.}\ \bibnamefont {Cross}},
  \bibinfo {author} {\bibfnamefont {C.~J.}\ \bibnamefont {Wood}}, \ and\
  \bibinfo {author} {\bibfnamefont {J.~M.}\ \bibnamefont {Gambetta}},\
  }\href@noop {} {\bibfield  {journal} {\bibinfo  {journal} {arXiv}\ }
  (\bibinfo {year} {2020})},\ \Eprint {http://arxiv.org/abs/2003.02354}
  {arXiv:2003.02354 [quant-ph]} \BibitemShut {NoStop}%
\bibitem [{\citenamefont {Erhard}\ \emph {et~al.}(2019)\citenamefont {Erhard},
  \citenamefont {Wallman}, \citenamefont {Postler}, \citenamefont {Meth},
  \citenamefont {Stricker}, \citenamefont {Martinez}, \citenamefont
  {Schindler}, \citenamefont {Monz}, \citenamefont {Emerson},\ and\
  \citenamefont {Blatt}}]{Erhard2019-wk}%
  \BibitemOpen
  \bibfield  {author} {\bibinfo {author} {\bibfnamefont {A.}~\bibnamefont
  {Erhard}}, \bibinfo {author} {\bibfnamefont {J.~J.}\ \bibnamefont {Wallman}},
  \bibinfo {author} {\bibfnamefont {L.}~\bibnamefont {Postler}}, \bibinfo
  {author} {\bibfnamefont {M.}~\bibnamefont {Meth}}, \bibinfo {author}
  {\bibfnamefont {R.}~\bibnamefont {Stricker}}, \bibinfo {author}
  {\bibfnamefont {E.~A.}\ \bibnamefont {Martinez}}, \bibinfo {author}
  {\bibfnamefont {P.}~\bibnamefont {Schindler}}, \bibinfo {author}
  {\bibfnamefont {T.}~\bibnamefont {Monz}}, \bibinfo {author} {\bibfnamefont
  {J.}~\bibnamefont {Emerson}}, \ and\ \bibinfo {author} {\bibfnamefont
  {R.}~\bibnamefont {Blatt}},\ }\href {\doibase 10.1038/s41467-019-13068-7}
  {\bibfield  {journal} {\bibinfo  {journal} {Nat. Commun.}\ }\textbf {\bibinfo
  {volume} {10}},\ \bibinfo {pages} {5347} (\bibinfo {year}
  {2019})}\BibitemShut {NoStop}%
\bibitem [{\citenamefont {Harper}\ \emph {et~al.}(2019)\citenamefont {Harper},
  \citenamefont {Flammia},\ and\ \citenamefont {Wallman}}]{Harper2019-ab}%
  \BibitemOpen
  \bibfield  {author} {\bibinfo {author} {\bibfnamefont {R.}~\bibnamefont
  {Harper}}, \bibinfo {author} {\bibfnamefont {S.~T.}\ \bibnamefont {Flammia}},
  \ and\ \bibinfo {author} {\bibfnamefont {J.~J.}\ \bibnamefont {Wallman}},\
  }\href@noop {} {\bibfield  {journal} {\bibinfo  {journal} {arXiv}\ }
  (\bibinfo {year} {2019})},\ \Eprint {http://arxiv.org/abs/1907.13022}
  {arXiv:1907.13022 [quant-ph]} \BibitemShut {NoStop}%
\bibitem [{\citenamefont {Proctor}\ \emph {et~al.}(2017)\citenamefont
  {Proctor}, \citenamefont {Rudinger}, \citenamefont {Young}, \citenamefont
  {Sarovar},\ and\ \citenamefont {Blume-Kohout}}]{Proctor2017-io}%
  \BibitemOpen
  \bibfield  {author} {\bibinfo {author} {\bibfnamefont {T.}~\bibnamefont
  {Proctor}}, \bibinfo {author} {\bibfnamefont {K.}~\bibnamefont {Rudinger}},
  \bibinfo {author} {\bibfnamefont {K.}~\bibnamefont {Young}}, \bibinfo
  {author} {\bibfnamefont {M.}~\bibnamefont {Sarovar}}, \ and\ \bibinfo
  {author} {\bibfnamefont {R.}~\bibnamefont {Blume-Kohout}},\ }\href {\doibase
  10.1103/PhysRevLett.119.130502} {\bibfield  {journal} {\bibinfo  {journal}
  {Phys. Rev. Lett.}\ }\textbf {\bibinfo {volume} {119}},\ \bibinfo {pages}
  {130502} (\bibinfo {year} {2017})}\BibitemShut {NoStop}%
\bibitem [{\citenamefont {Kueng}\ \emph {et~al.}(2016)\citenamefont {Kueng},
  \citenamefont {Long}, \citenamefont {Doherty},\ and\ \citenamefont
  {Flammia}}]{Kueng2016-kj}%
  \BibitemOpen
  \bibfield  {author} {\bibinfo {author} {\bibfnamefont {R.}~\bibnamefont
  {Kueng}}, \bibinfo {author} {\bibfnamefont {D.~M.}\ \bibnamefont {Long}},
  \bibinfo {author} {\bibfnamefont {A.~C.}\ \bibnamefont {Doherty}}, \ and\
  \bibinfo {author} {\bibfnamefont {S.~T.}\ \bibnamefont {Flammia}},\ }\href
  {\doibase 10.1103/PhysRevLett.117.170502} {\bibfield  {journal} {\bibinfo
  {journal} {Phys. Rev. Lett.}\ }\textbf {\bibinfo {volume} {117}},\ \bibinfo
  {pages} {170502} (\bibinfo {year} {2016})}\BibitemShut {NoStop}%
\bibitem [{\citenamefont {Huang}\ \emph {et~al.}(2019)\citenamefont {Huang},
  \citenamefont {Doherty},\ and\ \citenamefont {Flammia}}]{Huang2019-bh}%
  \BibitemOpen
  \bibfield  {author} {\bibinfo {author} {\bibfnamefont {E.}~\bibnamefont
  {Huang}}, \bibinfo {author} {\bibfnamefont {A.~C.}\ \bibnamefont {Doherty}},
  \ and\ \bibinfo {author} {\bibfnamefont {S.}~\bibnamefont {Flammia}},\ }\href
  {\doibase 10.1103/PhysRevA.99.022313} {\bibfield  {journal} {\bibinfo
  {journal} {Phys. Rev. A}\ }\textbf {\bibinfo {volume} {99}},\ \bibinfo
  {pages} {022313} (\bibinfo {year} {2019})}\BibitemShut {NoStop}%
\bibitem [{\citenamefont {Rudinger}\ \emph {et~al.}(2019)\citenamefont
  {Rudinger}, \citenamefont {Proctor}, \citenamefont {Langharst}, \citenamefont
  {Sarovar}, \citenamefont {Young},\ and\ \citenamefont
  {Blume-Kohout}}]{rudinger2019probing}%
  \BibitemOpen
  \bibfield  {author} {\bibinfo {author} {\bibfnamefont {K.}~\bibnamefont
  {Rudinger}}, \bibinfo {author} {\bibfnamefont {T.}~\bibnamefont {Proctor}},
  \bibinfo {author} {\bibfnamefont {D.}~\bibnamefont {Langharst}}, \bibinfo
  {author} {\bibfnamefont {M.}~\bibnamefont {Sarovar}}, \bibinfo {author}
  {\bibfnamefont {K.}~\bibnamefont {Young}}, \ and\ \bibinfo {author}
  {\bibfnamefont {R.}~\bibnamefont {Blume-Kohout}},\ }\href@noop {} {\bibfield
  {journal} {\bibinfo  {journal} {Physical Review X}\ }\textbf {\bibinfo
  {volume} {9}},\ \bibinfo {pages} {021045} (\bibinfo {year}
  {2019})}\BibitemShut {NoStop}%
\bibitem [{\citenamefont {Dumitrescu}\ and\ \citenamefont
  {Humble}(2016)}]{Dumitrescu2016-so}%
  \BibitemOpen
  \bibfield  {author} {\bibinfo {author} {\bibfnamefont {E.}~\bibnamefont
  {Dumitrescu}}\ and\ \bibinfo {author} {\bibfnamefont {T.~S.}\ \bibnamefont
  {Humble}},\ }\href {\doibase 10.1103/PhysRevA.94.042107} {\bibfield
  {journal} {\bibinfo  {journal} {Phys. Rev. A}\ }\textbf {\bibinfo {volume}
  {94}},\ \bibinfo {pages} {042107} (\bibinfo {year} {2016})},\ \Eprint
  {http://arxiv.org/abs/1607.05292} {arXiv:1607.05292 [quant-ph]} \BibitemShut
  {NoStop}%
\bibitem [{\citenamefont {Mohseni}\ \emph {et~al.}(2008)\citenamefont
  {Mohseni}, \citenamefont {Rezakhani},\ and\ \citenamefont
  {Lidar}}]{mohseni2008quantum}%
  \BibitemOpen
  \bibfield  {author} {\bibinfo {author} {\bibfnamefont {M.}~\bibnamefont
  {Mohseni}}, \bibinfo {author} {\bibfnamefont {A.}~\bibnamefont {Rezakhani}},
  \ and\ \bibinfo {author} {\bibfnamefont {D.}~\bibnamefont {Lidar}},\
  }\href@noop {} {\bibfield  {journal} {\bibinfo  {journal} {Physical Review
  A}\ }\textbf {\bibinfo {volume} {77}},\ \bibinfo {pages} {032322} (\bibinfo
  {year} {2008})}\BibitemShut {NoStop}%
\bibitem [{\citenamefont {Merkel}\ \emph {et~al.}(2013)\citenamefont {Merkel},
  \citenamefont {Gambetta}, \citenamefont {Smolin}, \citenamefont {Poletto},
  \citenamefont {C{\'o}rcoles}, \citenamefont {Johnson}, \citenamefont {Ryan},\
  and\ \citenamefont {Steffen}}]{merkel2013self}%
  \BibitemOpen
  \bibfield  {author} {\bibinfo {author} {\bibfnamefont {S.~T.}\ \bibnamefont
  {Merkel}}, \bibinfo {author} {\bibfnamefont {J.~M.}\ \bibnamefont
  {Gambetta}}, \bibinfo {author} {\bibfnamefont {J.~A.}\ \bibnamefont
  {Smolin}}, \bibinfo {author} {\bibfnamefont {S.}~\bibnamefont {Poletto}},
  \bibinfo {author} {\bibfnamefont {A.~D.}\ \bibnamefont {C{\'o}rcoles}},
  \bibinfo {author} {\bibfnamefont {B.~R.}\ \bibnamefont {Johnson}}, \bibinfo
  {author} {\bibfnamefont {C.~A.}\ \bibnamefont {Ryan}}, \ and\ \bibinfo
  {author} {\bibfnamefont {M.}~\bibnamefont {Steffen}},\ }\href@noop {}
  {\bibfield  {journal} {\bibinfo  {journal} {Physical Review A}\ }\textbf
  {\bibinfo {volume} {87}},\ \bibinfo {pages} {062119} (\bibinfo {year}
  {2013})}\BibitemShut {NoStop}%
\bibitem [{\citenamefont {Solgun}\ \emph {et~al.}(2019)\citenamefont {Solgun},
  \citenamefont {DiVincenzo},\ and\ \citenamefont
  {Gambetta}}]{solgun2019simple}%
  \BibitemOpen
  \bibfield  {author} {\bibinfo {author} {\bibfnamefont {F.}~\bibnamefont
  {Solgun}}, \bibinfo {author} {\bibfnamefont {D.~P.}\ \bibnamefont
  {DiVincenzo}}, \ and\ \bibinfo {author} {\bibfnamefont {J.~M.}\ \bibnamefont
  {Gambetta}},\ }\href@noop {} {\bibfield  {journal} {\bibinfo  {journal} {IEEE
  transactions on microwave theory and techniques}\ }\textbf {\bibinfo {volume}
  {67}},\ \bibinfo {pages} {928} (\bibinfo {year} {2019})}\BibitemShut
  {NoStop}%
\bibitem [{\citenamefont {Nielsen}\ \emph
  {et~al.}(2020{\natexlab{a}})\citenamefont {Nielsen}, \citenamefont {Gamble},
  \citenamefont {Rudinger}, \citenamefont {Scholten}, \citenamefont {Young},\
  and\ \citenamefont {Blume-Kohout}}]{Nielsen2020-th}%
  \BibitemOpen
  \bibfield  {author} {\bibinfo {author} {\bibfnamefont {E.}~\bibnamefont
  {Nielsen}}, \bibinfo {author} {\bibfnamefont {J.~K.}\ \bibnamefont {Gamble}},
  \bibinfo {author} {\bibfnamefont {K.}~\bibnamefont {Rudinger}}, \bibinfo
  {author} {\bibfnamefont {T.}~\bibnamefont {Scholten}}, \bibinfo {author}
  {\bibfnamefont {K.}~\bibnamefont {Young}}, \ and\ \bibinfo {author}
  {\bibfnamefont {R.}~\bibnamefont {Blume-Kohout}},\ }\href@noop {} {\bibfield
  {journal} {\bibinfo  {journal} {arXiv}\ } (\bibinfo {year}
  {2020}{\natexlab{a}})},\ \Eprint {http://arxiv.org/abs/2009.07301}
  {arXiv:2009.07301 [quant-ph]} \BibitemShut {NoStop}%
\bibitem [{\citenamefont {Nielsen}\ \emph {et~al.}(2021)\citenamefont
  {Nielsen}, \citenamefont {Rudinger}, \citenamefont {Proctor}, \citenamefont
  {Young},\ and\ \citenamefont {Blume-Kohout}}]{Nielsen2021-Onion}%
  \BibitemOpen
  \bibfield  {author} {\bibinfo {author} {\bibfnamefont {E.}~\bibnamefont
  {Nielsen}}, \bibinfo {author} {\bibfnamefont {K.}~\bibnamefont {Rudinger}},
  \bibinfo {author} {\bibfnamefont {T.}~\bibnamefont {Proctor}}, \bibinfo
  {author} {\bibfnamefont {K.}~\bibnamefont {Young}}, \ and\ \bibinfo {author}
  {\bibfnamefont {R.}~\bibnamefont {Blume-Kohout}},\ }\href@noop {} {\
  (\bibinfo {year} {2021})},\ \Eprint {http://arxiv.org/abs/2103.02188}
  {arXiv:2103.02188 [quant-ph]} \BibitemShut {NoStop}%
\bibitem [{AQT()}]{AQT-wp}%
  \BibitemOpen
  \href@noop {} {\enquote {\bibinfo {title} {{AQT@LBL - SC Qubit Testbed}},}\
  }\bibinfo {howpublished} {\url{https://aqt.lbl.gov/}},\ \bibinfo {note}
  {accessed: 2020-03-01}\BibitemShut {NoStop}%
\bibitem [{\citenamefont {Clark}\ \emph {et~al.}(2020)\citenamefont {Clark},
  \citenamefont {Hogle}, \citenamefont {Young},\ and\ \citenamefont
  {Stick}}]{Clark2020-jr}%
  \BibitemOpen
  \bibfield  {author} {\bibinfo {author} {\bibfnamefont {S.~M.}\ \bibnamefont
  {Clark}}, \bibinfo {author} {\bibfnamefont {C.~W.}\ \bibnamefont {Hogle}},
  \bibinfo {author} {\bibfnamefont {K.}~\bibnamefont {Young}}, \ and\ \bibinfo
  {author} {\bibfnamefont {D.~L.}\ \bibnamefont {Stick}},\ }\href {\doibase
  10.2172/1670515} {\enquote {\bibinfo {title} {Demonstrating robustness of
  analogue quantum simulators},}\ } (\bibinfo {year} {2020})\BibitemShut
  {NoStop}%
\bibitem [{QSC()}]{QSCOUT-wp}%
  \BibitemOpen
  \href@noop {} {\enquote {\bibinfo {title} {{Quantum Scientific Computing Open
  User Testbed (QSCOUT)}},}\ }\bibinfo {howpublished}
  {\url{https://www.sandia.gov/quantum/Projects/QSCOUT.html}},\ \bibinfo {note}
  {accessed: 2020-03-01}\BibitemShut {NoStop}%
\bibitem [{\citenamefont {Nielsen}\ \emph {et~al.}(2019)\citenamefont
  {Nielsen}, \citenamefont {Blume-Kohout}, \citenamefont {Rudinger},
  \citenamefont {Proctor}, \citenamefont {Saldyt} \emph
  {et~al.}}]{nielsen2019python}%
  \BibitemOpen
  \bibfield  {author} {\bibinfo {author} {\bibfnamefont {E.}~\bibnamefont
  {Nielsen}}, \bibinfo {author} {\bibfnamefont {R.~J.}\ \bibnamefont
  {Blume-Kohout}}, \bibinfo {author} {\bibfnamefont {K.~M.}\ \bibnamefont
  {Rudinger}}, \bibinfo {author} {\bibfnamefont {T.~J.}\ \bibnamefont
  {Proctor}}, \bibinfo {author} {\bibfnamefont {L.}~\bibnamefont {Saldyt}},
  \emph {et~al.},\ }\href@noop {} {\emph {\bibinfo {title} {Python GST
  Implementation (PyGSTi) v. 0.9}}},\ \bibinfo {type} {Tech. Rep.}\ (\bibinfo
  {institution} {Sandia National Lab.(SNL-NM), Albuquerque, NM (United
  States)},\ \bibinfo {year} {2019})\BibitemShut {NoStop}%
\bibitem [{\citenamefont {Nielsen}\ \emph
  {et~al.}(2020{\natexlab{b}})\citenamefont {Nielsen}, \citenamefont
  {Rudinger}, \citenamefont {Proctor}, \citenamefont {Russo}, \citenamefont
  {Young},\ and\ \citenamefont {Blume-Kohout}}]{nielsen2020probing}%
  \BibitemOpen
  \bibfield  {author} {\bibinfo {author} {\bibfnamefont {E.}~\bibnamefont
  {Nielsen}}, \bibinfo {author} {\bibfnamefont {K.}~\bibnamefont {Rudinger}},
  \bibinfo {author} {\bibfnamefont {T.}~\bibnamefont {Proctor}}, \bibinfo
  {author} {\bibfnamefont {A.}~\bibnamefont {Russo}}, \bibinfo {author}
  {\bibfnamefont {K.}~\bibnamefont {Young}}, \ and\ \bibinfo {author}
  {\bibfnamefont {R.}~\bibnamefont {Blume-Kohout}},\ }\href@noop {} {\bibfield
  {journal} {\bibinfo  {journal} {Quantum Science and Technology}\ }\textbf
  {\bibinfo {volume} {5}},\ \bibinfo {pages} {044002} (\bibinfo {year}
  {2020}{\natexlab{b}})}\BibitemShut {NoStop}%
\bibitem [{\citenamefont {Wilks}(1938)}]{wilks1938large}%
  \BibitemOpen
  \bibfield  {author} {\bibinfo {author} {\bibfnamefont {S.~S.}\ \bibnamefont
  {Wilks}},\ }\href@noop {} {\bibfield  {journal} {\bibinfo  {journal} {The
  annals of mathematical statistics}\ }\textbf {\bibinfo {volume} {9}},\
  \bibinfo {pages} {60} (\bibinfo {year} {1938})}\BibitemShut {NoStop}%
\bibitem [{\citenamefont {Akaike}(1974)}]{Akaike1974-xt}%
  \BibitemOpen
  \bibfield  {author} {\bibinfo {author} {\bibfnamefont {H.}~\bibnamefont
  {Akaike}},\ }\href {\doibase 10.1109/TAC.1974.1100705} {\bibfield  {journal}
  {\bibinfo  {journal} {IEEE Trans. Automat. Contr.}\ }\textbf {\bibinfo
  {volume} {19}},\ \bibinfo {pages} {716} (\bibinfo {year} {1974})}\BibitemShut
  {NoStop}%
\bibitem [{\citenamefont {Blume-Kohout}\ \emph {et~al.}(2020)\citenamefont
  {Blume-Kohout}, \citenamefont {Rudinger}, \citenamefont {Nielsen},
  \citenamefont {Proctor},\ and\ \citenamefont {Young}}]{blume2020wildcard}%
  \BibitemOpen
  \bibfield  {author} {\bibinfo {author} {\bibfnamefont {R.}~\bibnamefont
  {Blume-Kohout}}, \bibinfo {author} {\bibfnamefont {K.}~\bibnamefont
  {Rudinger}}, \bibinfo {author} {\bibfnamefont {E.}~\bibnamefont {Nielsen}},
  \bibinfo {author} {\bibfnamefont {T.}~\bibnamefont {Proctor}}, \ and\
  \bibinfo {author} {\bibfnamefont {K.}~\bibnamefont {Young}},\ }\href@noop {}
  {\bibfield  {journal} {\bibinfo  {journal} {arXiv preprint arXiv:2012.12231}\
  } (\bibinfo {year} {2020})}\BibitemShut {NoStop}%
\bibitem [{\citenamefont {Kitaev}(1997)}]{kitaev1997quantum}%
  \BibitemOpen
  \bibfield  {author} {\bibinfo {author} {\bibfnamefont {A.~Y.}\ \bibnamefont
  {Kitaev}},\ }\href@noop {} {\bibfield  {journal} {\bibinfo  {journal}
  {Uspekhi Matematicheskikh Nauk}\ }\textbf {\bibinfo {volume} {52}},\ \bibinfo
  {pages} {53} (\bibinfo {year} {1997})}\BibitemShut {NoStop}%
\bibitem [{\citenamefont {Watrous}(2009)}]{watrous2009semidefinite}%
  \BibitemOpen
  \bibfield  {author} {\bibinfo {author} {\bibfnamefont {J.}~\bibnamefont
  {Watrous}},\ }\href@noop {} {\bibfield  {journal} {\bibinfo  {journal} {arXiv
  preprint arXiv:0901.4709}\ } (\bibinfo {year} {2009})}\BibitemShut {NoStop}%
\bibitem [{\citenamefont {Sanders}\ \emph {et~al.}(2015)\citenamefont
  {Sanders}, \citenamefont {Wallman},\ and\ \citenamefont
  {Sanders}}]{sanders2015bounding}%
  \BibitemOpen
  \bibfield  {author} {\bibinfo {author} {\bibfnamefont {Y.~R.}\ \bibnamefont
  {Sanders}}, \bibinfo {author} {\bibfnamefont {J.~J.}\ \bibnamefont
  {Wallman}}, \ and\ \bibinfo {author} {\bibfnamefont {B.~C.}\ \bibnamefont
  {Sanders}},\ }\href@noop {} {\bibfield  {journal} {\bibinfo  {journal} {New
  Journal of Physics}\ }\textbf {\bibinfo {volume} {18}},\ \bibinfo {pages}
  {012002} (\bibinfo {year} {2015})}\BibitemShut {NoStop}%
\bibitem [{\citenamefont {Blume-Kohout}\ \emph {et~al.}(2021)\citenamefont
  {Blume-Kohout}, \citenamefont {da~Silva}, \citenamefont {Nielsen},
  \citenamefont {Proctor}, \citenamefont {Rudinger}, \citenamefont {Sarovar},\
  and\ \citenamefont {Young}}]{Blume-Kohout2021-Taxonomy}%
  \BibitemOpen
  \bibfield  {author} {\bibinfo {author} {\bibfnamefont {R.}~\bibnamefont
  {Blume-Kohout}}, \bibinfo {author} {\bibfnamefont {M.~P.}\ \bibnamefont
  {da~Silva}}, \bibinfo {author} {\bibfnamefont {E.}~\bibnamefont {Nielsen}},
  \bibinfo {author} {\bibfnamefont {T.}~\bibnamefont {Proctor}}, \bibinfo
  {author} {\bibfnamefont {K.}~\bibnamefont {Rudinger}}, \bibinfo {author}
  {\bibfnamefont {M.}~\bibnamefont {Sarovar}}, \ and\ \bibinfo {author}
  {\bibfnamefont {K.}~\bibnamefont {Young}},\ }\href@noop {} {\  (\bibinfo
  {year} {2021})},\ \Eprint {http://arxiv.org/abs/2103.01928} {arXiv:2103.01928
  [quant-ph]} \BibitemShut {NoStop}%
\bibitem [{\citenamefont {Rudnicki}\ \emph {et~al.}(2018)\citenamefont
  {Rudnicki}, \citenamefont {Pucha{\l}a},\ and\ \citenamefont
  {Zyczkowski}}]{Rudnicki2018-bn}%
  \BibitemOpen
  \bibfield  {author} {\bibinfo {author} {\bibfnamefont {{\L}.}~\bibnamefont
  {Rudnicki}}, \bibinfo {author} {\bibfnamefont {Z.}~\bibnamefont
  {Pucha{\l}a}}, \ and\ \bibinfo {author} {\bibfnamefont {K.}~\bibnamefont
  {Zyczkowski}},\ }\href@noop {} {\bibfield  {journal} {\bibinfo  {journal}
  {Quantum}\ }\textbf {\bibinfo {volume} {2}},\ \bibinfo {pages} {60} (\bibinfo
  {year} {2018})}\BibitemShut {NoStop}%
\bibitem [{\citenamefont {Lin}\ \emph {et~al.}(2019)\citenamefont {Lin},
  \citenamefont {Buonacorsi}, \citenamefont {Laflamme},\ and\ \citenamefont
  {Wallman}}]{Lin2019-ck}%
  \BibitemOpen
  \bibfield  {author} {\bibinfo {author} {\bibfnamefont {J.}~\bibnamefont
  {Lin}}, \bibinfo {author} {\bibfnamefont {B.}~\bibnamefont {Buonacorsi}},
  \bibinfo {author} {\bibfnamefont {R.}~\bibnamefont {Laflamme}}, \ and\
  \bibinfo {author} {\bibfnamefont {J.~J.}\ \bibnamefont {Wallman}},\ }\href
  {\doibase 10.1088/1367-2630/ab075a} {\bibfield  {journal} {\bibinfo
  {journal} {New J. Phys.}\ }\textbf {\bibinfo {volume} {21}},\ \bibinfo
  {pages} {023006} (\bibinfo {year} {2019})}\BibitemShut {NoStop}%
\bibitem [{\citenamefont {Magesan}\ \emph {et~al.}(2011)\citenamefont
  {Magesan}, \citenamefont {Gambetta},\ and\ \citenamefont
  {Emerson}}]{magesan2011scalable}%
  \BibitemOpen
  \bibfield  {author} {\bibinfo {author} {\bibfnamefont {E.}~\bibnamefont
  {Magesan}}, \bibinfo {author} {\bibfnamefont {J.~M.}\ \bibnamefont
  {Gambetta}}, \ and\ \bibinfo {author} {\bibfnamefont {J.}~\bibnamefont
  {Emerson}},\ }\href@noop {} {\bibfield  {journal} {\bibinfo  {journal}
  {Physical review letters}\ }\textbf {\bibinfo {volume} {106}},\ \bibinfo
  {pages} {180504} (\bibinfo {year} {2011})}\BibitemShut {NoStop}%
\bibitem [{\citenamefont {Proctor}\ \emph {et~al.}(2019)\citenamefont
  {Proctor}, \citenamefont {Carignan-Dugas}, \citenamefont {Rudinger},
  \citenamefont {Nielsen}, \citenamefont {Blume-Kohout},\ and\ \citenamefont
  {Young}}]{proctor2019direct}%
  \BibitemOpen
  \bibfield  {author} {\bibinfo {author} {\bibfnamefont {T.~J.}\ \bibnamefont
  {Proctor}}, \bibinfo {author} {\bibfnamefont {A.}~\bibnamefont
  {Carignan-Dugas}}, \bibinfo {author} {\bibfnamefont {K.}~\bibnamefont
  {Rudinger}}, \bibinfo {author} {\bibfnamefont {E.}~\bibnamefont {Nielsen}},
  \bibinfo {author} {\bibfnamefont {R.}~\bibnamefont {Blume-Kohout}}, \ and\
  \bibinfo {author} {\bibfnamefont {K.}~\bibnamefont {Young}},\ }\href@noop {}
  {\bibfield  {journal} {\bibinfo  {journal} {Physical review letters}\
  }\textbf {\bibinfo {volume} {123}},\ \bibinfo {pages} {030503} (\bibinfo
  {year} {2019})}\BibitemShut {NoStop}%
\bibitem [{\citenamefont {McKay}\ \emph {et~al.}(2017)\citenamefont {McKay},
  \citenamefont {Wood}, \citenamefont {Sheldon}, \citenamefont {Chow},\ and\
  \citenamefont {Gambetta}}]{McKay2017-fq}%
  \BibitemOpen
  \bibfield  {author} {\bibinfo {author} {\bibfnamefont {D.~C.}\ \bibnamefont
  {McKay}}, \bibinfo {author} {\bibfnamefont {C.~J.}\ \bibnamefont {Wood}},
  \bibinfo {author} {\bibfnamefont {S.}~\bibnamefont {Sheldon}}, \bibinfo
  {author} {\bibfnamefont {J.~M.}\ \bibnamefont {Chow}}, \ and\ \bibinfo
  {author} {\bibfnamefont {J.~M.}\ \bibnamefont {Gambetta}},\ }\href {\doibase
  10.1103/PhysRevA.96.022330} {\bibfield  {journal} {\bibinfo  {journal} {Phys.
  Rev. A}\ }\textbf {\bibinfo {volume} {96}},\ \bibinfo {pages} {022330}
  (\bibinfo {year} {2017})}\BibitemShut {NoStop}%
\bibitem [{\citenamefont {Patterson}\ \emph {et~al.}(2019)\citenamefont
  {Patterson}, \citenamefont {Rahamim}, \citenamefont {Tsunoda}, \citenamefont
  {Spring}, \citenamefont {Jebari}, \citenamefont {Ratter}, \citenamefont
  {Mergenthaler}, \citenamefont {Tancredi}, \citenamefont {Vlastakis},
  \citenamefont {Esposito} \emph {et~al.}}]{patterson2019calibration}%
  \BibitemOpen
  \bibfield  {author} {\bibinfo {author} {\bibfnamefont {A.}~\bibnamefont
  {Patterson}}, \bibinfo {author} {\bibfnamefont {J.}~\bibnamefont {Rahamim}},
  \bibinfo {author} {\bibfnamefont {T.}~\bibnamefont {Tsunoda}}, \bibinfo
  {author} {\bibfnamefont {P.}~\bibnamefont {Spring}}, \bibinfo {author}
  {\bibfnamefont {S.}~\bibnamefont {Jebari}}, \bibinfo {author} {\bibfnamefont
  {K.}~\bibnamefont {Ratter}}, \bibinfo {author} {\bibfnamefont
  {M.}~\bibnamefont {Mergenthaler}}, \bibinfo {author} {\bibfnamefont
  {G.}~\bibnamefont {Tancredi}}, \bibinfo {author} {\bibfnamefont
  {B.}~\bibnamefont {Vlastakis}}, \bibinfo {author} {\bibfnamefont
  {M.}~\bibnamefont {Esposito}},  \emph {et~al.},\ }\href@noop {} {\bibfield
  {journal} {\bibinfo  {journal} {Physical Review Applied}\ }\textbf {\bibinfo
  {volume} {12}},\ \bibinfo {pages} {064013} (\bibinfo {year}
  {2019})}\BibitemShut {NoStop}%
\bibitem [{\citenamefont {Olmschenk}\ \emph {et~al.}(2007)\citenamefont
  {Olmschenk}, \citenamefont {Younge}, \citenamefont {Moehring}, \citenamefont
  {Matsukevich}, \citenamefont {Maunz},\ and\ \citenamefont
  {Monroe}}]{Olmschenk2007}%
  \BibitemOpen
  \bibfield  {author} {\bibinfo {author} {\bibfnamefont {S.}~\bibnamefont
  {Olmschenk}}, \bibinfo {author} {\bibfnamefont {K.~C.}\ \bibnamefont
  {Younge}}, \bibinfo {author} {\bibfnamefont {D.~L.}\ \bibnamefont
  {Moehring}}, \bibinfo {author} {\bibfnamefont {D.~N.}\ \bibnamefont
  {Matsukevich}}, \bibinfo {author} {\bibfnamefont {P.}~\bibnamefont {Maunz}},
  \ and\ \bibinfo {author} {\bibfnamefont {C.}~\bibnamefont {Monroe}},\
  }\href@noop {} {\bibfield  {journal} {\bibinfo  {journal} {Phys. Rev. A}\
  }\textbf {\bibinfo {volume} {76}},\ \bibinfo {pages} {052314} (\bibinfo
  {year} {2007})}\BibitemShut {NoStop}%
\bibitem [{\citenamefont {Maunz}(2016)}]{Maunz2016}%
  \BibitemOpen
  \bibfield  {author} {\bibinfo {author} {\bibfnamefont {P.}~\bibnamefont
  {Maunz}},\ }\href@noop {} {\emph {\bibinfo {title} {High Optical Access Trap
  2.0}}},\ \bibinfo {type} {Tech. Rep.}\ \bibinfo {number} {SAND2016-0796R}\
  (\bibinfo  {institution} {Sandia National Laboratories},\ \bibinfo {address}
  {Albuquerque, NM},\ \bibinfo {year} {2016})\BibitemShut {NoStop}%
\bibitem [{\citenamefont {Islam}\ \emph {et~al.}(2014)\citenamefont {Islam},
  \citenamefont {Campbell}, \citenamefont {Choi}, \citenamefont {Clark},
  \citenamefont {Conover}, \citenamefont {Debnath}, \citenamefont {Edwards},
  \citenamefont {Fields}, \citenamefont {Hayes}, \citenamefont {Hucul},
  \citenamefont {Inlek}, \citenamefont {Johnson}, \citenamefont {Korenblit},
  \citenamefont {Lee}, \citenamefont {Lee}, \citenamefont {Manning},
  \citenamefont {Matsukevich}, \citenamefont {Mizrahi}, \citenamefont
  {Quraishi}, \citenamefont {Senko}, \citenamefont {Smith},\ and\ \citenamefont
  {Monroe}}]{Islam2014}%
  \BibitemOpen
  \bibfield  {author} {\bibinfo {author} {\bibfnamefont {R.}~\bibnamefont
  {Islam}}, \bibinfo {author} {\bibfnamefont {W.~C.}\ \bibnamefont {Campbell}},
  \bibinfo {author} {\bibfnamefont {T.}~\bibnamefont {Choi}}, \bibinfo {author}
  {\bibfnamefont {S.~M.}\ \bibnamefont {Clark}}, \bibinfo {author}
  {\bibfnamefont {C.~W.~S.}\ \bibnamefont {Conover}}, \bibinfo {author}
  {\bibfnamefont {S.}~\bibnamefont {Debnath}}, \bibinfo {author} {\bibfnamefont
  {E.~E.}\ \bibnamefont {Edwards}}, \bibinfo {author} {\bibfnamefont
  {B.}~\bibnamefont {Fields}}, \bibinfo {author} {\bibfnamefont
  {D.}~\bibnamefont {Hayes}}, \bibinfo {author} {\bibfnamefont
  {D.}~\bibnamefont {Hucul}}, \bibinfo {author} {\bibfnamefont {I.~V.}\
  \bibnamefont {Inlek}}, \bibinfo {author} {\bibfnamefont {K.~G.}\ \bibnamefont
  {Johnson}}, \bibinfo {author} {\bibfnamefont {S.}~\bibnamefont {Korenblit}},
  \bibinfo {author} {\bibfnamefont {A.}~\bibnamefont {Lee}}, \bibinfo {author}
  {\bibfnamefont {K.~W.}\ \bibnamefont {Lee}}, \bibinfo {author} {\bibfnamefont
  {T.~A.}\ \bibnamefont {Manning}}, \bibinfo {author} {\bibfnamefont {D.~N.}\
  \bibnamefont {Matsukevich}}, \bibinfo {author} {\bibfnamefont
  {J.}~\bibnamefont {Mizrahi}}, \bibinfo {author} {\bibfnamefont
  {Q.}~\bibnamefont {Quraishi}}, \bibinfo {author} {\bibfnamefont
  {C.}~\bibnamefont {Senko}}, \bibinfo {author} {\bibfnamefont
  {J.}~\bibnamefont {Smith}}, \ and\ \bibinfo {author} {\bibfnamefont
  {C.}~\bibnamefont {Monroe}},\ }\href {\doibase 10.1364/OL.39.003238}
  {\bibfield  {journal} {\bibinfo  {journal} {Opt. Lett.}\ }\textbf {\bibinfo
  {volume} {39}},\ \bibinfo {pages} {3238} (\bibinfo {year}
  {2014})}\BibitemShut {NoStop}%
\bibitem [{\citenamefont {Wimperis}(1994)}]{Wimperis1994-fd}%
  \BibitemOpen
  \bibfield  {author} {\bibinfo {author} {\bibfnamefont {S.}~\bibnamefont
  {Wimperis}},\ }\href {\doibase 10.1006/jmra.1994.1159} {\bibfield  {journal}
  {\bibinfo  {journal} {J. Magn. Reson. A}\ }\textbf {\bibinfo {volume}
  {109}},\ \bibinfo {pages} {221} (\bibinfo {year} {1994})}\BibitemShut
  {NoStop}%
\bibitem [{\citenamefont {Nielsen}(2020)}]{nielsen2020efficient}%
  \BibitemOpen
  \bibfield  {author} {\bibinfo {author} {\bibfnamefont {E.}~\bibnamefont
  {Nielsen}},\ }\href@noop {} {\emph {\bibinfo {title} {Efficient Scalable
  Tomography of Many-Qubit Quantum Processors.}}},\ \bibinfo {type} {Tech.
  Rep.}\ (\bibinfo  {institution} {Sandia National Lab.(SNL-NM), Albuquerque,
  NM (United States), url={https://www.osti.gov/servlets/purl/1673168}},\
  \bibinfo {year} {2020})\BibitemShut {NoStop}%
\end{thebibliography}%
\end{document}